\documentclass[sn-mathphys,Numbered,iicol]{sn-jnl}

\usepackage{graphicx}%
\usepackage{multirow}%
\usepackage{amsmath,amssymb,amsfonts}%
\usepackage{amsthm}%
\usepackage{mathrsfs}%
\usepackage[title]{appendix}%
\usepackage{xcolor}%
\usepackage{textcomp}%
\usepackage{manyfoot}%
\usepackage{booktabs}%

\usepackage{subcaption}

\newcommand{\beqy}{\begin{eqnarray}}
\newcommand{\eeqy}{\end{eqnarray}}

\newcommand{\Dn}{\Delta_n}
\newcommand{\Tbase}{T_{\rm b}}
\newcommand{\Tcore}{T_{\rm core}}
\newcommand{\Qimp}{Q_{\rm imp}}
\newcommand{\Vn}{\mathbb{V}_n}
\newcommand{\VLn}{\mathbb{V}_{Ln}}
\newcommand{\Vcn}{\mathbb{V}^{(0)}_{cn}}
\newcommand{\Tcn}{T_{cn}^{(0)}}
\newcommand{\Singlet}{$^1S_0$ }
\newcommand{\rb}{\pmb{r}}
\newcommand{\Euler}{\text{e}}
\newcommand{\DoSn}{\mathscr{D}_n\left(\mathcal{E},\Delta_n(T,\mathbb{V}_n)\right)}
\newcommand{\QuasipartEnergyn}{\mathfrak{E}_{\pmb{k}}^{(n)}}
\newcommand{\DNn}{\mathscr{D}_N^{(n)}(0)}
\newcommand{\mnOplus}{m_n^{\oplus}}
\newcommand{\sech}{{\rm sech}}
\newcommand{\Cnormn}{c_N^{(n)}}
\newcommand{\expgamma}{\text{e}^{\gamma}}
\newcommand{\MNS}{{\rm M}_{\rm NS}}
\newcommand{\RNS}{{\rm R}_{\rm NS}}
\newcommand{\MSol}{{\rm M}_{\odot}}
\newcommand{\mEdd}{\dot{m}_{\rm Edd}}

\def\apj{Astrophys.~J.}  
\def\apjl{Astrophys.~J.~Lett.} 
\def\apjs{Astrophys.~J.~Suppl.~S.} 
\def\apss{Astrophys.~Space~Sci.} 
\def\prl{Phys.~Rev.~Lett.} 
\def\prb{Phys.~Rev.~B} 
\def\prc{Phys.~Rev.~C} 
\def\prd{Phys.~Rev.~D} 
\def\aap{Astron.~Astrophys.}   
\def\nphysa{Nucl.~Phys.~A}   
\def\mnras{Mon.~Not.~R.~Astron.~Soc.}             
\def\epja{Eur.~Phys.~J.~A}
\def\jaa{J.~Astrophys.~Astron.} 
\def\nat{Nature.} 
\def\arep{Astron.~Rep.} 
\def\currsc{Curr.~Sci.~India} 
\def\revmodphys{Rev.~Mod.~Phys.} 
\def\japa{J.~Astrophys.~Astron.}  
\def\RepProgPhys{Rep.~Prog.~Phys.} 

\begin{document}

\title[Article Title]{Gapless neutron superfluidity in the crust of the accreting neutron stars KS~1731$-$260 and MXB~1659$-$29}

\author[1]{\fnm{Valentin} \sur{Allard}}\email{valentin.allard@ulb.be}

\author*[1]{\fnm{Nicolas} \sur{Chamel}}\email{nicolas.chamel@ulb.be}

\affil[1]{\orgdiv{Institute of Astronomy and Astrophysics}, \orgname{Universit\'e Libre de Bruxelles}, \orgaddress{\street{Boulevard du Triomphe}, \city{Brussels}, \postcode{B-1050}, \country{Belgium}}}

\abstract{The interpretation of the thermal evolution of the transiently accreting neutron stars MXB~1659$-$29 and 
KS~1731$-$260 after an outburst is challenging, both within the traditional deep-crustal heating paradigm and the 
thermodynamically consistent approach of Gusakov and Chugunov  that accounts for neutron diffusion throughout the
crust. All these studies assume that the neutron superfluid in the crust is at rest. However, we have recently shown 
that a finite superflow could exist and could lead to a new gapless superfluid phase if quantized vortices are pinned. 
We have revisited the cooling of MXB~1659$-$29 and KS~1731$-$260 and we have found that gapless superfluidity could 
naturally explain their late time cooling. We pursue here our investigation by performing new simulations of the thermal 
relaxation of the crust of MXB~1659$-$29 and KS~1731$-$260 within a Markov Chain Monte Carlo method accounting for 
neutron diffusion and allowing for gapless superfluidity. We have varied the global structure of the neutron star, the composition of the heat-blanketing envelope, and the mass accretion rate. In all cases, observations are best fitted by models with gapless 
superfluidity. Finally, we make predictions that could be tested by future observations.}


\maketitle

\section{Introduction}\label{sec:introduction}

Quasipersistent soft X-ray transients (SXTs) are neutron stars episodically accreting material from a stellar companion in 
a low-mass X-ray binary~\cite{Bahramian2023}. Accretion outbursts can last years to decades, a time long enough for the crust to be heated 
out of thermal equilibrium with the core due to compression-induced nuclear reactions, mainly pycnonuclear reactions taking 
place in the deep layers of the inner crust at densities of order $10^{12}-10^{13}$~g~cm$^{-3}$~\cite{HaenselZdunik1990} (for 
this reason, this process is usually referred to as ``deep crustal heating''~\cite{Brown1998}). In quiescence, heat is 
transported to the surface thus cooling down the neutron star until the crust-core thermal equilibrium is restored. This 
leads to a soft component in the X-ray spectrum (below a few keV), first detected in KS~1731$-$260~\cite{Wijnands2001} in 
March 2001, a few months after the end of an outburst of 12.5 years. Subsequent observations~\cite{wijnands2002,cackett2006} 
were consistent with the expected thermal emission from the neutron star surface~\cite{rutledge2002,shternin2007,BrownCumming2009}. 
However, the star appeared colder than predicted when it was observed again in May 2009~\cite{Cackett2010}. Observations taken 
six years later suggested that the crust had thermally relaxed~\cite{Merritt2016}. Observations are summarized in Tab.~\ref{KSPoints}. The second source for which the thermal 
emission was detected is MXB~1659$-$29. This source went into quiescence also in 2001 after an accretion period of 2.5 years
and was regularly observed~\cite{Wijnands2003,wijnands2004,cackett2006,cackett2008,Cackett2013} (see Tab.~\ref{MXBPoints}) before entering into a 
new accretion episode in 2015~\cite{negoro2015,sanchez2015}. This outburst ended in 2017 after 1.7 years and the X-ray emission 
during quiescence was observed up to about 500 days~\cite{Parikh2019} (see Tab.~\ref{MXBPointsII}). At the time of 
this writing, KS~1731$-$260 and MXB~1659$-$29 are still in quiescence. Several other SXTs have been monitored (see, e.g., Ref.~\cite{Wijnands2017} and references therein). These observations directly probe the properties of the inner crust and neutron 
superfluidity~\cite{page2012,Chaikin2018}.

\begin{table*}
\centering
\caption{Inferred effective surface temperature $T_{\rm eff}^{\infty}$ (as seen by an observer at infinity) of KS~1731$-$260 at different times after outburst from Ref.~\cite{Merritt2016}. Time is given in terms of modified Julian date (MJD). The outburst ended at time $t_0 =51930.5$. The temperature and the uncertainties at the 68\% confidence level are given in electronvolts.}
\label{KSPoints}
\begin{tabular}{@{}ccccccccc@{}}
\toprule

\textbf{} &  & Observatory &  & Obs ID          &  & MJD     &  & $k_{\rm B}T_{\rm eff}^{\infty}$ (eV) \\ \midrule
1         &  & Chandra     &  & 2468            &  & 51995.1 &  & 104.6$\pm$1.3     \\
2         &  & XMM-Newton  &  & 013795201/301   &  & 52165.7 &  & 89.5$\pm$1.03     \\
3         &  & Chandra     &  & 3796            &  & 52681.6 &  & 76.4$\pm$1.8      \\
4         &  & Chandra     &  & 3797            &  & 52859.5 &  & 73.8$\pm$1.9      \\
5         &  & XMM-Newton  &  & 0202680101      &  & 53430.5 &  & 71.7$\pm$1.4      \\
6         &  & Chandra     &  & 6279/5486       &  & 53512.9 &  & 70.3$\pm$1.9      \\
7         &  & Chandra     &  & 10037/10911     &  & 54969.7 &  & 64.5$\pm$1.8      \\
8         &  & Chandra     &  & 16734/17706/707 &  & 57242.1 &  & 64.4$\pm$1.2      \\ \bottomrule
\end{tabular}
\end{table*}

\begin{table*}
\centering
\caption{Same as Table~\ref{KSPoints} after the end of the 1999-2001 outburst of MXB~1659$-$29 (referred to as outburst I). The first six points come from  Ref.~\cite{Parikh2019} while the last point is from the best spectral fits of Ref.~\cite{Cackett2013} assuming a fixed hydrogen column-density and considering a neutron-star atmosphere plus power-law model with an index $\Gamma=2$. The outburst ended at time $t_0 =52162$. Uncertainties are given at the 90\% confidence level.}
\label{MXBPoints}
\begin{tabular}{@{}ccccccccc@{}}
\toprule
\textbf{} &  & Observatory &  & Obs ID      &  & MJD     &  & $k_{\rm B}T_{\rm eff}^{\infty}$ (eV) \\ 
\midrule
1         &  & Chandra     &  & 2688        &  & 52197.7 &  & 111.1$\pm$1.3     \\
2         &  & Chandra     &  & 3794        &  & 52563.0 &  & 79.5$\pm$1.6      \\
3         &  & XMM-Newton  &  & 0153190101  &  & 52711.6 &  & 73.0$\pm$1.9      \\
4         &  & Chandra     &  & 3795        &  & 52768.7 &  & 67.8$\pm$2.1      \\
5         &  & Chandra     &  & 5469/6337   &  & 53566.4 &  & 55.5$\pm$2.4      \\
6         &  & Chandra     &  & 8984        &  & 54583.8 &  & 54.8$\pm$3.2      \\
7        &  & Chandra     &  & 13711/14453 &  & 56113   &  & 43.0$\pm$5.0        \\ \bottomrule
\end{tabular}
\end{table*}

\begin{table*}
\centering
\caption{Same as Table~\ref{KSPoints} for MXB~1659$-$29 after the end of the 2015-2017 outburst (referred to as outburst II) using the data from Ref.~\cite{Parikh2019}. The outburst ended at time $t_0 =57809.7$. Uncertainties are given at the 90\% confidence level.}
\label{MXBPointsII}
\begin{tabular}{@{}ccccccccc@{}}
\toprule
\textbf{} &  & Observatory &  & Obs ID      &  & MJD     &  & $k_{\rm B}T_{\rm eff}^{\infty}$ (eV) \\ 
\midrule
1   && Swift  && Interval 1 && 57822.0 && 91.5$\pm$8.8     \\
2 && XMM-Newton && 0803640301 && 57835.8 &  & 87.9$\pm$1.4      \\
3 && Chandra && 19599  && 57868.0 && 
 82.7$\pm$2.0      \\
4 && Chandra && 19600 && 57937.6 && 74.8$\pm$2.5      \\
5 && XMM-Newton && 0803640401 && 57987.3 && 75.1$\pm$2.4      \\
6 && Chandra && 19601 && 58151.5 && 66.0$\pm$3.0      \\
7 && Chandra && 19602 && 58314.4 && 56.3$\pm$4.2        \\ \bottomrule
\end{tabular}
\end{table*}

The standard cooling paradigm has been challenged by these observations. First, some additional heat sources located in the  
shallow layers of the crust are needed to explain the thermal evolution during the first months following the end of an
outburst~\cite{BrownCumming2009} (shallow heating is also expected to play a role in the spin down of accreting neutron stars~\cite{Singh2020}). The required heat sources are compiled in Ref.~\cite{ChamelFantina2020}. 
Moreover, the late time 
cooling after $10^3-10^4$ days revealed unexpected features, especially the dimming of the X-ray emission after the first 
outburst of MXB~1659$-$29 (see the last row in Tab.~\ref{MXBPoints}), suggesting a further cooling of the 
crust~\cite{Cackett2013}.
Running classical molecular dynamics simulations, Horowitz et al.~\cite{horowitz2015} 
proposed that the densest layers of the crust have a low thermal conductivity. But this was not confirmed by quantum molecular 
dynamics simulations performed later~\cite{nandi2018}. Alternatively, Turlione et al.~\cite{Turlione2015} obtained reasonably good fits to the 
cooling data by artificially fine-tuning the \Singlet neutron pairing gaps. The impact of neutron superfluidity was further 
investigated by Deibel et al.~\cite{Deibel2017}. They managed to reproduce the observations quite well by extrapolating the 
neutron pairing gaps obtained from quantum Monte Carlo calculations~\cite{Gandolfi2008} such that free neutrons in the deep crust 
are not superfluid. But more recent calculations based on various many-body approaches have disproved
this extrapolation~\cite{Gandolfi2022,drissi2022,Krotscheck2023}. The subsequent cooling phase following the second outburst of MXB~1659$-$29 was studied by Parikh et al.~\cite{Parikh2019}, 
who also reanalyzed the data from the first outburst. In contrast to Refs.~\cite{horowitz2015,Deibel2017}, the last 
observation reported by Cackett et al.~\cite{Cackett2013} after the end of the first outburst was discarded. This observation was also ignored in 
more recent studies~\cite{potekhin2021,Lu2022}. This omission was motivated by an alternative spectral fit obtained by 
Cackett et al.~\cite{Cackett2013}, who showed that the observed dimming of the X-ray emission could be reproduced by an increase 
of the hydrodygen column density $N_H$ without any significant change of the temperature. Cackett et al. suggested that this 
variation of $N_H$ could be due to the precession of the accretion disk. If this interpretation were correct, further variations of 
$N_H$ should have occurred afterwards on a timescale of a few years. However, Parikh et al.~\cite{Parikh2019} did not find any significant 
variations of $N_H$ throughout the two outbursts (see their footnote 9, page 5). Their analysis therefore turns out to be inconsistent with 
the scenario of a precession of the accretion disk.

All these calculations were based on the model of accreted neutron-star crusts developed by Haensel and Zdunik~\cite{HaenselZdunik1990}
or some recent versions~\cite{HaenselZdunik2003,HaenselZdunik2008,fantina2018}. This approach implicitly assumes that as matter 
accumulates onto the neutron-star surface, nuclear clusters in the inner crust sink very slowly into deeper layers together 
with the neutrons they emit or capture. As first noticed in Ref.~\cite{Steiner2012} and further discussed in Ref.~\cite{ChamelFantina2015}, 
this assumption leads to 
a discontinuous change of the neutron chemical potential at the interface between the outer and inner crusts suggesting that 
free neutrons should diffuse upward to restore equilibrium thus lowering the neutron-drip density. 
This was explicitly studied in Ref.~\cite{ChugunovShchechilin2020}. Accreted
crust models accounting for the diffusion of superfluid neutrons have been developed in Refs.~\cite{GusakovChugunov2020,GusakovChugunov2021,GusakovKantor2021} and detailed calculations have been presented in 
Refs.~\cite{Shchechilin2021,Shchechilin2022,Shchechilin2023}. The thermal evolution of MXB~1659$-$29 has been recently simulated 
in Ref.~\cite{potekhin2023}. The cooling curves are qualitatively similar to those obtained in the standard cooling paradigm. 
In other words, the late-time observations cannot be explained by the diffusion of superfluid neutrons in the crust. 

One important aspect that has not been considered so far in neutron-star crust cooling simulations 
is that the neutron superfluid is weakly coupled to the crust and therefore can flow with a different velocity. 
The accretion of material from a stellar companion accelerates the neutron star (the crust and all particles strongly coupled 
to it constituting most of the star~\cite{alpar1984}) up to very high frequencies, 524~Hz and 567~Hz for KS~1731$-$260 
and MXB~1659$-$29 respectively, as inferred from observations of X-ray burst oscillations~\cite{Smith1997,Wijnands2001,Wijnands2003,Galloway2008}. 
This so called 'recycling' scenario~\cite{alpar1982,radhakrishnan1982} has been recently confirmed by the discovery of 
accreting millisecond X-ray pulsars~\cite{patruno2021,Salvo2022} and transitional millisecond pulsars~\cite{papitto2022}. Due 
to the pinning of quantized vortices, the neutron superfluid tends to lag slightly behind the rest of the star. In turn, the 
induced relative superflow in the crust exerts a Magnus force on individual vortices~\cite{SourieChamel2020} that opposes the 
pinning force, and therefore limits the lag. The existence of such a lag in isolated neutron stars is supported by observations 
of pulsar frequency glitches~\cite{antonopoulou2022}, during which a large collection of vortices are thought to suddenly unpin~\cite{anderson1975,alpar1985}. Although these phenomena are much more difficult to detect in accreting neutron stars owing 
to the lack of accurate timing, they have been seen in a few systems~\cite{Galloway2004,serim2017}. 

Neutron superfluidity impacts the thermal evolution of transiently accreting neutron stars through the specific heat. 
In the absence of superflow as has been tacitly assumed so far, the neutron specific heat is exponentially suppressed due to 
the presence of a gap in the energy spectrum of quasiparticle excitations; this is a well-known consequence of the Bardeen-Cooper-Schrieffer (BCS) 
theory. However, a finite superflow is expected from the pinning of quantized neutron vortices. If the lag between the superfluid 
and the crust lies within a certain range, the neutron superfluid can become gapless~\cite{AllardChamel2023PartI}. The neutron 
specific heat is then comparable to that in the normal phase and represents the main contribution to the crustal specific heat. We 
have recently shown that gapless superfluidity provides a natural explanation for the observed late-time cooling of KS~1731$-$260 and 
MXB~1659$-$29~\cite{AllardChamel2023Letter}. In any case, shallow heating is still needed to explain the early cooling.

In this paper, we present results from additional simulations of the thermal evolution of KS~1731$-$260 and MXB~1659$-$29, varying 
the global structure of the neutron star as well as the accretion rate. We also study more closely the role of the heat blanketing 
envelope, whose composition is not well determined. 
After briefly recalling the main properties of gapless superfluidity in Section~\ref{sec:gapless-superfluidity}, we describe our models of 
neutron-star crust cooling in Section~\ref{sec:cooling-model}. Our results are presented and discussed in Section~\ref{sec:results}.

\section{Gapless neutron superfluidity}
\label{sec:gapless-superfluidity}

\subsection{Order parameter}

The stationary state of a neutron superfluid with number density $n_n$ flowing with superfluid velocity $\pmb{V_n}$ with respect to the normal frame at finite 
temperature $T$ can be 
calculated exactly in the framework of the nuclear energy-density functional theory~\cite{ChamelAllard2020}. 
Excitations can be described 
by quasiparticles with momentum $\hbar\pmb{k}$ and energy given by
($\hbar$ is the Planck-Dirac constant) 
\begin{align}\label{eq:QuasiparticleEnergy}
    \QuasipartEnergyn=\hbar\pmb{k}\cdot\pmb{\Vn} + \sqrt{\varepsilon^{(n)2}_{\pmb{k}} +  \Dn^2}\, ,
\end{align}
where we have introduced the neutron \emph{effective} superfluid velocity
\begin{align}\label{eq:EffectiveSuperfluidVelocity}
\pmb{\Vn}\equiv \frac{m_n}{\mnOplus}\pmb{V_n}+\frac{\pmb{I_n}}{\hbar}\, , 
\end{align}
and 
\begin{align}
    \label{eq:varepsilon}
\varepsilon^{(n)}_{\pmb{k}} =\frac{\hbar^2 k^2}{2\mnOplus}  - \mu_n 
\end{align}
is the single-particle energy. 
Here $m_n$ is the (bare) neutron mass, $\mnOplus$ is the neutron effective mass, and 
$\pmb{I_n}$ is a vector mean-field potential arising from current-current interactions. 
The quantity $\Dn$ is related to the local (complex) order parameter of the 
neutron superfluid phase at position $\rb$ given by
\begin{align}\label{eq:OrderParameterHom}
\Psi_n(\rb)=\frac{\Dn}{v^{\pi n}}  \exp\left( \frac{2 i m_n\pmb{V_n}\cdot \rb}{\hbar}\right)\, ,
\end{align}
and is determined by the solution of the self-consistent equation
\begin{align}
\label{eq:GapEquation}
    \Dn = -\frac{v^{\pi n}}{2V} \sum_{\pmb{k}} \frac{\Dn}{\sqrt{\varepsilon_{\pmb{k}}^{(n)2} +\Dn^2}}\tanh\left(\frac{\beta}{2}\QuasipartEnergyn\right)\, ,
\end{align}
with $v^{\pi n}$ the neutron pairing strength, $V$ the normalization volume and $\beta\equiv\left(k_{\rm B}T\right)^{-1}$ 
(with $k_{\rm B}$, the Boltzmann constant) and it is understood that the summation must be regularized to avoid ultraviolet divergences
(see, e.g., Refs.~\cite{bulgac2002,schunck2019} for more discussions). 
The reduced neutron chemical potential $\mu_n$ is fixed by the particle number conservation 
\begin{align}
\label{eq:DensityHomogeneous}
n_n=
\frac{1}{V}\sum_{\pmb{k}}\left[1-\frac{\varepsilon^{(n)}_{\pmb{k}}}{\sqrt{\varepsilon^{(n)2}_{\pmb{k}} +  \Dn^2}}\tanh\left(\frac{\beta}{2}\QuasipartEnergyn\right)\right]\, . 
\end{align}

As can be seen from Eqs.~\eqref{eq:QuasiparticleEnergy} and \eqref{eq:GapEquation}, $\Dn=\Dn(T,\pmb{\Vn})$ and $\mu_n=\mu_n(T,\pmb{\Vn})$
will generally depend on $T$ and $\pmb{\Vn}$.  

\subsection{Low-temperature limit and weak-coupling approximations}

At low temperatures $T\ll \Tcn$ (with $\Tcn$, the temperature above which superfluidity is destroyed in the absence of superflow), $\Dn$ and $\mu_n$ are almost independent of $T$ and can be accurately calculated taking the limit $T\longrightarrow 0$ in Eqs.~\eqref{eq:GapEquation} and~\eqref{eq:DensityHomogeneous}. As shown in Ref.~\cite{AllardChamel2023PartI}, $\Dn$ and $\mu_n$ are unaffected by the presence of 
superflow if $\Vn<\VLn$ and we will use the notations $\Dn (T=0,\Vn < \VLn)\equiv \Dn^{(0)}$ and $\mu_n (T=0,\Vn <\VLn) \equiv \mu_n^{(0)}$. 

The threshold velocity $\VLn$ is the nuclear analog of Landau's velocity originally introduced in the context of 
superfluid helium~\cite{Landau1941}, and is given by~\cite{AllardChamel2023PartI} 
\begin{align}\label{eq:LandauVelocity-exact}
    \VLn = V_{Fn}\sqrt{\frac{\mu_n^{(0)}}{2\varepsilon_{Fn}}\Biggl[ \sqrt{1+\left(\frac{\Dn^{(0)}}{\mu_n^{(0)}}\right)^2}-1\Biggr] } \, ,
\end{align}
where $V_{Fn}=\hbar k_{Fn}/\mnOplus$ denotes the Fermi velocity, $\varepsilon_{Fn}=\hbar^2 k_{Fn}^2/2\mnOplus$ is the Fermi energy, 
and $k_{Fn}=(3\pi^2 n_n)^{1/3}$ is the Fermi wave number. 
No approximation has been made to derive Eq.~\eqref{eq:LandauVelocity-exact}. In the weak-coupling approximation (see Refs.~\cite{ChamelAllard2021,AllardChamel2023PartI} and Section 2.5 of Ref.~\cite{ChamelAllard2020} for more details) $\Vn\ll V_{Fn}$, $\mu_n^{(0)}\approx \varepsilon_{Fn}$ and $\Dn \ll \varepsilon_{Fn}$, Eq.~\eqref{eq:LandauVelocity-exact} reduces to a familiar expression 
derived within the BCS theory of superconductivity~\cite{Parmenter1962,bardeen1962}
\begin{align}\label{eq:LandauApprox}
    \VLn \approx \frac{\Dn^{(0)}}{\hbar k_{Fn}}\; . 
\end{align}

Neutron superfluidity does not immediately disappear for $\Vn=\VLn$. As $\Vn$ further increases, $\Dn$ remains finite but decreases. 
Superfluidity therefore persists until $\Vn$ reaches a certain critical velocity $\Vcn$, for which $\Dn=0$. In the weak-coupling 
approximation,  $\Vcn$ is given by~\cite{AllardChamel2023PartI}  
\begin{align}
    \Vcn \approx \frac{\Euler}{2}\VLn \approx 1.359 \VLn\; , 
\end{align}
with $\Euler\approx 2.718$ the Euler's number. 

\begin{figure}[h]
\centering\includegraphics[width=0.45\textwidth]{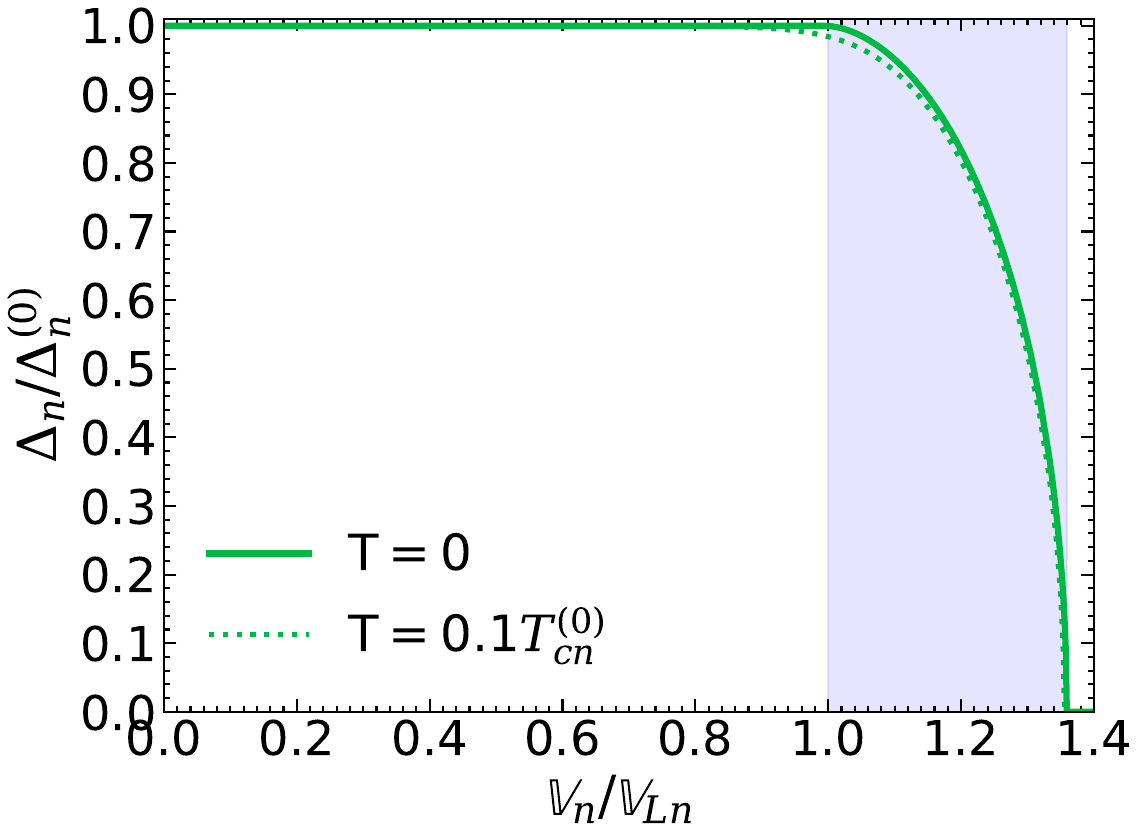}
\caption{Dimensionless ratio $\Dn^{(0)}/\Dn$ as a function of the normalized neutron effective superfluid velocity $\Vn/\VLn$ at $T=0$ (solid curve) and at $T=0.1\Tcn$ (dotted curve). Both curves have been obtained in the weak-coupling approximation. 
The shaded area indicates the gapless regime. See text for details.}
\label{fig:OrderParameterSuperflows}
\end{figure}

As shown in Ref.~\cite{ChamelAllard2021} the ratio $\Dn/\Dn^{(0)}$ at zero temperature is a universal function of $\Vn/\VLn$ (or equivalently of $\Vn/\Vcn$) in the weak-coupling approximation, and is well fitted by
\begin{align}\label{eq:GapInterpolation}
&\frac{\Dn(\VLn < \Vn \leq \Vcn)}{\Dn^{(0)}}=0.5081\sqrt{1-\frac{\Vn}{\Vcn}}\notag\\
& \qquad\times\left(3.312\frac{\Vn}{\Vcn}-3.811\sqrt{\frac{ \Vcn}{\Vn}} +5.842\right)\, . 
\end{align}
This expression remains fairly accurate at low temperatures, as can be seen in Fig.~\ref{fig:OrderParameterSuperflows}. 

The relation between the superfluid velocity $\pmb{V_n}$ and the effective superfluid velocity $\pmb{\Vn}$ is explicitly given by~\cite{ChamelAllard2021}
\begin{align}\label{eq:SuperfluidVelocityvsEffective}
\pmb{V_n}= \pmb{\Vn}\left[1+\left(\frac{\mnOplus}{m_n}-1\right)\mathcal{Y}_n\right]\, , 
\end{align}
where we have introduced the generalized Yosida function, which 
vanishes for $\Vn<\VLn$, and is approximately given in the gapless regime by~\cite{AllardChamel2023PartII}
\begin{align}
\mathcal{Y}_n(\VLn\leq\Vn\leq \Vcn) \approx \left[1-\left(\frac{\Dn}{\Dn^{(0)}}\frac{\VLn}{\Vn}\right)^2\right]^{3/2} \, .
\end{align}
For densities 
prevailing in the inner crust of a neutron star, the neutron effective mass  $\mnOplus$ deviates from the bare neutron 
mass $m_n$ by a few percents only according 
to extended Brueckner-Hartree-Fock calculations with realistic potentials~\cite{cao2006}. 
According to Eq.~\eqref{eq:SuperfluidVelocityvsEffective}, $\pmb{\Vn}$ is therefore approximately 
given by $\pmb{V_n}$. 

\subsection{Superfluidity and quasiparticle energy gap}

\begin{figure}
\centering\includegraphics[width=0.45\textwidth]{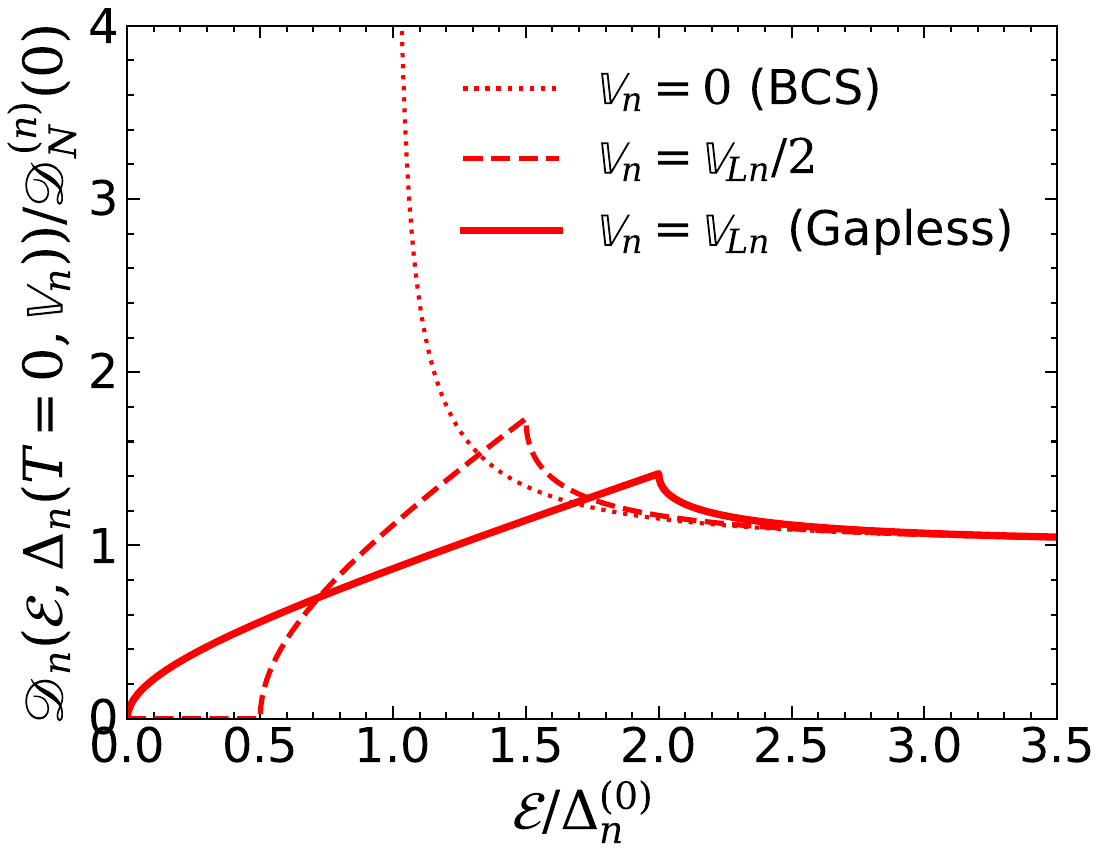}
\caption{Density of neutron quasiparticle states per spin $\mathscr{D}_n(\mathcal{E},\Dn(T=0,\Vn))$ at $T=0$ normalized by that in the normal phase $\DNn$ as a function of quasiparticle energy $\mathcal{E}$ in units of $\Dn^{(0)}$ for different effective superfluid velocities $\Vn$. In the BCS regime for which the superfluid is at rest (dotted line), no quasiparticle exists for $\mathcal{E}\leq\Dn^{(0)}$. In the presence of superflow, this energy gap shrinks with increasing $\Vn$ and disappears at Landau's velocity $\VLn$ (gapless regime). }
\label{fig:DoS-0toVLq}
\end{figure}

The presence of a neutron superflow with $\VLn\leq \Vn \leq \Vcn$ leads to profound changes in the energy spectrum of neutron quasiparticle excitations. 
This can be seen from the density of quasiparticle states per spin defined by
\begin{align}\label{eq:DoS-Def}
\DoSn=
\int \frac{\text{d}^3\pmb{k}}{(2\pi)^3}\delta(\mathcal{E}-\QuasipartEnergyn)\, .
\end{align}
In the weak-coupling approximation, this reduces to~\cite{AllardChamel2023PartI} 
\hfill
\begin{align}\label{eq:QuasiparticleDoS}
&\DoSn = \frac{\DNn}{2\hbar k_{Fn}\Vn}\notag\\
&\quad \times H(\mathcal{E}+\hbar k_{Fn}\Vn-\Dn)\Bigl[\sqrt{(\mathcal{E}+\hbar k_{Fn}\Vn)^2-\Dn^2}\notag\\
&\quad -H(\mathcal{E}-\hbar k_{Fn}\Vn-\Dn) \sqrt{(\mathcal{E}-\hbar k_{Fn}\Vn)^2-\Dn^2}\Bigr]\, ,
\end{align}
where $\DNn=k_{Fn}\mnOplus/\pi^2\hbar^2$ is the density of quasiparticle states in the normal phase and $H$ denotes the Heaviside distribution. For illustration purposes, Eq.~\eqref{eq:QuasiparticleDoS} has been plotted in Fig.~\ref{fig:DoS-0toVLq}, for three different effective superfluid velocities. 

In absence of superflow (i.e. $\Vn=0$), the density of quasiparticle states is identically zero for energies $\mathcal{E}\leq \Dn^{(0)}$ (as seen 
in Fig.~\ref{fig:DoS-0toVLq}) so that $\Dn(\Vn=0)=\Dn^{(0)}$ represents the gap in the quasiparticle energy spectrum and this quantity is
therefore generally called ``pairing gap". The existence of such a gap is often believed to be a necessary condition for neutron superfluidity.
However, this is not always true and we insist on making the distinction between $\Dn$, which is related to the order 
parameter~\eqref{eq:OrderParameterHom}, and the quasiparticle energy gap. Indeed, as shown in Fig.~\ref{fig:DoS-0toVLq}, the energy gap 
shrinks with increasing $\Vn$ and vanishes at Landau's velocity $\VLn$ while the modulus of the order parameter $|\Psi_n|$ remains unchanged (we 
still have $\Dn=\Dn^{(0)}$, as can be seen in Fig.~\ref{fig:OrderParameterSuperflows}). For higher velocities $\VLn\leq \Vn \leq \Vcn$, the 
gap disappears and is filled with (negative) quasiparticle energy states while $0\leq \Dn\leq\Dn^{(0)}$: neutrons remain superfluid
($|\Psi_n|>0$) despite the absence of a gap. In other words, superfluidity has become gapless. When the critical velocity $\Vcn$ is reached, 
the order parameter vanishes marking the transition to the normal phase.

\subsection{Critical temperature}

In absence of superflow, the neutron critical temperature $\Tcn$ is related to the neutron pairing gap through the well-known BCS relation 
(shown to remain valid for nuclear systems within the nuclear energy-density functional theory in Ref.~\cite{AllardChamel2023PartI})
\begin{align}\label{eq:Tc}
    k_{\rm B}\Tcn = \frac{\expgamma}{\pi}\Dn^{(0)}\approx 0.5669\Dn^{(0)}\, ,
\end{align}
where $\gamma\approx 0.5772$ denotes the Euler-Mascheroni constant. 

With increasing neutron effective superfluid velocity $\Vn$, the critical temperature decreases and vanishes for $\Vn=\Vcn$. 
This behavior can be approximately described by~\cite{ChamelAllard2021} 
\begin{align}\label{eq:Tc-Gapless}
    T_{cn} (0\leq \Vn\leq \Vcn) \approx \Tcn \left[1-\left(\frac{2\Vn}{\Euler\VLn}\right) ^2\right]^{2/5}\, . 
\end{align}
At the onset of gapless superfluidity, the critical temperature is only moderately reduced: $T_{cn}(\Vn=\VLn)\approx 0.7321\Tcn$. 

\subsection{Specific heat}

Neutron superfluidity impacts the late-time cooling of transiently accreting neutron stars through the neutron specific heat, 
which can be approximately expressed as~\cite{AllardChamel2023PartI}
\begin{multline}
\label{eq:SpecificHeat-LowT}
c_V^{(n)}(T\ll \Tcn, \Vn) \\ \qquad \approx \frac{1}{2}k_{\text{B}}\beta^2\int_{-\infty}^{+\infty} \text{d}E \; \mathscr{D}_n\left(E,\Dn (T\ll\Tcn, \Vn)\right)\\ \times E^2\sech^2\left(\frac{\beta}{2}E\right)\, . 
\end{multline}

In the standard cooling models ignoring superflow (the regime $\Vn=0$ will be referred to as BCS), Eqs.~\eqref{eq:SpecificHeat-LowT} and~\eqref{eq:QuasiparticleDoS} 
lead to the familiar exponential suppression of the neutron specific heat at low temperatures compared to that in the normal 
phase (see, e.g, Ref.~\cite{Abrikosov}); the reduction factor is approximately given by 
\begin{multline}\label{eq:BCS-Cv}
    R^{\rm BCS}_{00}(T\ll \Tcn) \equiv\frac{c_V^{(n)}(T\ll \Tcn,\Vn=0)}{\Cnormn (T)} \\
    \qquad\approx \frac{3\sqrt{2}}{\pi^{3/2}} \left(\frac{\Tcn}{T}\frac{\pi}{\expgamma}\right)^{5/2}\exp\left(-\frac{\Tcn}{T}\frac{\pi}{\expgamma}\right) \, .
\end{multline}
The neutron specific heat in the normal phase reads 
\begin{align}\label{eq:CV-normal}
\Cnormn(T)\approx \frac{\pi^2}{3}\DNn  k_{\text{B}}^2 T=\frac{k_{Fn}\mnOplus}{3\hbar^2} k_{\text{B}}^2 T\; . 
\end{align}

\begin{figure}
\centering\includegraphics[width=0.45\textwidth]{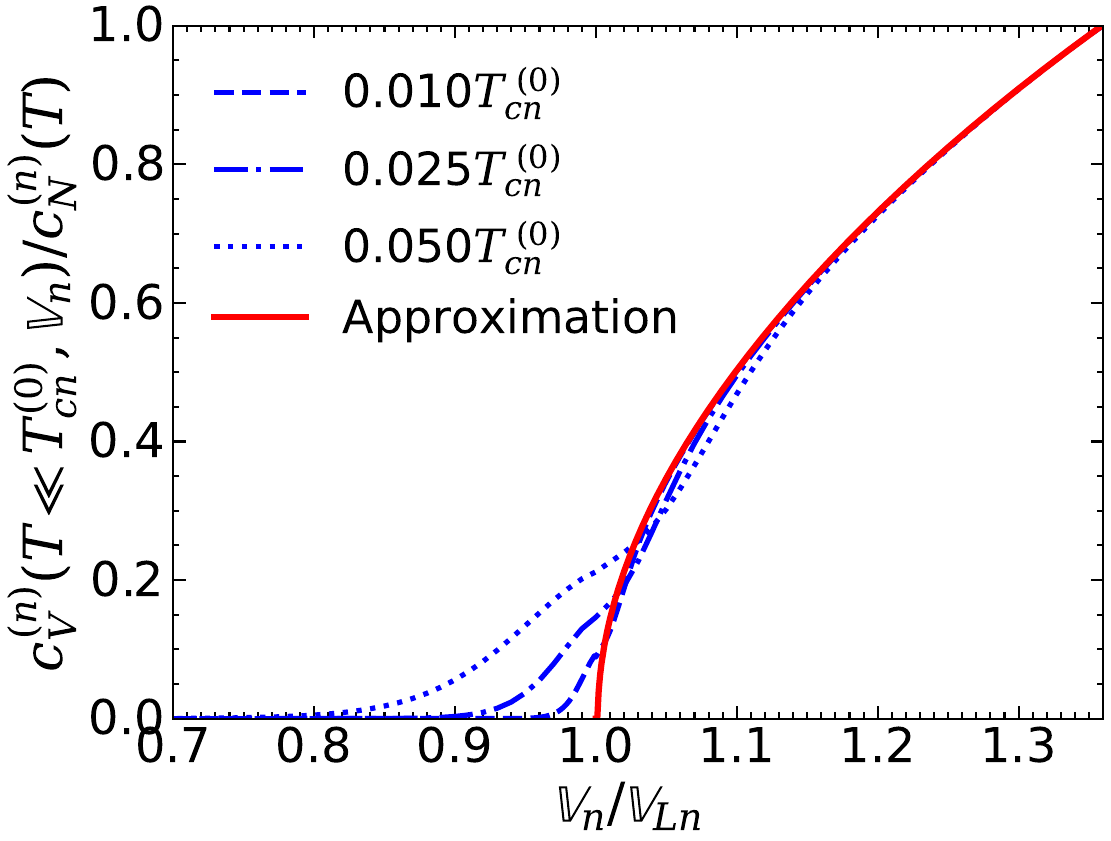}
\caption{Reduction factor of the neutron specific heat at low temperatures as a function of the normalized neutron effective superfluid velocity $\Vn/\VLn$. Results have been computed using Eqs.~\eqref{eq:SpecificHeat-LowT},~\eqref{eq:QuasiparticleDoS} and~\eqref{eq:GapInterpolation}. The reduction factor is exponentially suppressed when $\Vn<\VLn$ whereas it is well approximated by Eq.~\eqref{eq:SpecificHeat-LowT-HighV} (solid curve) for $\Vn>\VLn$. The highest velocity shown is the critical velocity $\Vcn$ for which the reduction factor is 1. See text for details.}
\label{fig:CvFacteur}
\end{figure}

In the \emph{subgapless} regime $0\leq \Vn < \VLn$, the neutron specific heat takes a more complicated form but still exhibits an 
exponential suppression at low temperatures which is, however, mitigated as $\Vn$ increases (see Eq.~(75) of Ref.~\cite{AllardChamel2023PartI}), 
as illustrated in Fig.~\ref{fig:CvFacteur}. At the onset of the gapless regime ($\Vn=\VLn$), the reduction 
factor is approximately given by~\cite{AllardChamel2023PartI} 
\begin{align}\label{eq:CV-Landau}
R^{\rm Gapless}_{00}(T\ll \Tcn,\Vn=\VLn)
\approx 0.9328  \sqrt{\frac{T}{\Tcn}}\, .
\end{align}
The specific heat is thus moderately reduced compared to its counterpart in the normal phase~\eqref{eq:CV-normal}. In the gapless regime 
$\VLn<\Vn\leq\Vcn$, the neutron specific heat becomes comparable to that in the normal phase. As can be seen in Fig.~\ref{fig:CvFacteur}, 
the reduction factor is well approximated by~\cite{AllardChamel2023PartI} 
\begin{align}
    \label{eq:SpecificHeat-LowT-HighV}
R_{00}^{\rm Gapless}&(T\ll \Tcn,\VLn<\Vn\leq \Vcn)\notag \\
&\approx 
\sqrt{1-\left(\frac{\Dn}{\Dn^{(0)}}\frac{\VLn}{\Vn}\right)^2}\, .
\end{align}
Note that for $\Vn=\Vcn$, we have $\Dn=0$ therefore $R_{00}^{\rm Gapless}=1$: we recover the specific heat in the normal phase $\Cnormn(T)$, 
as expected.

A systematic study of the deviations between Eq.~\eqref{eq:SpecificHeat-LowT} and the approximate formulas 
can be found in Ref.~\cite{AllardChamel2023PartI}.

\section{Neutron-star cooling model}\label{sec:cooling-model}

\subsection{Model assumptions}

In principle, one should solve the full dissipative superfluid hydrodynamical equations 
involving two distinct dynamical components, the neutron superfluid and the ``normal" 
viscous part of the crust (see, e.g., Ref.~\cite{CarterChamel2005}). However, one should 
keep in mind that the lag $V_n \approx \Vn$ between the superfluid and the crust is typically very small and is limited by the maximum critical lag $V_{\rm cr}$ beyond which vortices are unpinned. The recent statistical analysis of 541 glitches in 177 pulsars~\cite{Melatos2023} has lead to 
$V_{\rm cr}\sim 10^5$~cm~s$^{-1}$. This estimate, however, is not model independent and 
systematic uncertainties are difficult to assess.
Estimating $V_{\rm cr}$ is not a trivial task and involves a suitable 
average of the local dynamics of individual vortices. 
In the \emph{snow plow model} of Ref.~\cite{pizzochero2011} assuming straight parallel vortices pinned to the crust, the critical lag is found to be approximately given by $V_{\rm cr}\approx 10^7(f_p/10^{18}$~dyn~cm$^{-1})$~cm~s$^{-1}$, where $f_p$ denotes the maximum mesoscopically averaged pinning force per 
unit length. The theoretical challenges to estimate this force are numerous
(see, e.g., Ref.~\cite{antonopoulou2022} for a recent review). 
The internal quantum structure of a vortex (see, e.g., Ref.~\cite{Pecak2021}) is locally 
modified in presence of nuclear clusters 
so that both should be described within a fully self-consistent quantum mechanical approach. The pinning force arising from a single 
cluster has been traditionally determined from static calculations of  
energy differences (see Ref.~\cite{klausner2023} and references). A more reliable approach proposed in Ref.~\cite{wlazlowski2016} consists in calculating the 
force dynamically. However, no such calculations have been systematically 
carried out so far. Calculations on a mesoscopic scale are even more 
uncertain. Results depend on the vortex tension and the structure of the crust~\cite{seveso2016,link2022}. Estimates of $f_p$ differ by orders of magnitude. Even its attractive or repulsive nature is a matter of debate. 
To add to the complexity, vortices are also expected to pin to proton 
fluxoids in the deepest layers of the crust~\cite{ZhaoWen2021} and in core~\cite{muslimov1985vortex,srinivasan1990novel,sauls1989superfluidity,ruderman1998neutron,alpar2017} (assuming protons are in a type II superconducting phase~\cite{baym1969,charbonneau2007,alford2008,wood2022}). Pinning to 
fluxoids is supported by observations of Crab and Vela pulsar glitches~\cite{Sourie2020}. The pinning of vortices in the core of neutron stars was not considered in the statistical study of Ref.~\cite{Melatos2023}. 
Adopting the estimate $10^{18}$~dyn~cm$^{-1}$ given in Ref.~\cite{antonopoulou2022} for the maximum pinning force $f_p$ yields $V_{\rm cr}\sim 10^7$~cm~s$^{-1}$. 

For comparison, Landau's velocity~\eqref{eq:LandauApprox} is approximately given by
\begin{align}\label{eq:VLEstimate}
    \VLn \approx 1.2\times 10^8~{\rm cm}~{\rm s}^{-1}\left(\frac{\Delta_n^{(0)}}{1~{\rm MeV}}\right) \notag \\ \times \left(\frac{10^{14}~{\rm g}~{\rm cm}^{-3}}{\rho Y_{\rm nf}}\right)^{1/3}\; , 
\end{align}
where $\rho$ is the average mass density and $Y_{\rm nf}$ is the fraction of free neutrons in the inner crust. The pinning force in any given crustal layer can be roughly estimated from the Magnus force (see, e.g., Ref.~\cite{SourieChamel2020KJ}) as 
\begin{align}\label{eq:Magnus}
    f_p(\rho) \approx 2.5\times 10^{19}~{\rm dyn}~{\rm cm}^{-1}
     \left(\frac{\Delta_n^{(0)}}{1~{\rm MeV}}\right) 
     \notag \\ \times \left(\frac{\rho Y_{\rm nf}}{10^{14}~{\rm g}~{\rm cm}^{-3}}\right)^{2/3}\; ,
\end{align}
where we have taken Landau's velocity~\eqref{eq:VLEstimate} as the characteristic velocity in the gapless regime. 

Landau's velocity $\VLn$ is thus found to be an order of magnitude higher than the critical lag $V_{\rm cr}$. However, these simple estimates of  $V_{\rm cr}$, $\VLn$, and $f_p$ were  
obtained ignoring the influence of nuclear clusters on the superflow. 
As shown in Ref.~\cite{Antonelli2017}, these effects increase 
the critical lag by a factor $(1-\epsilon_n)=m_n^*/m_n$, where 
$\epsilon_n$ is a parameter characterizing entrainment effects and 
$m_n^*/m_n$ is the dynamical effective mass introduced in Ref.~\cite{carterchamelhaensel2006}. According to the calculations of 
Ref.~\cite{chamel2012}, $m_n^*/m_n \approx 1-14$ depending on the crustal
layer therefore the maximum critical lag could be increased by an order of
magnitude, leading to $V_{\rm cr}\sim 10^8$~cm~s$^{-1}$. These entrainment  
effects are also expected to reduce $\VLn$ (therefore also our estimate of 
$f_p$) by the same amount~\cite{CarterChamelhaensel2005b}. This prediction 
found support from experiments using cold atoms~\cite{miller2007}. Therefore, 
$\VLn \lesssim V_{\rm cr}$ is not implausible.

The critical lag represents a few percents at most of the rotation velocity of the star. 
We expect the purely hydrodynamical 
multifluid effects on the cooling of transiently accreting neutron star crusts to be of the 
same order. Now, the effective surface temperatures inferred from observations are determined 
within a precision of about 1\% at best; the uncertainties generally grow with time, reaching 
about 10\% for the late time observations - the focus of this work. Therefore, to the current 
level of precision of astrophysical observations, the multifluid aspects can be neglected.  
In constrast, the neutron specific heat is increased by several orders of magnitude in the 
gapless state and brings the main contribution to the crustal specific heat. This can lead to 
significant modifications of the cooling curves. In particular, the thermal relaxation is delayed
recalling that the thermal relaxation time $\tau$ scales as~\cite{Page2013}
\begin{equation}\label{eq:ThermalTime}
\tau \approx \frac{c^{\rm crust}_V}{\kappa_{\rm crust}}\Delta R^2
\end{equation}
where $c^{\rm crust}_V$ is the crustal specific heat, $\kappa_{\rm crust}$ is the thermal conductivity
and $\Delta R$ is the crust thickness. This thermal relaxation time can be estimated as 
(see Appendix~\ref{appendix:ThermalRelaxationTime})
\begin{align}\label{eq:ThermalTimescaleGapless}
    \tau &\approx 3\times 10^4~{\rm days}\notag\\
    &\qquad\times\left[ \left(\frac{0.05 Y_{\rm nf}}{Y_e}\right)^{1/3} \left(\frac{\Qimp \Lambda_{eQ}}{\langle Z\rangle}\right)\left(\frac{\Delta R}{1~{\rm km}}\right)^2 \right]\notag\\
    &\qquad \times  R_{00}^{\rm Gapless}(T,\Vn/\VLn)\; , 
\end{align}
with $Y_e$ the fraction of electrons 
in the inner crust, 
$\langle Z\rangle$ the average proton number, and $\Lambda_{eQ}$ the Coulomb logarithm.

As previously shown in Ref.~\cite{Deibel2017}, the neutron contribution to the thermal conductivity of 
the crust remains smaller than that of electrons even if neutrons are not superfluid, except near the 
crust-core interface, where these two contributions can become comparable. However, this is mitigated by 
the increase of the crustal specific heat by about one or two orders of magnitude. Therefore, the influence 
of gapless superfluidity on the neutron thermal conductivity is unlikely to significantly alter the late-time cooling.

We consider here the same heat equation as in previous studies but we take into account the 
effect of gapless superfluidity on the neutron specific heat. The thermal evolution of KS~1731$-$260 and MXB~1659$-$29 after an outburst is computed using the 
\texttt{crustcool} code\footnote{\url{https://github.com/andrewcumming/crustcool}}, 
which solves the time-dependent equations for the temperature and luminosity in the neutron-star crust together with the 
hydrostatic structure equations in the plane-parallel approximation assuming a constant gravity 
(see Ref.~\cite{BrownCumming2009}). The underlying model of accreted neutron-star crust and heating is 
that of Haensel and Zdunik~\cite{HaenselZdunik1990,HaenselZdunik2003,HaenselZdunik2008}. 
We have modified this code to account 1) for neutron diffusion resulting in a different composition 
of the crust and a reduction of the heat sources, 2) for the change of neutron critical temperature 
and neutron specific heat in presence of superflow, both of which can be expressed as universal functions 
of $\Vn/\VLn$~\cite{AllardChamel2023PartI}. These two modifications are discussed in the next subsections.

\subsection{Neutron diffusion}

To account for the diffusion of free neutrons throughout the crust, we have modified the original \texttt{crustcool} 
code following the prescription given in Refs.~\cite{GusakovChugunov2020,GusakovChugunov2021,potekhin2023}, namely: 
\begin{itemize}
    \item For the outer crust we have taken the composition and the equation of state of accreted crusts 
    to be the same as that of the traditional Haensel and Zdunik model~\cite{HaenselZdunik1990} (which assumes 
    ashes of X-ray bursts made of pure $^{56}$Fe), down to the bottom of the outer-crust where the proton number 
    $Z$ becomes equal to 20~\cite{GusakovChugunov2021}.
    \item Due to shell effects, the proton number remains $Z=20$ in the inner crust. The mass number 
    $A_{\rm cl}$ of clusters is found assuming the proton fraction $Y_p=Z/A_{\rm cl}$ and the fraction of free 
    neutrons are the same as the ones found in the non-accreted crust model of Douchin and Haensel~\cite{douchinhaensel01}. 
    \item The heat released by electron captures and pycnonuclear reactions is reduced to $Q_{\rm nuc}=0.1$~MeV/nucleon in 
    the outer crust and $Q_{\rm nuc}=0.3$~MeV/nucleon in the inner crust~\cite{potekhin2023}. 
\end{itemize}

\subsection{Gapless neutron superfluidity and specific heat}

The specific heat of the crust is the sum of three contributions due electrons, ions and free neutrons (in the inner crust). 
The expressions of the first two are the same as the ones described in Ref.~\cite{BrownCumming2009}. For the neutron 
contribution, we have added the reduction factors~\eqref{eq:CV-Landau} and \eqref{eq:SpecificHeat-LowT-HighV} in the 
\texttt{crustcool} code to include the possibility of gapless superfluidity with $\VLn\leq\Vn\leq \Vcn$. In the absence of 
superflow  $\Vn=0$, the
reduction factor of Ref.~\cite{YakovlevandLevenfish1994} is used. We have not considered the intermediate subgapless 
regime $0<\Vn<\VLn$. Indeed, looking at Fig.~\ref{fig:CvFacteur} and Eq.~(75) of Ref.~\cite{AllardChamel2023PartI}, the 
neutron specific heat remains exponentially suppressed as in the absence of superflow, except for $\Vn$ very close to 
$\VLn$. Therefore, the cooling curves are not expected to be very different from those obtained for 
$\Vn=0$. At the onset of the gapless regime ($\Vn=\VLn$), the reduction factor~\eqref{eq:CV-Landau} depends on the neutron pairing 
gap $\Dn^{(0)}$ through the critical temperature~\eqref{eq:Tc}. For higher $\Vn$, the reduction 
factor~\eqref{eq:SpecificHeat-LowT-HighV} is expressed as a universal function of $\Vn/\VLn$. The neutron specific heat still
implicitly depends on $\Dn^{(0)}$ in the sense that superfluidity obviously disappears whenever $T>T_{cn}$ (the relevant
critical temperature given by Eq.~\eqref{eq:Tc-Gapless} is directly proportional to $\Dn^{(0)}$), in which case the neutron specific heat reduces to Eq.~\eqref{eq:CV-normal}. At each cooling stage, we check whether neutrons are superfluid or not
in each crustal layer. The neutron specific heat in the normal phase is given by Eq.~\eqref{eq:CV-normal}. 

In the code, the neutron
effective mass is set to $\mnOplus=m_n$. As can be seen from Eq.~\eqref{eq:SuperfluidVelocityvsEffective}, the effective superfluid velocity $\pmb{\Vn}$ coincides 
exactly with the superfluid velocity $\pmb{V_n}$ in this case. 

\begin{figure}[h]
\centering\includegraphics[width=0.45\textwidth]{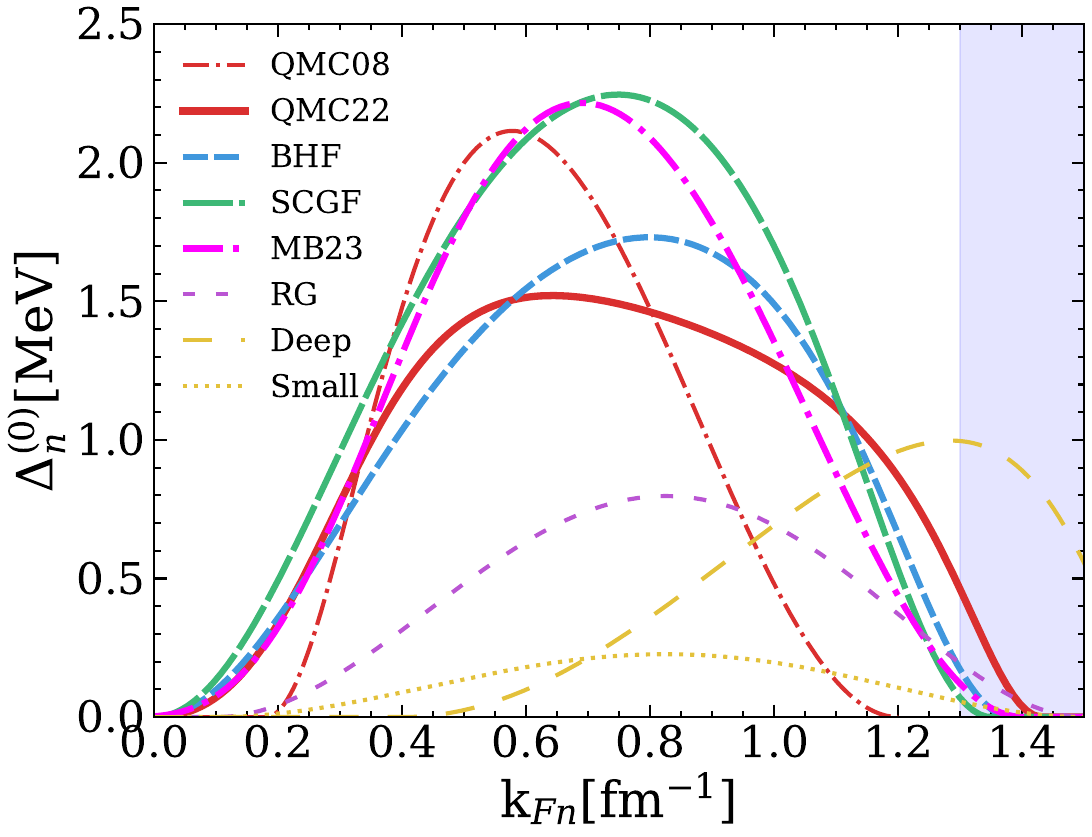}
\caption{Neutron pairing gap $\Dn^{(0)}$ (at zero temperature and in the absence of superflow) as a function 
of the neutron Fermi wave number $k_{Fn}$ as predicted by renormalization group~\cite{Schwenk2003} (RG), 
quantum Monte Carlo calculations from 2008~\cite{Gandolfi2008} (QMC08) and 2022~\cite{Gandolfi2022} (QMC22), 
Brueckner Hartree-Fock theory~\cite{cao2006} (BHF), self-consistent Green function theory~\cite{drissi2022} (SCGF), 
and other diagrammatic calculations~\cite{Krotscheck2023} (MB23). The dashed and dotted yellow curves correspond to 
the ``Deep'' and ``Small''  gaps fine-tuned in Ref.~\cite{Turlione2015} to fit the cooling data of SXTs. Shaded area indicates Fermi wave numbers reached in the outer core of neutron stars. 
}
\label{fig:OrderParameterNeutron}
\end{figure}

\begin{figure}[h]
\centering\includegraphics[width=0.45\textwidth]{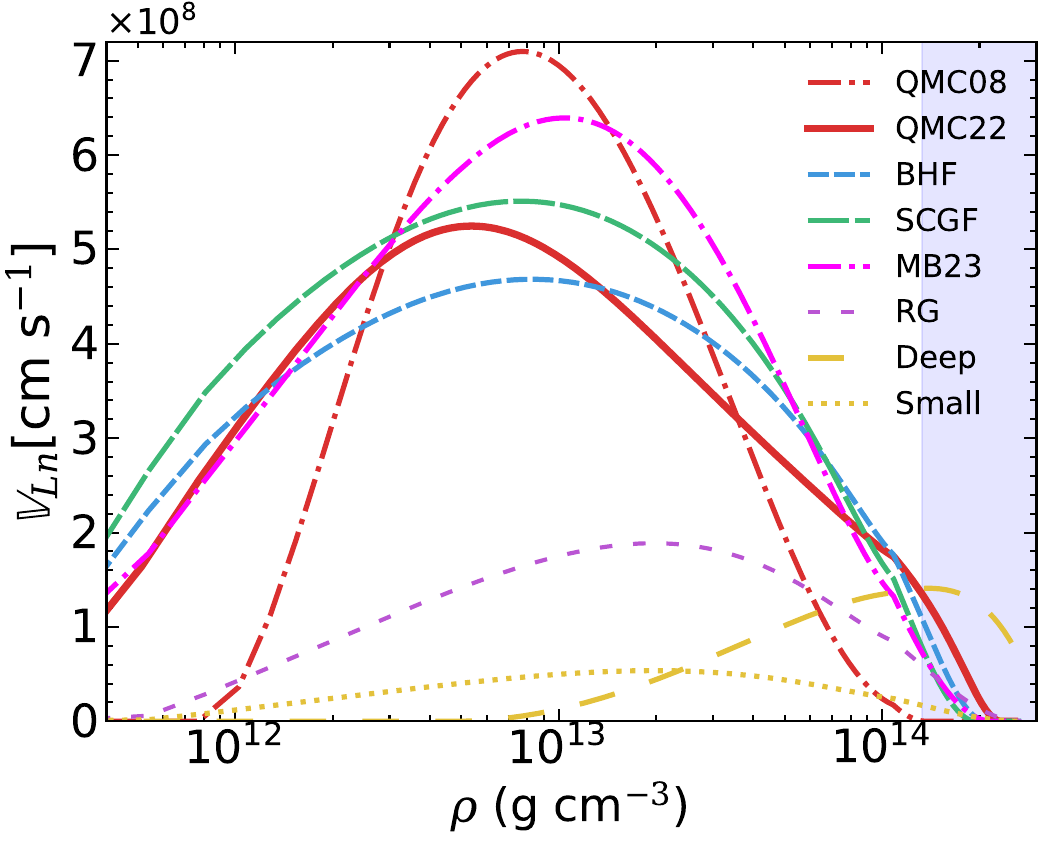}
\caption{Landau's velocity $\VLn$ (defined by~\eqref{eq:LandauApprox}, in cm~s$^{-1}$) as a function of the average mass density $\rho$ (in g~cm$^{-3}$) in the inner crust and core (shaded area) of accreted neutron stars using the composition of Refs.~\cite{GusakovChugunov2020,GusakovChugunov2021} and the neutron pairing gaps $\Delta_n^{(0)}$ as predicted in Fig.~\ref{fig:OrderParameterNeutron}. 
}
\label{fig:LandauVelocity}
\end{figure}

\begin{figure}[h]
\centering\includegraphics[width=0.45\textwidth]{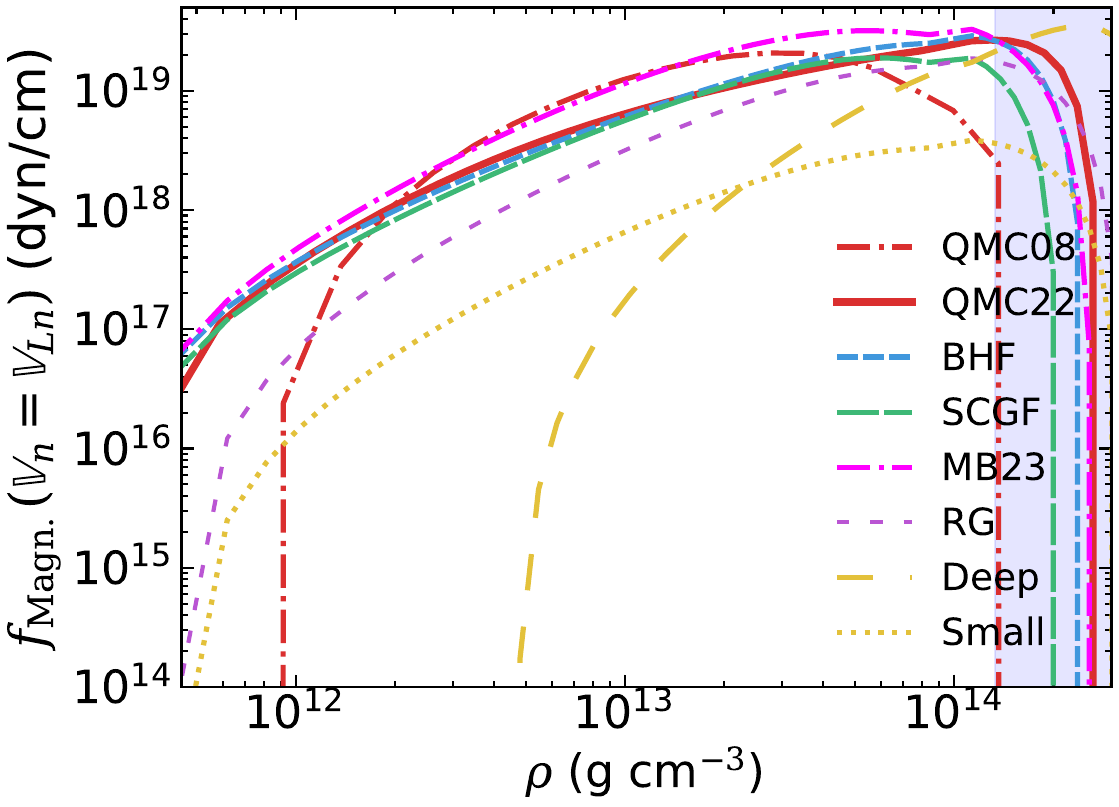}
\caption{Local pinning force per unit length (defined by~\eqref{eq:Magnus}, in dyn~cm$^{-1}$) as a function of the average mass density $\rho$ (in g~cm$^{-3}$) in the inner crust and core (shaded area) of accreted neutron stars using the composition of Refs.~\cite{GusakovChugunov2020,GusakovChugunov2021} and the neutron pairing gaps $\Delta_n^{(0)}$ as predicted in Fig.~\ref{fig:OrderParameterNeutron}. 
}
\label{fig:MagnusForce}
\end{figure}

The actual value of $\Vn/\VLn$ depends on the dynamical evolution of the neutron star and may vary with depth.
In this paper, we treat this ratio as a free constant parameter. 
The only microscopic inputs are therefore the neutron pairing gap $\Dn^{(0)}$ through the critical temperature, 
as seen from Eqs.~\eqref{eq:Tc-Gapless} and \eqref{eq:Tc}. We have implemented more recent microscopic 
calculations of $\Dn^{(0)}$ shown in Fig.~\ref{fig:OrderParameterNeutron}. 
For comparison, we have also plotted the ``Deep" 
and ``Small" gaps of Ref.~\cite{Turlione2015} obtained empirically by fitting the observations of 
SXTs. All recent calculations based on extended Brueckner Hartree-Fock theory~\cite{cao2006}, quantum 
Monte Carlo method~\cite{Gandolfi2022}, self-consistent Green function approach~\cite{drissi2022} and 
other diagrammatic methods~\cite{Krotscheck2023} agree on the extent of the neutron superfluid phase, 
which spans neutron Fermi wave numbers up to $k_{Fn}\lesssim 1.4$~fm$^{-1}$ (corresponding to average mass 
densities $\rho\lesssim 2\times 10^{14}$~g~cm$^{-3}$ beyond the crust-core transition). Note that the older 
quantum Monte Carlo calculations of Ref.~\cite{Gandolfi2008} predicted a closure of the gap at 
$k_{Fn}\approx 1.2$~fm$^{-1}$ at variance with other approaches. This discrepancy, which allowed Deibel 
et al.~\cite{Deibel2017} to fit the late-time cooling data of SXTs, has disappeared with the latest 
quantum Monte Carlo calculations~\cite{Gandolfi2022}. In what follows, we will present neutron-star cooling 
simulations using the calculations of $\Dn^{(0)}$ from Ref.~\cite{Gandolfi2022}. Landau's velocities~\eqref{eq:LandauApprox} and local pinning forces~\eqref{eq:Magnus} corresponding to the various 
gaps are plotted in Figs.~\ref{fig:LandauVelocity} and ~\ref{fig:MagnusForce} respectively.

\subsection{Markov Chain Monte Carlo simulations}

In principle, the mass $\MNS$ and radius $\RNS$ of the neutron star could be inferred from the spectral fit. 
However, the data are not accurate enough. In our previous work, we set $\MNS=1.62\MSol$ and $\RNS=11.2$~km, as 
in the original paper of Brown and Cumming~\cite{BrownCumming2009}. For 
comparison, we have also run simulations using the canonical values $\MNS=1.4\MSol$ and $\RNS=10$~km; those values
were considered in Refs.~\cite{cackett2006,Cackett2010,Merritt2016} to fit the X-ray spectrum of KS~1731$-$260 and in Refs.~\cite{cackett2006,cackett2008,Cackett2013} to fit the X-ray spectrum of MXB~1659$-$29. 

\begin{figure}[h]
\centering\includegraphics[width=0.45\textwidth]{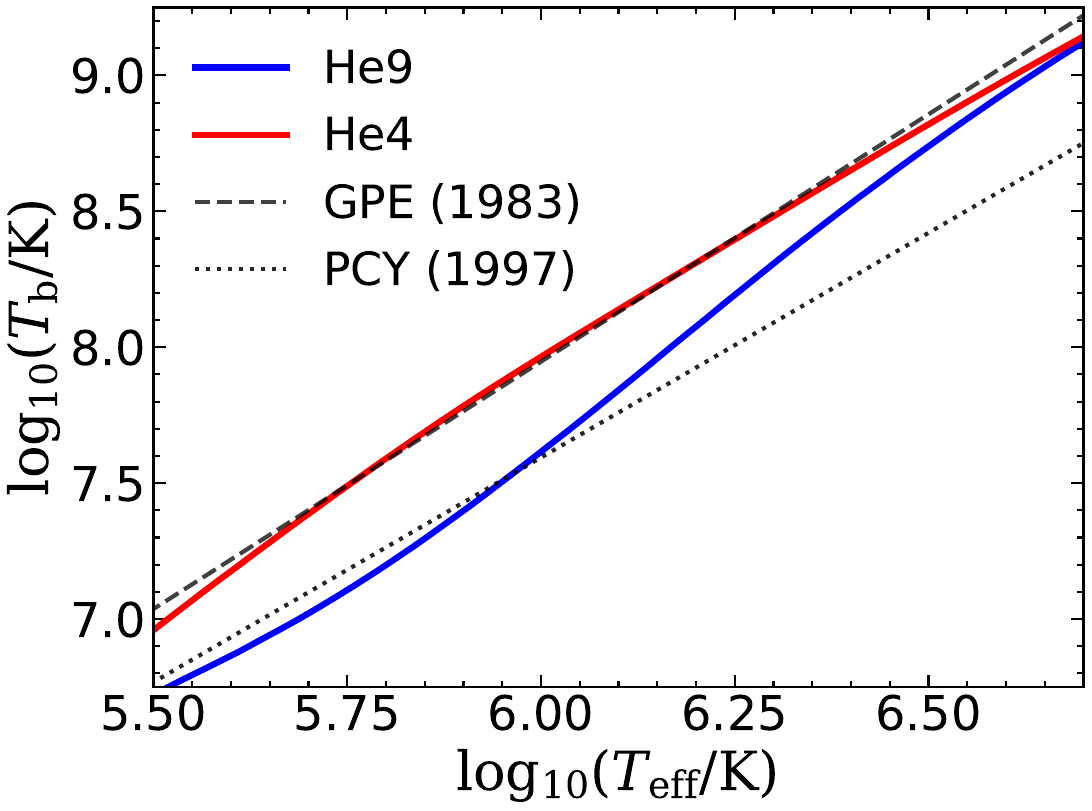}
\caption{Temperature at the bottom of the neutron star envelope $\Tbase$ at column depth $y_{\rm b}=10^{12}$~g~cm$^{-2}$ as a function 
of the effective surface temperature $T_{\rm eff}$ for two different models considered in this study consisting of He 
down to $y_{\rm He}$ and Fe beyond: He9~\cite{BrownCumming2009} with to $y_{\rm He}=10^9$~g~cm$^{-2}$ (blue curve) and 
He4~\cite{cumming2017} with $y_{\rm He}=10^4$~g~cm$^{-2}$ (red curve). The pure Fe envelope of Ref.~\cite{gudmundsson1983} (dashed curve) and the fully accreted envelope of Ref.~\cite{potekhin1997} (dotted curve) are also shown for comparison.}
\label{fig:TbTs}
\end{figure}

During each accretion episode, light elements (mainly composed of H and He) accumulate on the surface of the
neutron star and, after a critical threshold, they burn to heavier elements. Most H is expected to be consumed.
This whole process repeats until the end of the outburst~\cite{brown2002}. The column depth of the remaining 
light elements is not known. To study how this uncertainty impacts the cooling, we have considered two
different heat-blanketing envelope models, which we denote by He9 (previously employed in Ref.~\cite{BrownCumming2009}) 
and He4 for short and provided with the original 
\texttt{crustcool} code in the files \texttt{grid\_He9} and \texttt{grid\_He4} respectively. 
In both cases, the envelope is made of pure He down to the column depth $y_{\rm He}$, and pure Fe down to  
$y_{\rm b}=10^{12}$~g~cm$^{-2}$ marking the bottom of the envelope. The transition between He and Fe is set to 
$y_{\rm He}=10^9$~g~cm$^{-2}$ and
$y_{\rm He}=10^4$~g~cm$^{-2}$ for the models He9  and He4 
respectively. The associated relations between the temperature $\Tbase$ at the column depth $y_{\rm b}$ and 
the effective surface temperature $T_{\rm eff}$ are displayed in Fig.~\ref{fig:TbTs} for $\MNS=1.62\MSol$ and 
$\RNS=11.2$~km. Results obtained for $\MNS=1.4\MSol$ and $\RNS=10$~km are hardly distinguishable. The 
$\Tbase-T_{\rm eff}$ relation for the model He4 turns out to be very similar to the pure Fe envelope model 
of Ref.~\cite{gudmundsson1983}. For comparison, we have also plotted the fully accreted model of 
Ref.~\cite{potekhin1997}.

Unless stated otherwise, the accretion rate is fixed to its time-averaged value $0.1\mEdd$~\cite{Galloway2008} (with $\mEdd \simeq 1.1\times 10^{18}$~g~s$^{-1}$ being the Eddington accretion rate), consistent with previous studies~\cite{BrownCumming2009,horowitz2015,Merritt2016,Deibel2017,brown2018}.  

The fitting parameters of our neutron-star crust cooling model are: 1) the temperature of the neutron star 
core $\Tcore$, 2) the temperature $\Tbase$ at the bottom of the envelope at the column depth of 
$y_{\rm b}=10^{12}$~g~cm$^{-2}$ during outburst, which takes into account shallow heating in an effective way (see 
Refs.~\cite{Turlione2015,Deibel2015,cumming2017}), 3) the impurity parameter $\Qimp$ (assumed to be constant throughout 
the crust) and, 4) the normalized neutron effective superfluid velocity $\Vn/\VLn$ entering the calculation of the neutron
specific heat (simulations in the classical BCS regime are performed with the fixed value $\Vn/\VLn=0$). Note that in 
Refs.~\cite{BrownCumming2009,Deibel2017}, $\Tcore$ was set to the last observation thus assuming that the thermal evolution lasted 
long enough for the crust-core thermal equilibrium to be restored. We do not make this assumption here and treat $\Tcore$ 
as a free parameter. 

To determine the values of these fitting parameters and their uncertainties, we have performed extensive neutron-star cooling 
simulations within a Markov Chain Monte Carlo (MCMC) method  using the Python \texttt{emcee} package\footnote{\url{https://github.com/dfm/emcee}}. 
We have found that 25 chains with a length of $3000$ steps (corresponding to 
$7.5\times 10^4$ samples) are typically enough to reach convergence. We have run multiple MCMC simulations with different 
initial guess values for the parameters to check the reliability of our results. 
The ``best'' value of each parameter is 
determined by the median of the corresponding marginalized 1D posterior probability distribution. Errors are 
estimated at the 68\% uncertainty level.

\section{Results}\label{sec:results}

\subsection{KS~1731$-$260}

\subsubsection{With BCS superfluidity}

\begin{figure*}
    \centering
    \begin{subfigure}[b]{0.45\textwidth}
        \centering\includegraphics[width=\textwidth]{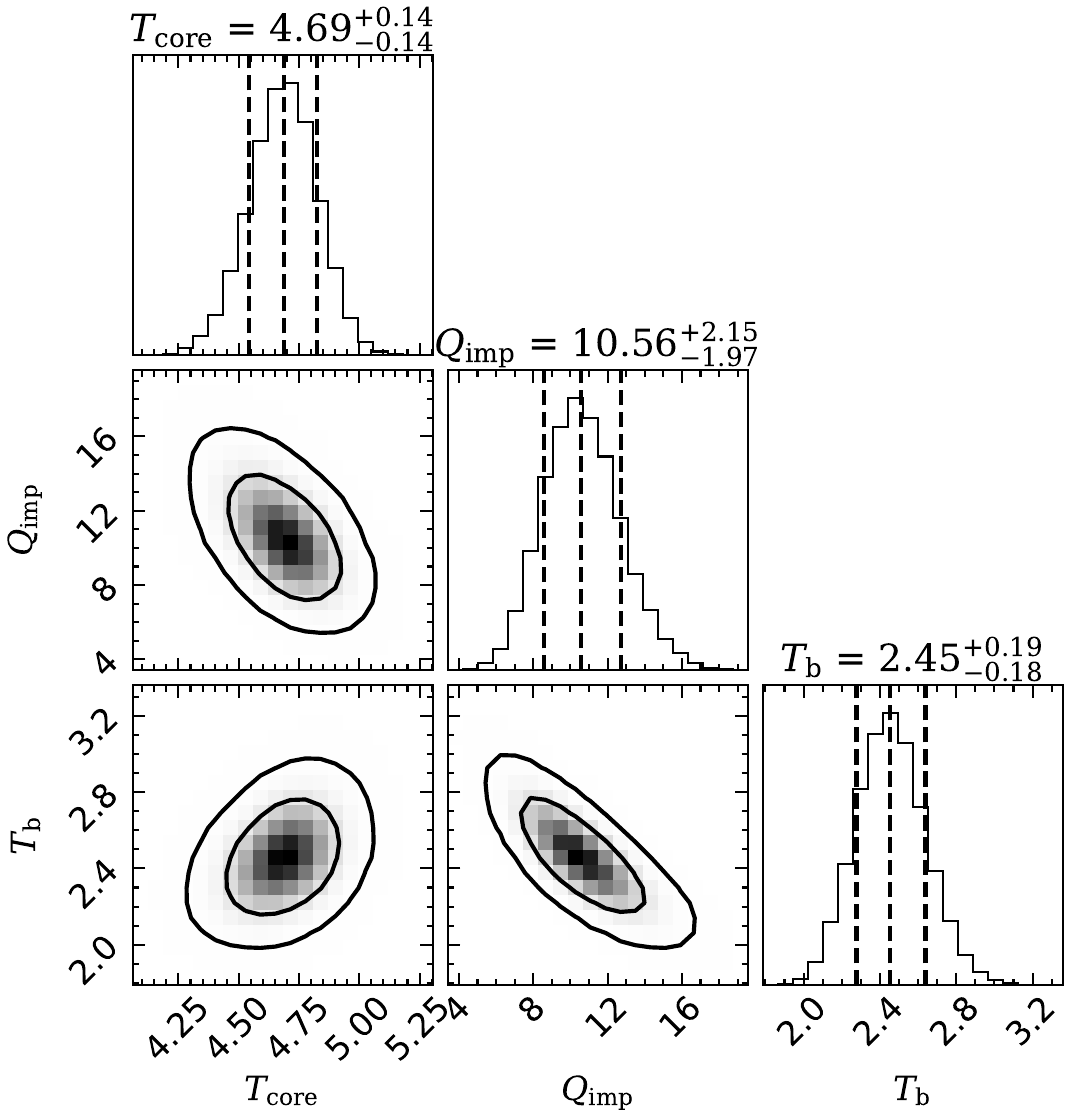}
        \caption{With He9 envelope.}
    \end{subfigure}
    \begin{subfigure}[b]{0.45\textwidth}
        \centering\includegraphics[width=\textwidth]{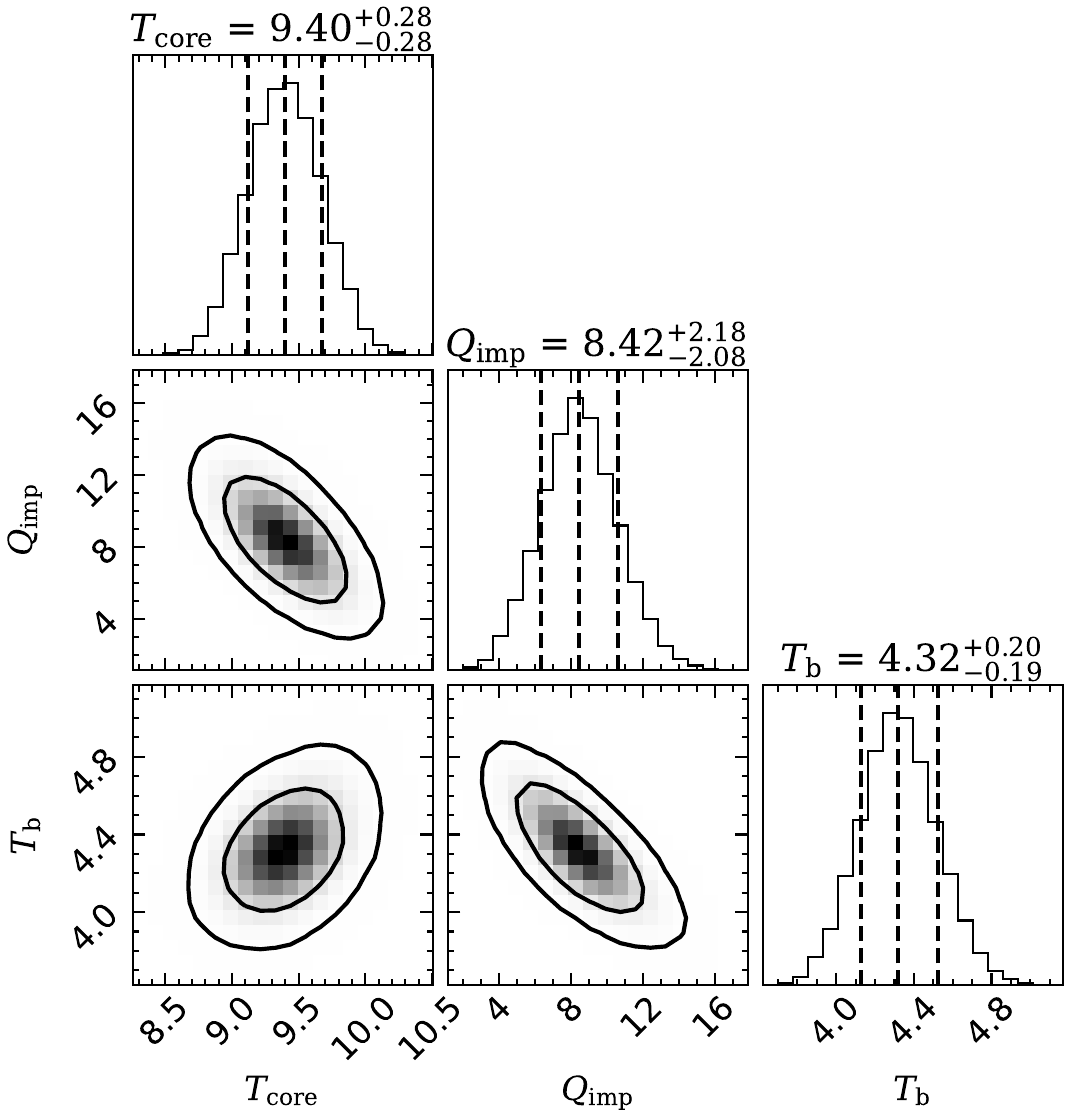}
        \caption{With He4 envelope.}
    \end{subfigure}
    \caption{Marginalized 1-D and 2-D probability distributions for the parameters of our cooling model of KS~1731$-$260 within the model of Gusakov\&Chugunov~\cite{GusakovChugunov2020,GusakovChugunov2021} of accreted neutron stars in the absence of superflow (BCS regime) using the realistic neutron pairing calculations of Ref.~\cite{Gandolfi2022}. Results were obtained setting $\MNS=1.62\MSol$ and $\RNS=11.2$~km with the envelope model He9 (left panel) or He4 (right panels). $\Tcore$ and $\Tbase$ are expressed in units of $10^7$~K and $10^8$~K respectively. The dotted lines in the histograms mark the median value and the 68\% uncertainty level while the contours in the 2-D probability distributions correspond to 68\% and 95\% confidence ranges. The associated cooling curves, using the median values, are displayed in Fig.~\ref{fig:KSnHD_Cooling}.}
    \label{fig:Corner_KS-BCS}
\end{figure*}

\begin{figure*}
    \centering
    \begin{subfigure}[b]{0.45\textwidth}
        \centering\includegraphics[width=\textwidth]{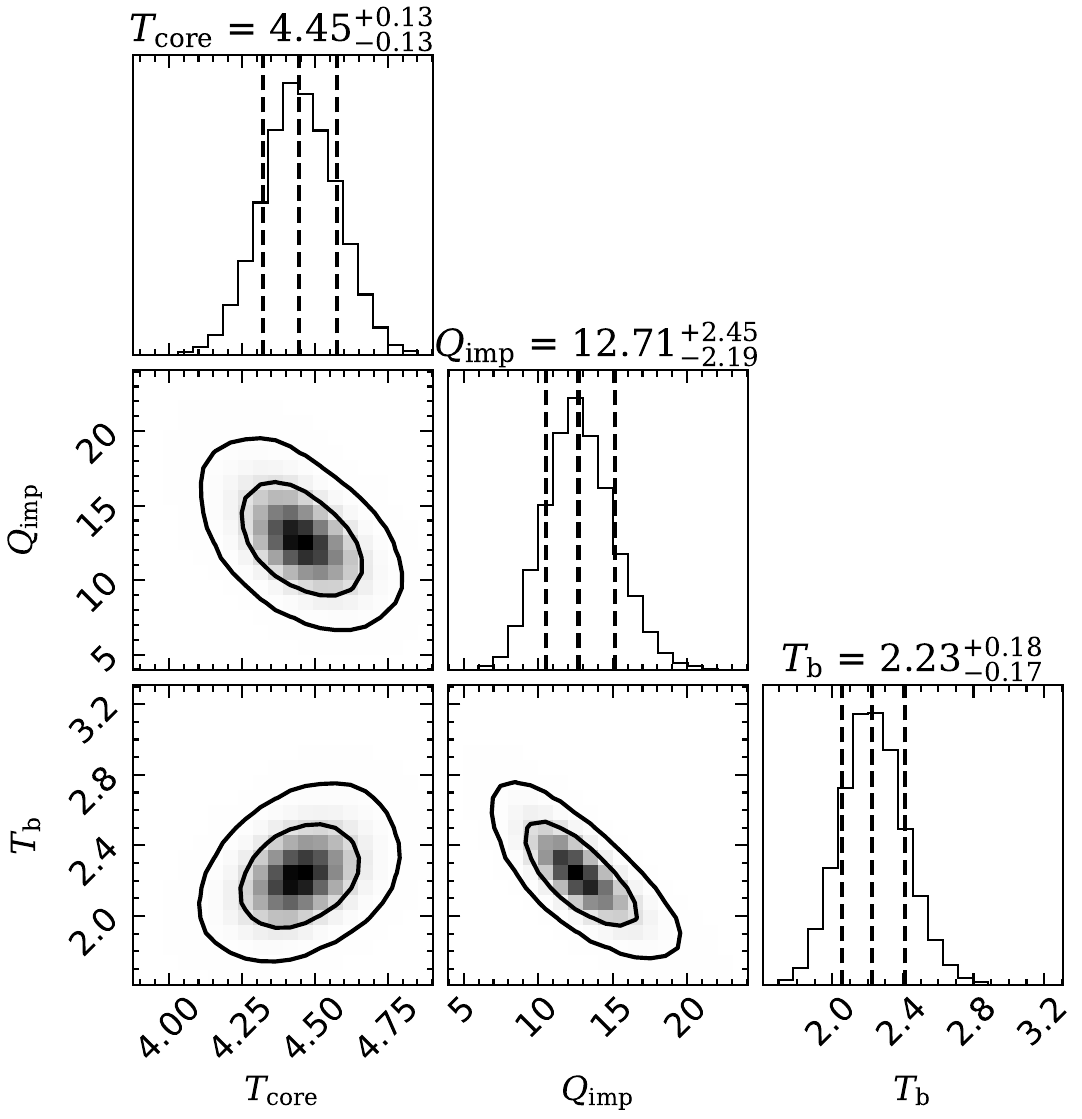}
        \caption{With He9 envelope.}
    \end{subfigure}
    \begin{subfigure}[b]{0.45\textwidth}
        \centering\includegraphics[width=\textwidth]{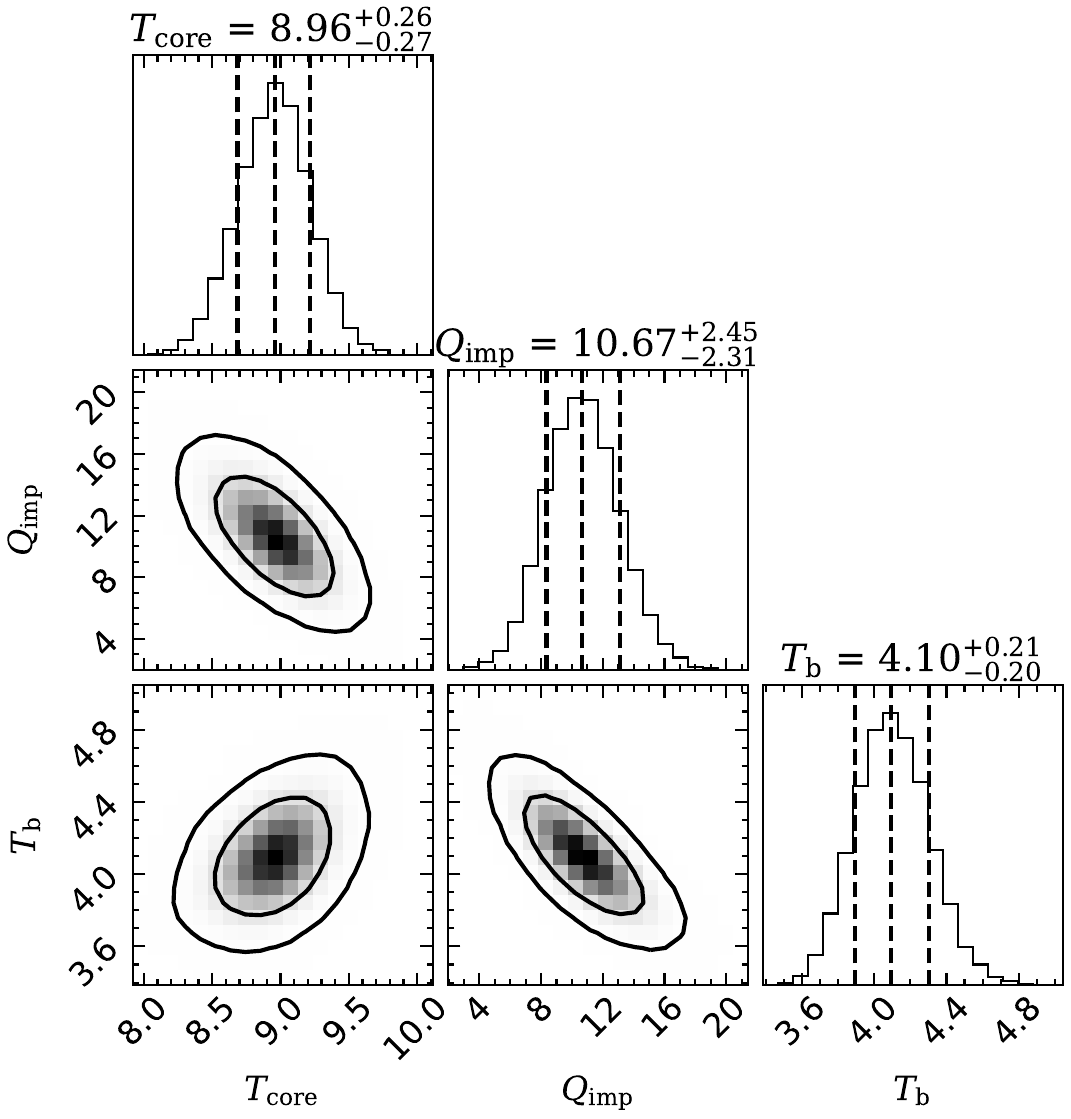}
        \caption{With He4 envelope.}
    \end{subfigure}
    \caption{Same as Fig.~\ref{fig:Corner_KS-BCS} for $\MNS=1.4\MSol$ and $\RNS=10$~km.}
    \label{fig:Corner_KS-BCS-canonical}
\end{figure*}

\begin{figure*}[h]
 \centering
    \begin{subfigure}[b]{0.45\textwidth}
        \centering\includegraphics[width=\textwidth]{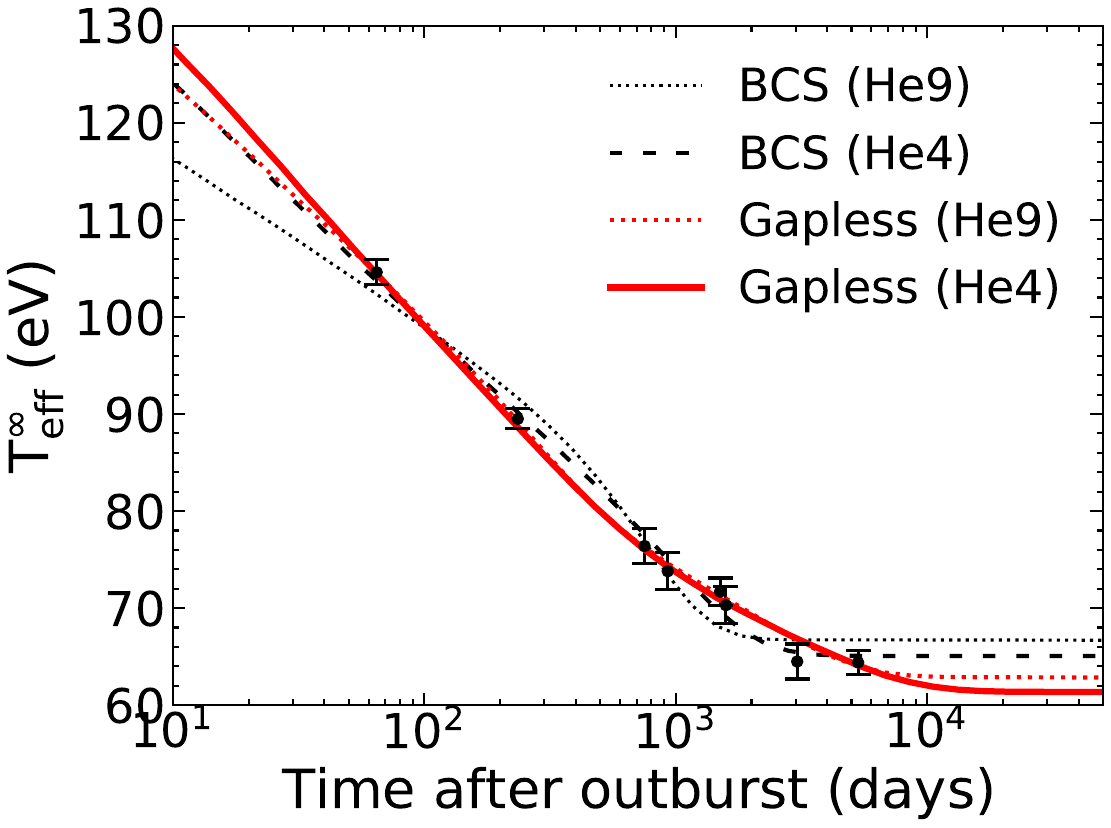}
        \caption{$\MNS=1.62\MSol$ and $\RNS=11.2$~km.}
    \end{subfigure}
    \begin{subfigure}[b]{0.45\textwidth}
        \centering\includegraphics[width=\textwidth]{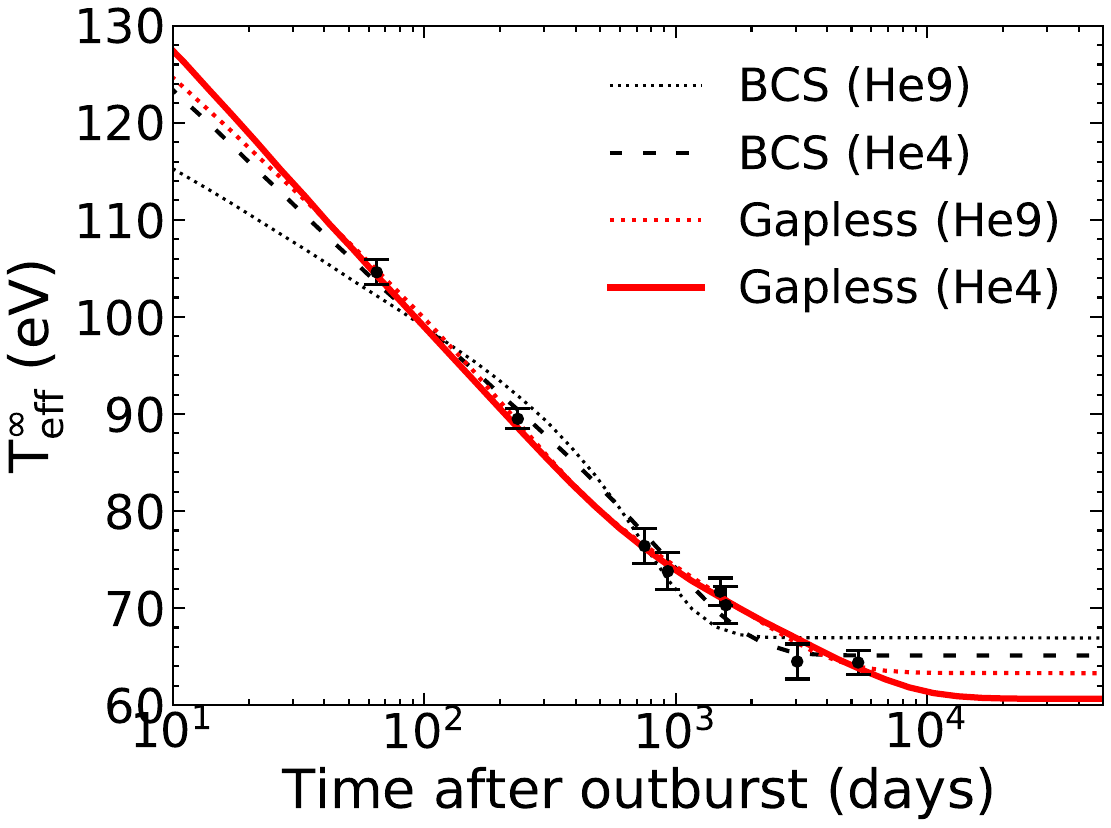}
        \caption{$\MNS=1.4\MSol$ and $\RNS=10$~km.}
    \end{subfigure}
\caption{Evolution of the effective surface temperature of KS~1731$-$260 in electronvolts (as seen by an observer at infinity) as a function of the time in days after the end of the outburst within the model of Gusakov\&Chugunov~\cite{GusakovChugunov2020,GusakovChugunov2021} of accreted neutron star with $\MNS=1.62\MSol$ and $\RNS=11.2$~km (left panel) or the canonical values $\MNS=1.4\MSol$ and $\RNS=10$~km (right panel). Symbols represent observational data with error bars. The black and red lines are models considering BCS and gapless superfluidity, respectively. Dotted curves represent results obtained assuming the He9 envelope model while the black-dashed and red-solid curves assumed the He4 envelope model. All the calculations were made using the realistic neutron pairing calculations of Ref.~\cite{Gandolfi2022}.}
\label{fig:KSnHD_Cooling}
\end{figure*}

Figures~\ref{fig:Corner_KS-BCS} and \ref{fig:Corner_KS-BCS-canonical} show the marginalized posterior probability distributions of the model parameters for KS~1731$-$260 within the accreted 
crust model of Gusakov and Chugunov~\cite{GusakovChugunov2020,GusakovChugunov2021} in the absence of superflow (BCS regime). 
The former figure corresponds to models with $\MNS=1.62\MSol$ and $\RNS=11.2$~km as in Ref.~\cite{BrownCumming2009} while 
the latter corresponds to models with the canonical values $\MNS=1.4\MSol$ and $\RNS=10$~km. Results have been obtained assuming two different envelope models, He9 (left panels) and He4 (right panels). The associated
cooling curves are displayed in Fig.~\ref{fig:KSnHD_Cooling} and the parameters are given in Table~\ref{tab:KS}. 
The two sets of $\MNS$ and $\RNS$ lead to comparable values for $\Tcore$, $\Tbase$ and $\Qimp$. The associated cooling curves are hardly distinguishable. 
The composition of the envelope is found to play a more important role. As discussed previously, the envelope of the He4 model containing more Fe is therefore 
more opaque and leads to a hotter crust compared to the He9 envelope model for the same effective surface temperature (see Fig.~\ref{fig:TbTs}). The fit 
to the cooling data therefore yields higher temperatures $\Tcore$ and $\Tbase$ whereas the impurity parameters $\Qimp$ remain  comparable. 
As can be seen in Fig.~\ref{fig:TbTs}, the He4 envelope model provides a better fit to the cooling data. These results are consistent with those previously 
reported in Ref.~\cite{cumming2017} within the traditional deep crustal heating paradigm. The core temperatures we obtain are comparable to those found in
Refs.~\cite{Merritt2016,cumming2017,Deibel2017} without taking into account neutron diffusion. 
In all cases, our models predict that KS~1731$-$260 returned to thermal equilibrium about 2000 (3000) days after the end of the outburst for the He9 (He4)
envelope model in agreement with Ref.~\cite{Merritt2016}. 

\subsubsection{With gapless superfluidity}

\begin{figure*}
    \centering
    \begin{subfigure}[b]{0.45\textwidth}
        \centering\includegraphics[width=\textwidth]{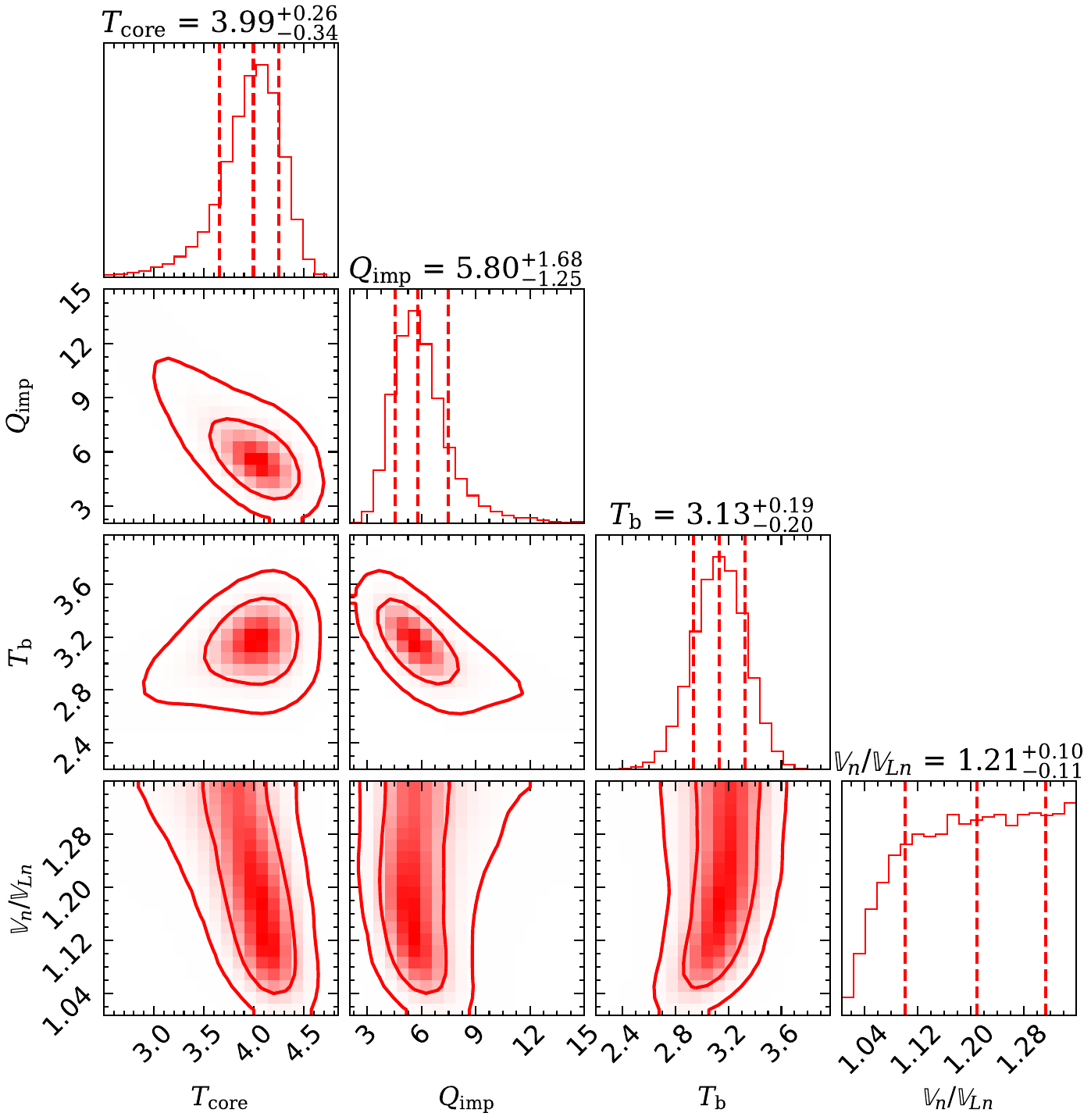}
        \caption{With He9 envelope.}
    \end{subfigure}
    \begin{subfigure}[b]{0.45\textwidth}
        \centering\includegraphics[width=\textwidth]{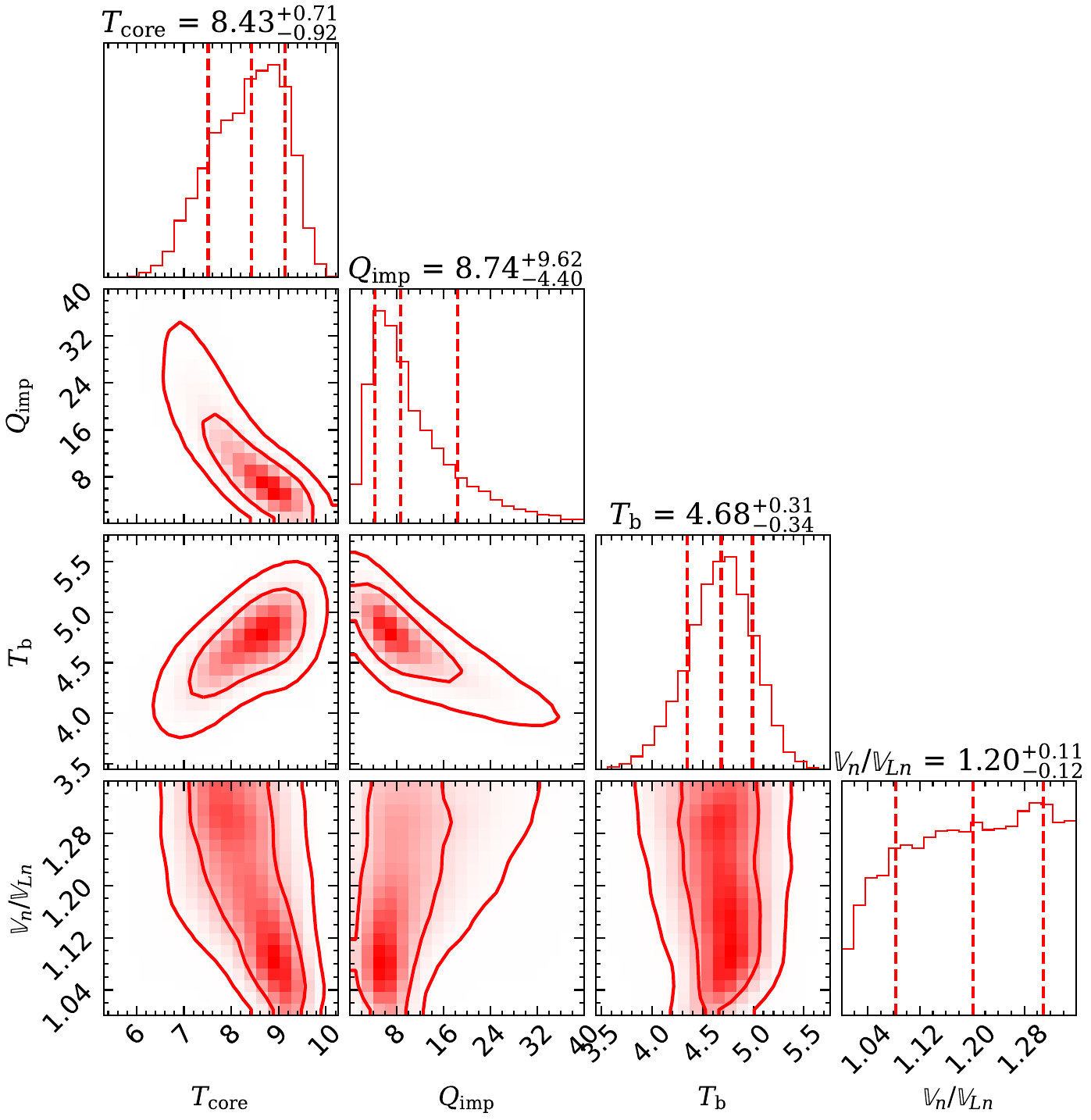}
        \caption{With He4 envelope.}
    \end{subfigure}
    \caption{Marginalized 1-D and 2-D probability distributions for the parameters of our cooling model of KS~1731$-$260 within the model of Gusakov\&Chugunov~\cite{GusakovChugunov2020,GusakovChugunov2021} of accreted neutron stars but considering gapless superfluidity using the realistic neutron pairing calculations of Ref.~\cite{Gandolfi2022}.  Results were obtained setting $\MNS=1.62\MSol$ and $\RNS=11.2$~km with the envelope model He9 (left panel) or He4 (right panel). $\Tcore$ and $\Tbase$ are expressed in units of $10^7$~K and $10^8$~K respectively. The dotted lines in the histograms mark the median value and the 68\% uncertainty level while the contours in the 2-D probability distributions correspond to 68\% and 95\% confidence ranges. The associated cooling curves, using the median values, are displayed in Fig~\ref{fig:KSnHD_Cooling}.}
    \label{fig:Corner_KS-Gapless}
\end{figure*}

\begin{figure*}
    \centering
    \begin{subfigure}[b]{0.45\textwidth}
        \centering\includegraphics[width=\textwidth]{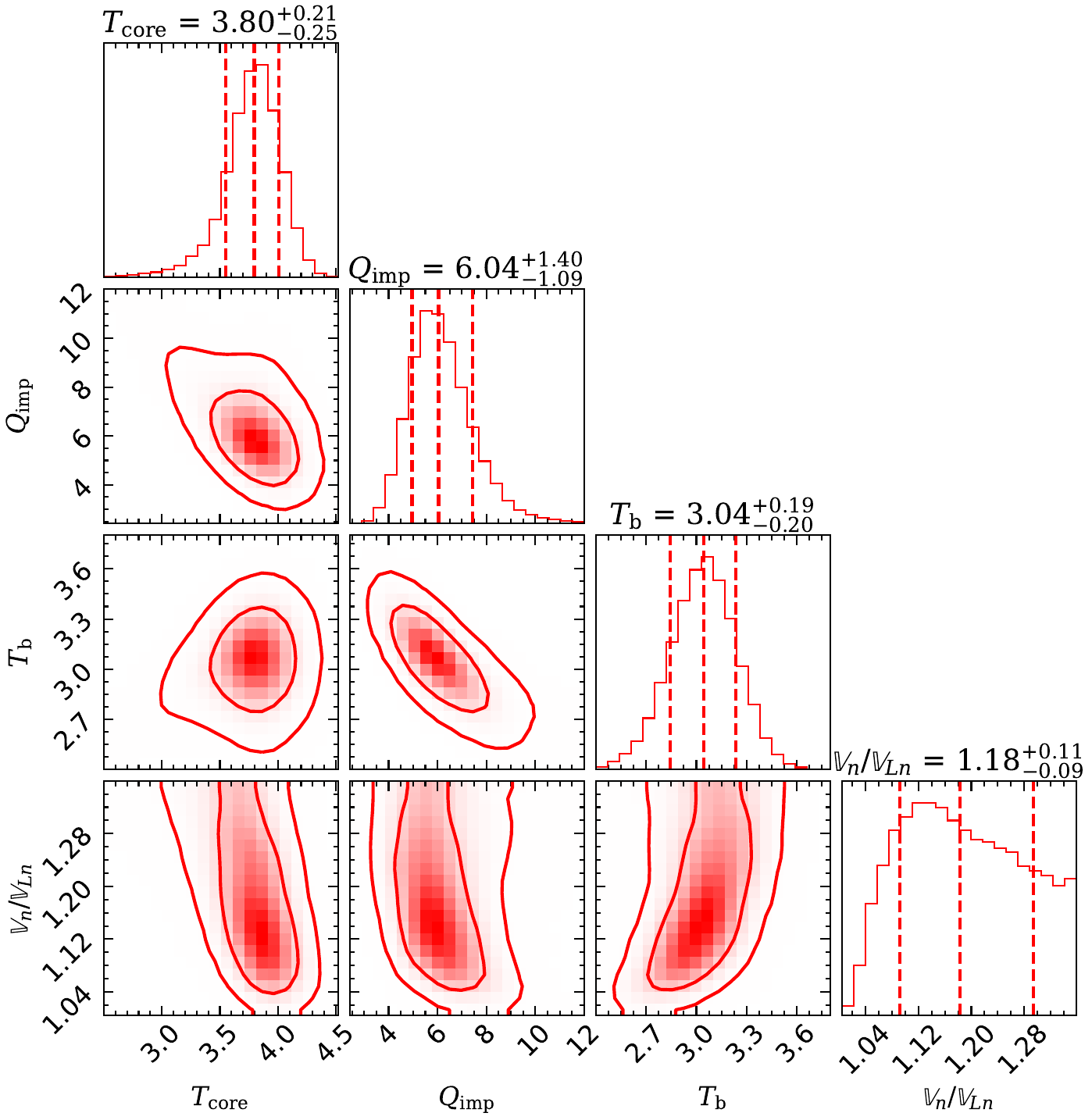}
        \caption{With He9 envelope.}
    \end{subfigure}
    \begin{subfigure}[b]{0.45\textwidth}
        \centering\includegraphics[width=\textwidth]{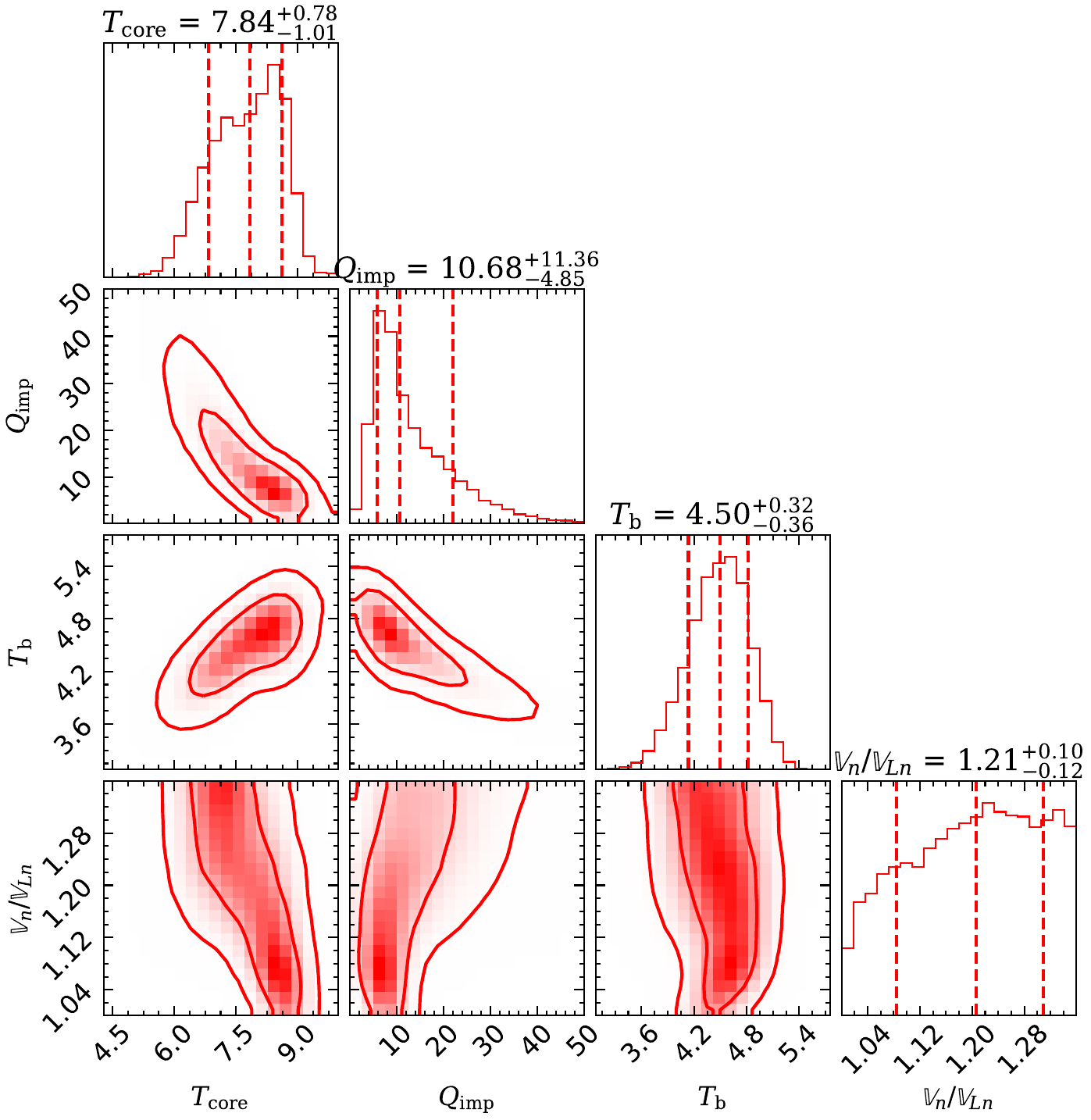}
        \caption{With He4 envelope.}
    \end{subfigure}
    \caption{Same as Fig.~\ref{fig:Corner_KS-Gapless} for $\MNS=1.4\MSol$ and $\RNS=10$~km.}
    \label{fig:Corner_KS-Gapless-canonical}
\end{figure*}

Figures~\ref{fig:Corner_KS-Gapless} and \ref{fig:Corner_KS-Gapless-canonical} show the marginalized posterior probability distributions of the model parameters for KS~1731$-$260 
now allowing for 
the presence of superflow in gapless regime. The associated cooling curves are plotted in Fig.~\ref{fig:KSnHD_Cooling} and the model parameters are collected 
in Table~\ref{tab:KS}. As in the BCS regime, the thermal evolution is not sensitive to the adopted values for $\MNS$ and $\RNS$, and 
the He4 envelope yields higher $\Tcore$ and $\Tbase$. Gapless superfluidity leads to a reduction of $\Tcore$ and $\Qimp$ and an 
increase of $\Tbase$, but the changes are only significant for the He9 envelope models. 

In all considered cases, gapless superfluidity provides a better fit to the cooling data as shown in Fig.~\ref{fig:KSnHD_Cooling}. However, 
the marginalized posterior probability distribution of the normalized effective superfluid velocity is rather broad: the existing cooling data  
are not very constraining to accurately determine the value of $\Vn/\VLn$. Contrary to the cooling models assuming BCS superfluidity, both He4 and He9 
envelope models fit the data equally well: small deviations appear only at early and late times but lie within the current observational uncertainties.  At variance with models based on BCS superfluidity, models allowing for gapless superfluidity predict that KS~1731$-$260 has been further cooling since the last observation in 2016~\cite{Merritt2016} and thermal equilibrium will be reach within the next decade or so depending 
on the composition of the envelope.

\begin{table*}
\caption{Parameters of our best cooling models for KS~1731$-$260 within the model of Gusakov\&Chugunov~\cite{GusakovChugunov2020,GusakovChugunov2021} of accreted neutron stars considering BCS (i.e. $\Vn/\VLn=0$) or gapless (i.e. $\Vn/\VLn> 1$) superfluidity. Results are displayed with $68\%$ uncertainty level. The labels ``He9'' and ``He4'' refer to the envelope composition. See text for details.}
\label{tab:KS}
\begin{tabular}{@{}ccccccccccccccccc@{}}
\toprule
Envelope && ($\MNS$, $\RNS$) && $\Tcore$ (10$^7$K) && $\Qimp$ && $\Tbase$ (10$^8$K) && $\Vn/\VLn$ \\ \midrule
   He9  && (1.62$\MSol$,~11.2~km)  && 4.69$\pm$0.14&& 10.56$^{+2.15}_{-1.97}$ && 2.45$^{+0.19}_{-0.18}$ && 0  \\
   He4  && (1.62$\MSol$,~11.2~km)  && 9.40$\pm$0.28&& 8.42$^{+2.18}_{-2.08}$ && 4.32$^{+0.20}_{-0.19}$ && 0  \\
   \midrule
   He9  && (1.4$\MSol$,~10~km)  && 4.45$\pm$0.13&& 12.71$^{+2.45}_{-2.19}$ && 2.23$^{+0.18}_{-0.17}$ && 0  \\
   He4  && (1.4$\MSol$,~10~km)  && 8.96$^{+0.26}_{-0.27}$&& 10.67$^{+2.45}_{-2.31}$ && 4.10$^{+0.21}_{-0.20}$ && 0  \\
   \midrule
   \midrule
   He9  && (1.62$\MSol$,~11.2~km)  && 3.99$^{+0.26}_{-0.34}$&& 5.80$^{+1.68}_{-1.25}$ && 3.13$^{+0.19}_{-0.20}$ && 1.21$^{+0.10}_{-0.11}$  \\
   He4  && (1.62$\MSol$,~11.2~km)  && 8.43$^{+0.71}_{-0.92}$ && 8.74$^{+9.62}_{-4.40}$ && 4.68$^{+0.31}_{-0.34}$ && 1.20$^{+0.11}_{-0.12}$  \\
   \midrule
   He9  && (1.4$\MSol$,~10~km)  && 3.80$^{+0.21}_{-0.25}$&& 6.04$^{+1.40}_{-1.09}$ && 3.04$^{+0.19}_{-0.20}$ && 1.18$^{+0.11}_{-0.09}$  \\
   He4  && (1.4$\MSol$,~10~km)  && 7.84$^{+0.78}_{-1.01}$&& 10.68$^{+11.36}_{-4.85}$ && 4.50$^{+0.32}_{-0.36}$ && 1.21$^{+0.10}_{-0.12}$  \\
\bottomrule
\end{tabular}
\end{table*}

\subsection{MXB~1659$-$29 with BCS superfluidity}

\subsubsection{Outburst I}

\begin{figure*}
    \centering
  \begin{subfigure}[b]{0.45\textwidth}
        \centering\includegraphics[width=\textwidth]{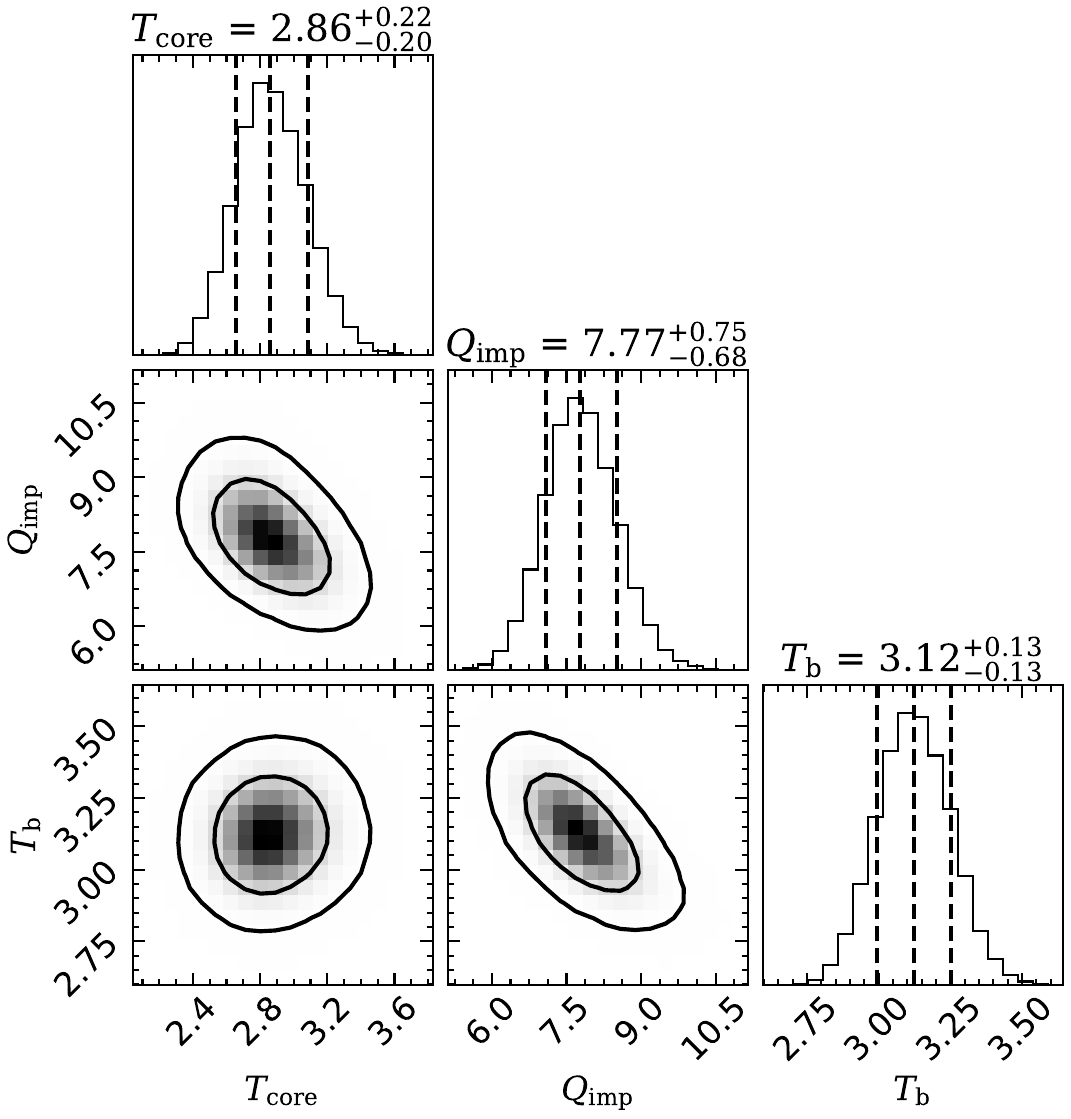}
        \caption{With He9 envelope.}
    \end{subfigure}
    \begin{subfigure}[b]{0.45\textwidth}
        \centering\includegraphics[width=\textwidth]{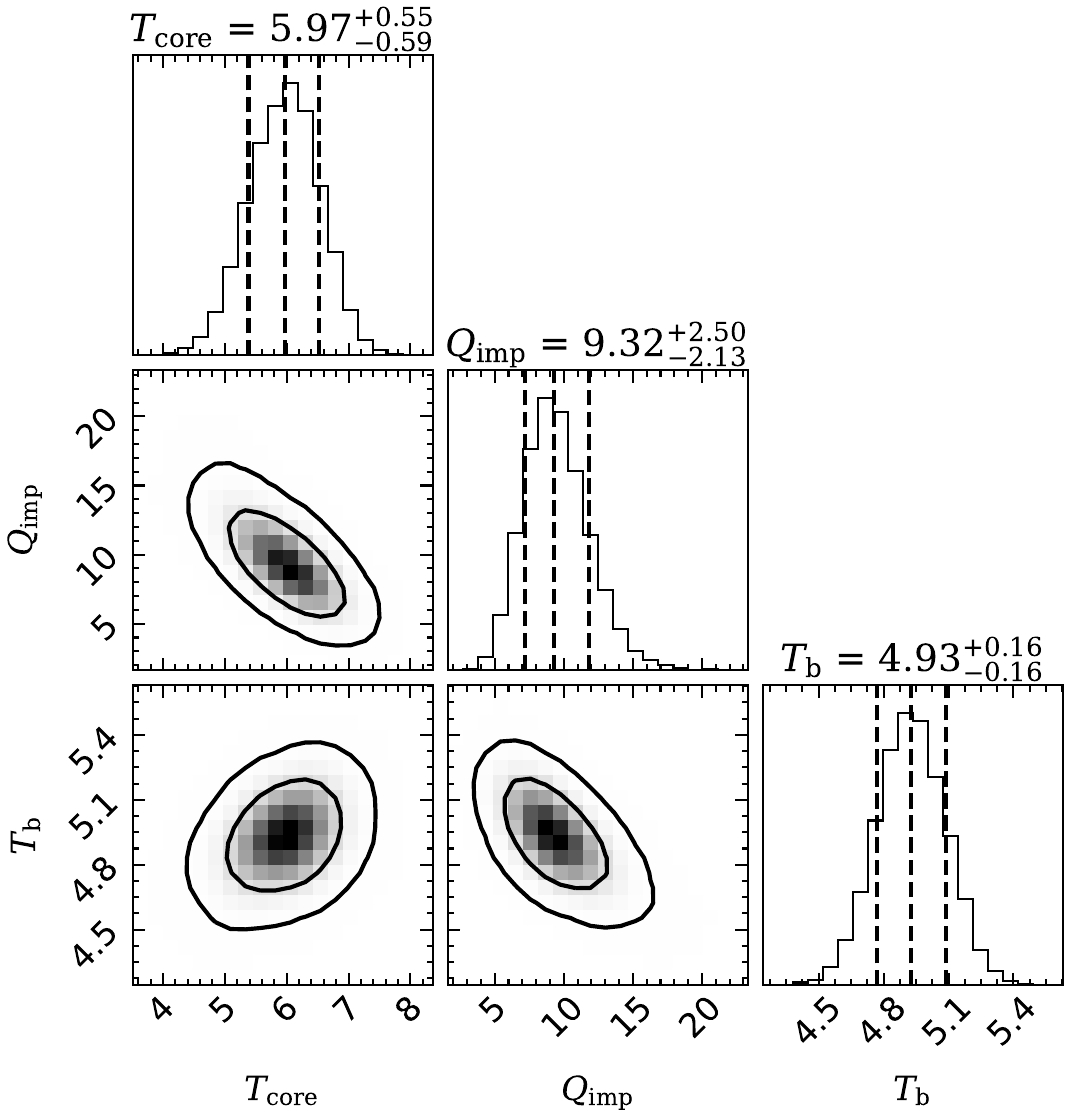}
        \caption{With He4 envelope.}
    \end{subfigure}
    \caption{Marginalized 1-D and 2-D probability distributions for the parameters of our cooling model of MXB~1659$-$29 (outburst I) within the model of Gusakov\&Chugunov~\cite{GusakovChugunov2020,GusakovChugunov2021} of accreted neutron stars in the absence of superflow (BCS regime) using the realistic neutron pairing calculations of Ref.~\cite{Gandolfi2022}. Results were obtained setting $\MNS=1.62\MSol$ and $\RNS=11.2$~km with the envelope model He9 (left panel) or He4 (right panel). $\Tcore$ and $\Tbase$ are expressed in units of $10^7$~K and $10^8$~K respectively. The dotted lines in the histograms mark the median value and the 68\% uncertainty level while the contours in the 2-D probability distributions correspond to 68\% and 95\% confidence ranges. The associated cooling curves, using the median values, are displayed in Fig~\ref{fig:MXBOutburstInHD_Cooling}.}
    \label{fig:Corner_MXBOutburstI-BCS}
\end{figure*}

\begin{figure*}
    \centering
    \begin{subfigure}[b]{0.45\textwidth}
        \centering\includegraphics[width=\textwidth]{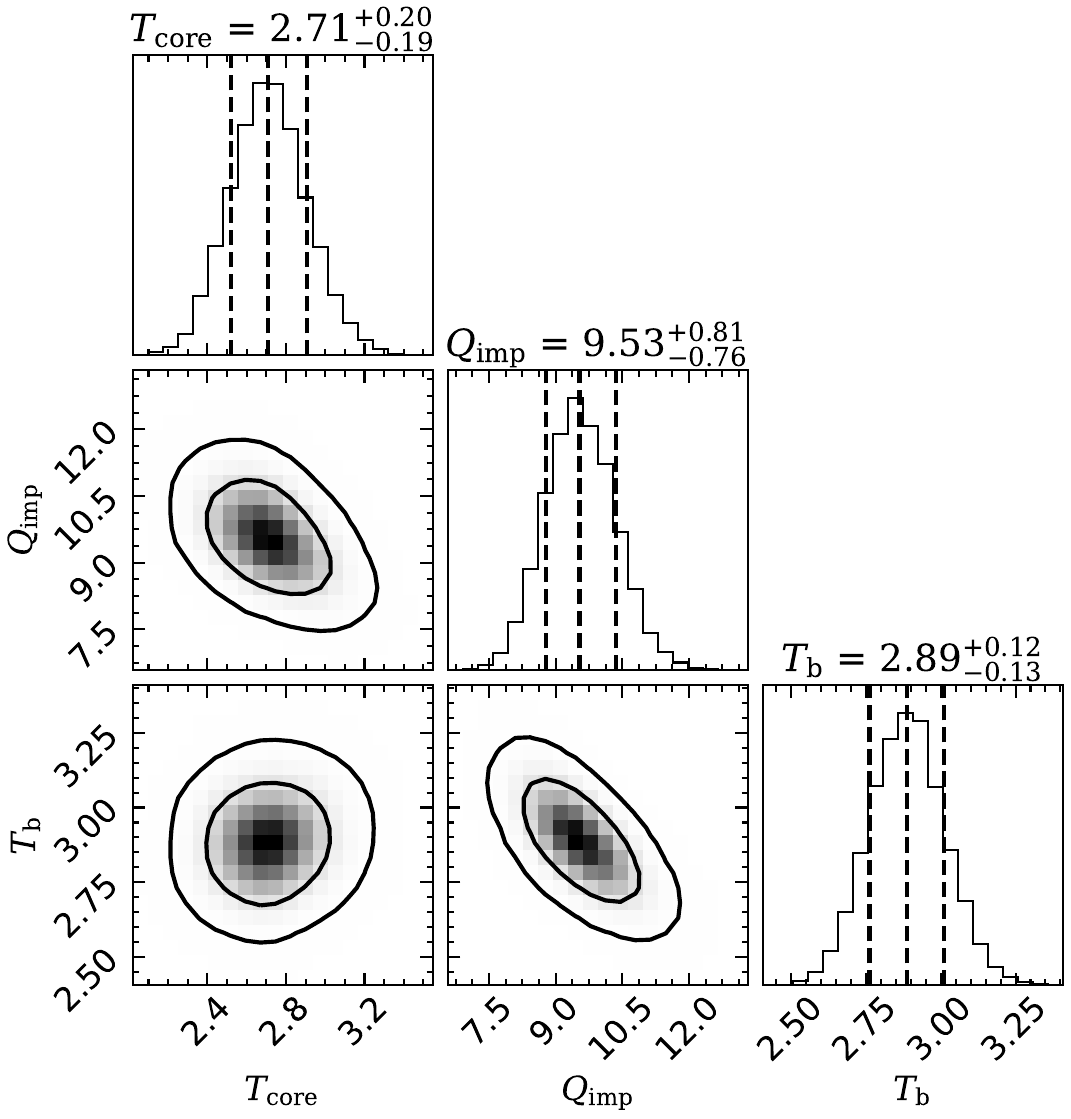}
        \caption{With He9 envelope.}
    \end{subfigure}
    \begin{subfigure}[b]{0.45\textwidth}
        \centering\includegraphics[width=\textwidth]{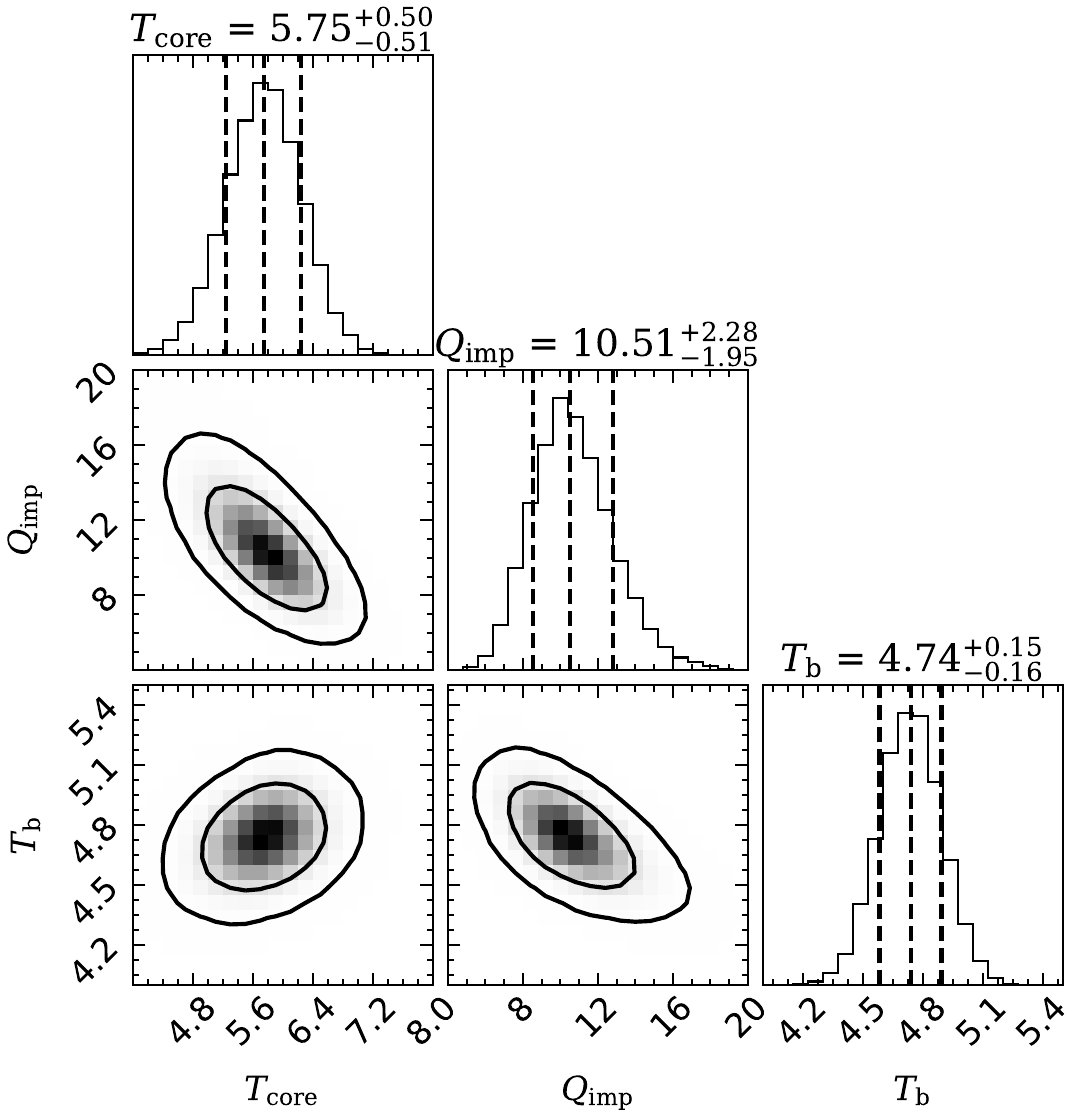}
        \caption{With He4 envelope.}
    \end{subfigure}
    \caption{Same as Fig.~\ref{fig:Corner_MXBOutburstI-BCS} for $\MNS=1.4\MSol$ and $\RNS=10$~km.}
    \label{fig:Corner_MXBOutburstI-BCS-canonical}
\end{figure*}

\begin{figure*}[h]
 \centering
    \begin{subfigure}[b]{0.45\textwidth}
        \centering\includegraphics[width=\textwidth]{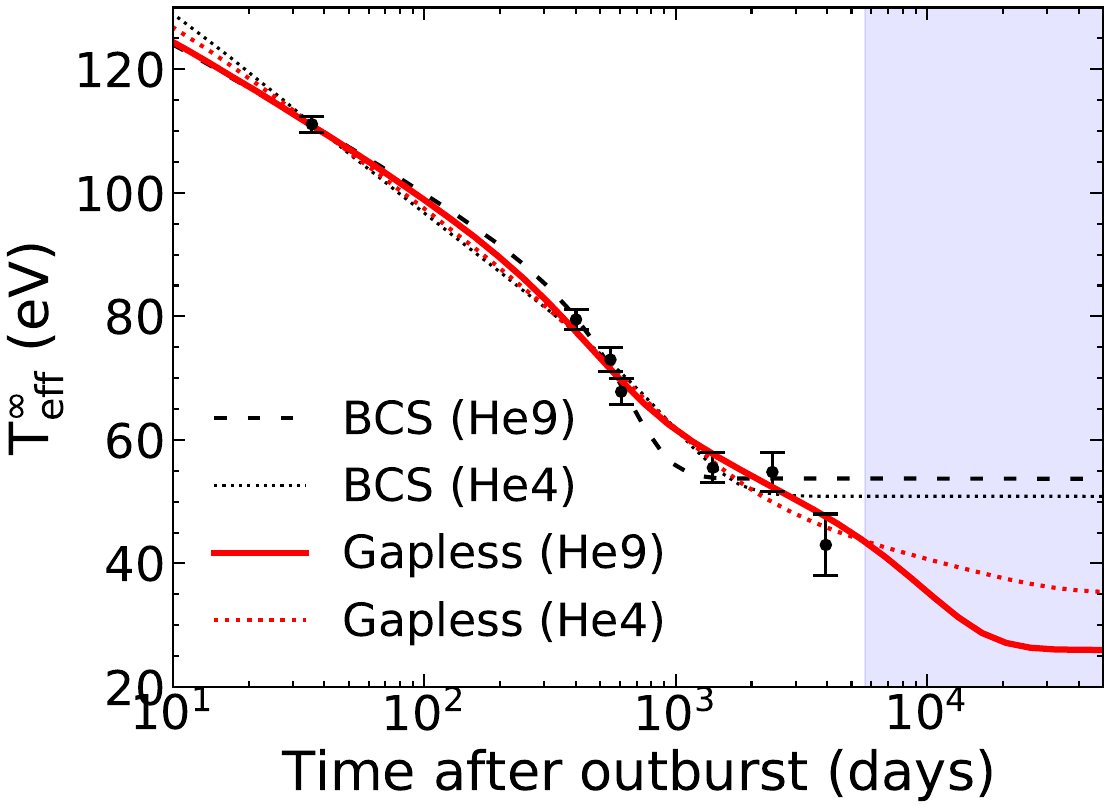}
        \caption{$\MNS=1.62\MSol$ and $\RNS=11.2$~km.}
    \end{subfigure}
    \begin{subfigure}[b]{0.45\textwidth}
        \centering\includegraphics[width=\textwidth]{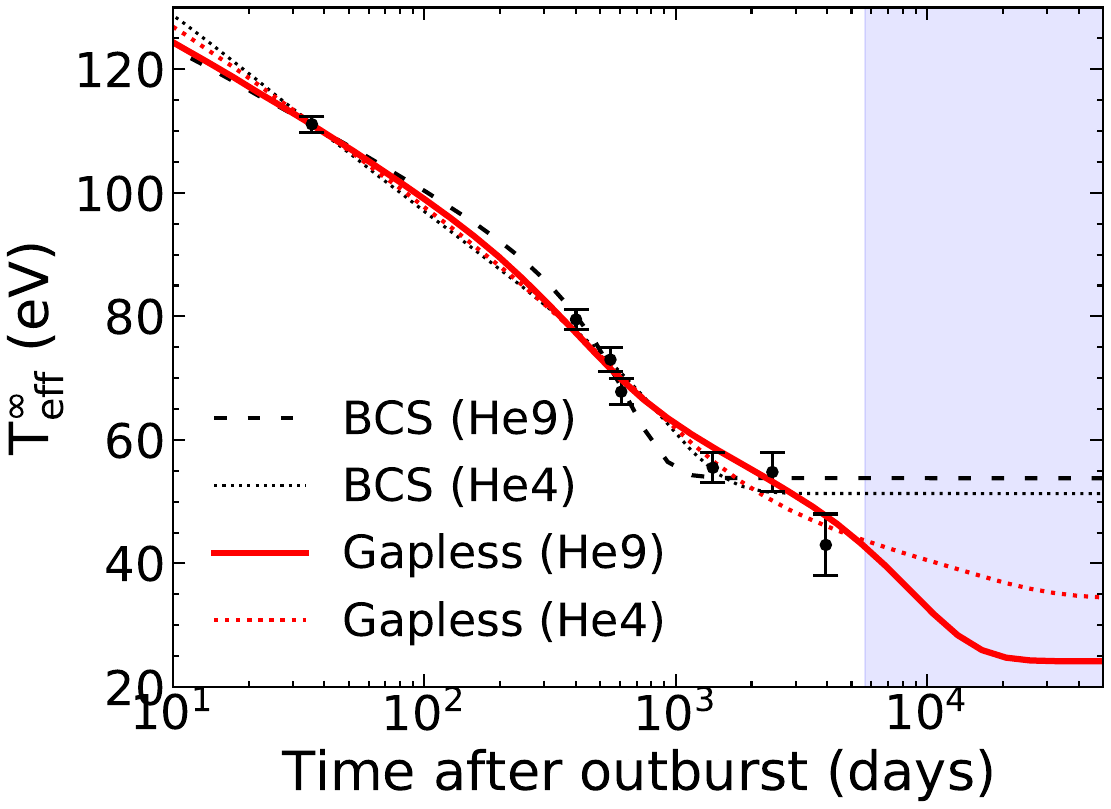}
        \caption{$\MNS=1.4\MSol$ and $\RNS=10$~km.}
    \end{subfigure}
\caption{Evolution of the effective surface temperature of MXB~1659$-$29 in electronvolts (as seen by an observer at infinity) as a function of the time in days after the end of outburst I within the model of Gusakov\&Chugunov~\cite{GusakovChugunov2020,GusakovChugunov2021} and assuming a neutron star with $\MNS=1.62\MSol$ and $\RNS=11.2$~km (left panel) or the canonical values $\MNS=1.4\MSol$ and $\RNS=10$~km (right panel). Symbols represent observational data with error bars. The black and red lines are models considering BCS and gapless superfluidity, respectively. Dotted curves represent results obtained assuming the He9 envelope model while the black-dashed and red-solid curves assumed the He4 envelope model. All the calculations were made using the realistic neutron pairing calculations of Ref.~\cite{Gandolfi2022}. The shaded area corresponds to the second accretion phase (which occurred in 2015) and its subsequent cooling phase: the cooling curves within this region depict the expected behavior had the outburst II not occurred.}
\label{fig:MXBOutburstInHD_Cooling}
\end{figure*}

Figures~\ref{fig:Corner_MXBOutburstI-BCS} and \ref{fig:Corner_MXBOutburstI-BCS-canonical} show the 
marginalized posterior probability distributions of the model parameters for the cooling of MXB~1659$-$29 after the accretion phase from 1999 to 2001 (referred to as outburst I) within the accreted crust
model of Gusakov and Chugunov~\cite{GusakovChugunov2020,GusakovChugunov2021} in the absence of superflow (BCS regime). 
The former figure corresponds to models with $\MNS=1.62\MSol$ and $\RNS=11.2$~km as in Ref.~\cite{BrownCumming2009} while 
the latter corresponds to models with the canonical values $\MNS=1.4\MSol$ and $\RNS=10$~km. The left (right) panels correspond to the He9 (He4) envelope models. The associated cooling curves are displayed in Fig.~\ref{fig:MXBOutburstInHD_Cooling} and the parameters are given in Table~\ref{tab:MXBI}. 

As found for KS~1731$-$260, adopting canonical values for $\MNS$ and $\RNS$ has little effect on the cooling curves. 
Models based on the He9 envelope yield a slightly better fit to the cooling data consistently with the analysis of Ref.~\cite{cumming2017} within the 
traditional deep crustal heating cooling paradigm. The associated core temperatures are compatible with the values obtained in Refs.~\cite{Turlione2015,Parikh2019,Lu2022} 
without considering neutron diffusion (recalling that the last data point was discarded in these studies). However, none of these models can explain the 
observed late-time cooling. With neutron superfluidity in the BCS regime, the neutron star is cooling so rapidly that thermal equilibrium is restored about 1000 and 2000 days only after the end of the outburst for the He9 and He4 envelope models respectively.

\begin{table*}[h]
\caption{Parameters of our best cooling models for outburst I of MXB~1659$-$29 within the model of Gusakov\&Chugunov~\cite{GusakovChugunov2020,GusakovChugunov2021} of accreted neutron stars considering BCS (i.e. $\Vn/\VLn=0$) or gapless (i.e. $\Vn/\VLn>1 0$) superfluidity. Results are displayed with $68\%$ uncertainty level. The labels ``He9'' and ``He4'' refer to the envelope composition. See text for details.}
\label{tab:MXBI}
\begin{tabular}{@{}ccccccccccccccccc@{}}
\toprule
Envelope && ($\MNS$, $\RNS$) && $\Tcore$ (10$^7$K) && $\Qimp$ && $\Tbase$ (10$^8$K) && $\Vn/\VLn$ \\ \midrule
   He9  && (1.62$\MSol$,~11.2~km)  && 2.86$^{+0.22}_{-0.20}$&& 7.77$^{+0.75}_{-0.68}$ && 3.12$\pm$0.13 && 0  \\
   He4  && (1.62$\MSol$,~11.2~km)  &&  5.97$^{+0.55}_{-0.59}$ && 9.32$^{+2.50}_{-2.13}$ && 4.93$\pm$0.16 && 0  \\
   \midrule
   He9  && (1.4$\MSol$,~10~km)  && 2.71$^{+0.20}_{-0.19}$  && 9.53$^{+0.81}_{-0.76}$ && 2.89$^{+0.12}_{-0.13}$ && 0  \\
   He4  && (1.4$\MSol$,~10~km)  && 5.75$^{+0.50}_{-0.51}$ && 10.51$^{+2.28}_{-1.95}$ && 4.74$^{+0.15}_{-0.16}$ && 0  \\
   \midrule
   \midrule
   He9  && (1.62$\MSol$,~11.2~km)  && 0.65$^{+0.63}_{-0.45}$&& 16.35$^{+3.68}_{-3.58}$ && 3.14$^{+0.16}_{-0.15}$ && 1.20$\pm$0.11  \\
   He4  && (1.62$\MSol$,~11.2~km)  && 2.90$^{+1.23}_{-1.53}$ && 42.31$^{+12.40}_{-10.50}$ && 4.51$\pm$0.20 && 1.18$^{+0.12}_{-0.11}$ \\
   \midrule
   He9  && (1.4$\MSol$,~10~km)  && 0.55$^{+0.68}_{-0.39}$&& 16.88$^{+4.24}_{-4.03}$ && 2.98$\pm$0.16 && 1.17$^{+0.13}_{-0.10}$  \\
   He4  && (1.4$\MSol$,~10~km)  && 2.59$^{+1.15}_{-1.51}$ && 45.69$^{+12.28}_{-11.44}$ && 4.33$^{+0.21}_{-0.20}$ && 1.19$\pm$ 0.12  \\
\bottomrule
\end{tabular}
\end{table*}

\subsubsection{Outburst II}

\begin{figure*}
    \centering
    \begin{subfigure}[b]{0.45\textwidth}
        \centering\includegraphics[width=\textwidth]{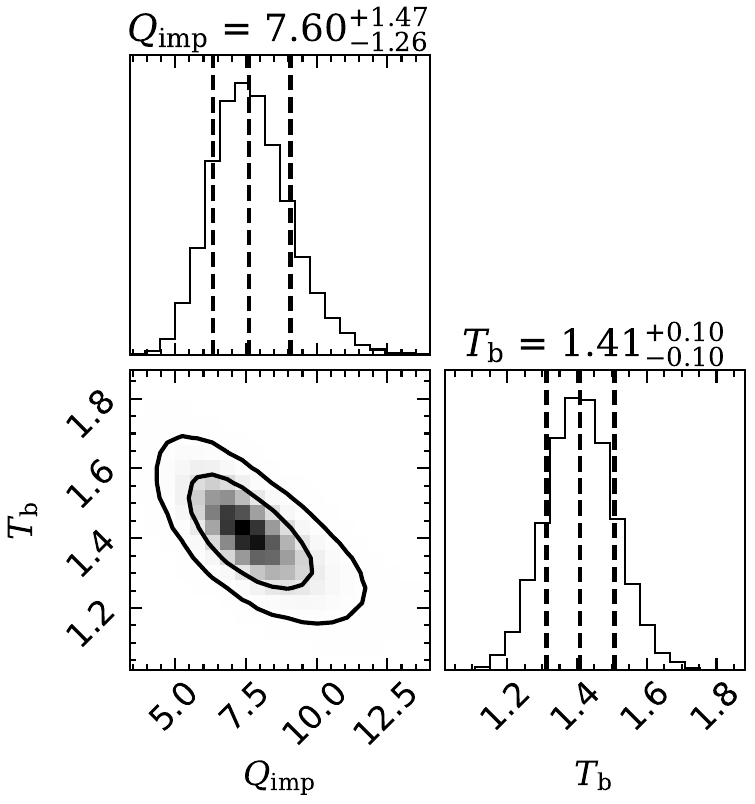}
        \caption{With He9 envelope.}
    \end{subfigure}
    \begin{subfigure}[b]{0.45\textwidth}
        \centering\includegraphics[width=\textwidth]{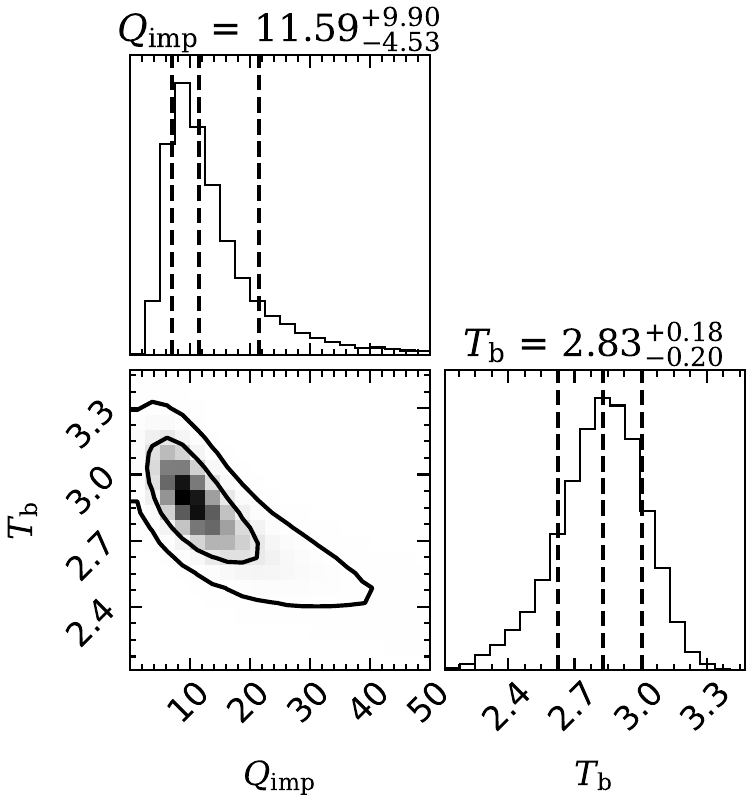}
        \caption{With He4 envelope.}
    \end{subfigure}
    \caption{Marginalized 1-D and 2-D probability distributions for the parameters of our cooling model of MXB~1659$-$29 (outburst II) within the model of Gusakov\&Chugunov~\cite{GusakovChugunov2020,GusakovChugunov2021} of accreted neutron stars in the absence of superflow (BCS regime), using the realistic neutron pairing calculations of Ref.~\cite{Gandolfi2022}. Results were obtained setting $\MNS=1.62\MSol$ and $\RNS=11.2$~km for the envelope model He9 (left panel) or He4 (right panel). $\Tbase$ are expressed in units of $10^8$~K while $\Tcore$ is fixed to the best-fit value obtained for the associated outburst I. The dotted lines in the histograms mark the median value and the 68\% uncertainty level while the contours in the 2-D probability distributions correspond to 68\% and 95\% confidence ranges. The associated cooling curves, using the median values, are displayed in Fig~\ref{fig:MXBOutburstIInHD_Cooling}.}
    \label{fig:Corner_MXBOutburstII-BCS}
\end{figure*}

\begin{figure*}
    \centering
    \begin{subfigure}[b]{0.45\textwidth}
        \centering\includegraphics[width=\textwidth]{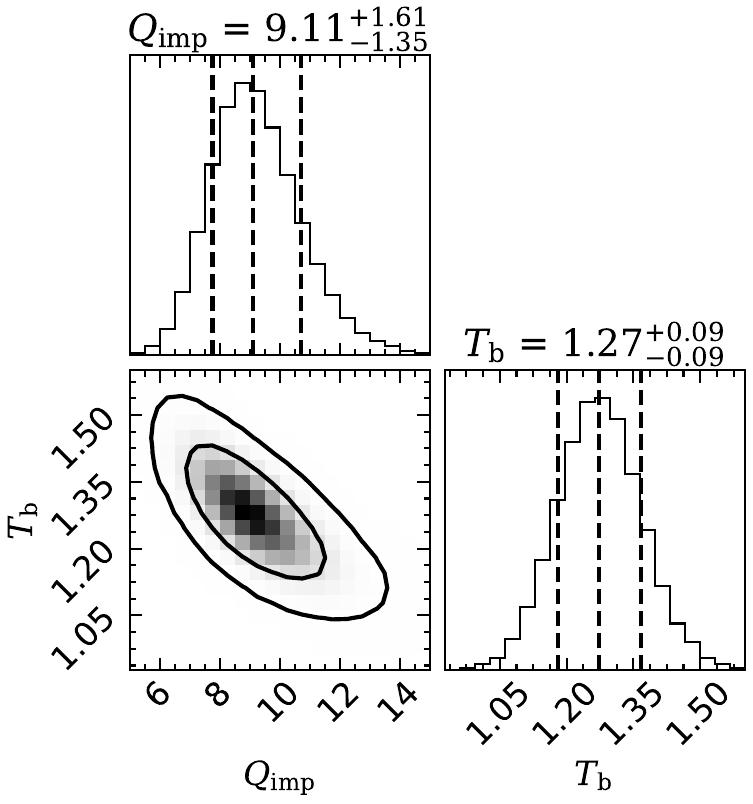}
        \caption{With He9 envelope.}
    \end{subfigure}
    \begin{subfigure}[b]{0.45\textwidth}
        \centering\includegraphics[width=\textwidth]{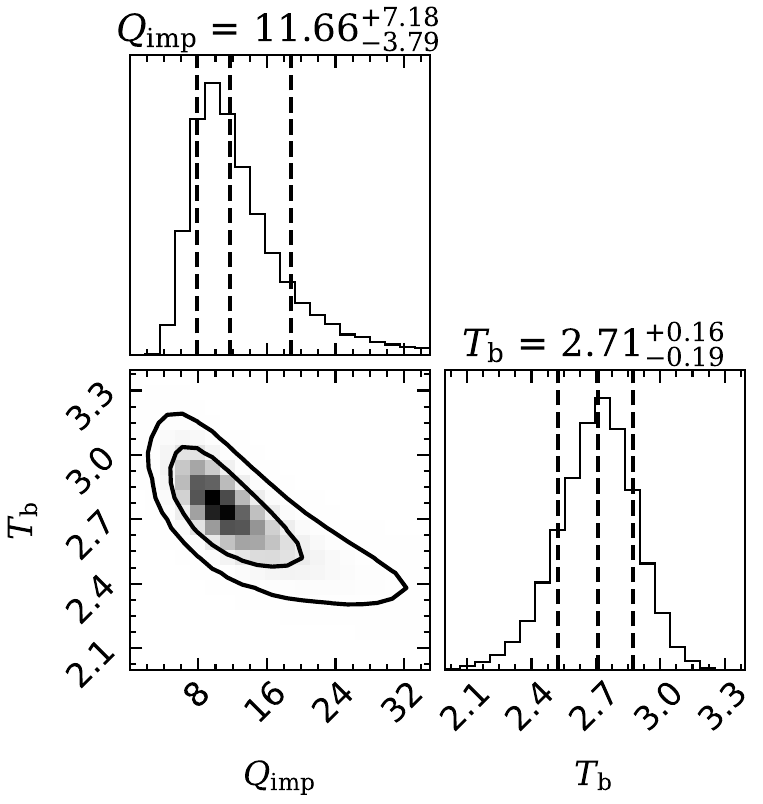}
        \caption{With He4 envelope.}
    \end{subfigure}
    \caption{Same as Fig.~\ref{fig:Corner_MXBOutburstII-BCS} for $\MNS=1.4\MSol$ and $\RNS=10$~km .}
    \label{fig:Corner_MXBOutburstII-BCS-canonical}
\end{figure*}

\begin{figure*}[h]
 \centering
    \begin{subfigure}[b]{0.45\textwidth}
        \centering\includegraphics[width=\textwidth]{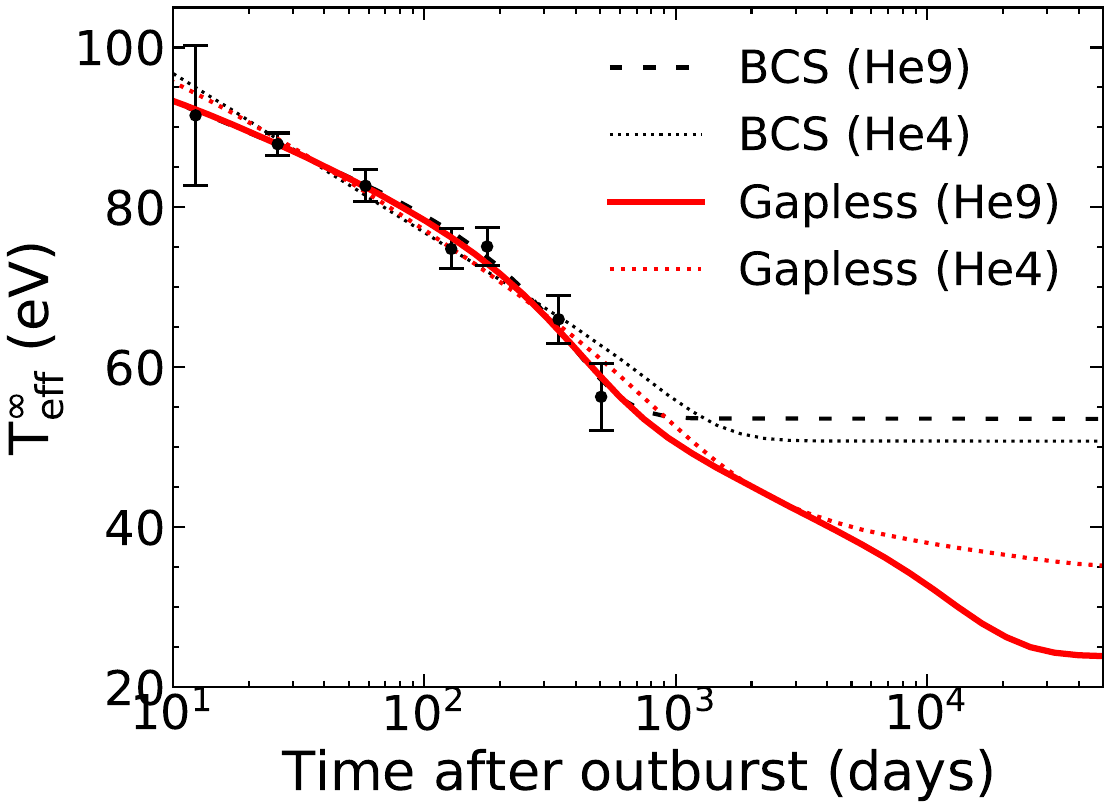}
        \caption{$\MNS=1.62\MSol$ and $\RNS=11.2$~km.}
    \end{subfigure}
    \begin{subfigure}[b]{0.45\textwidth}
        \centering\includegraphics[width=\textwidth]{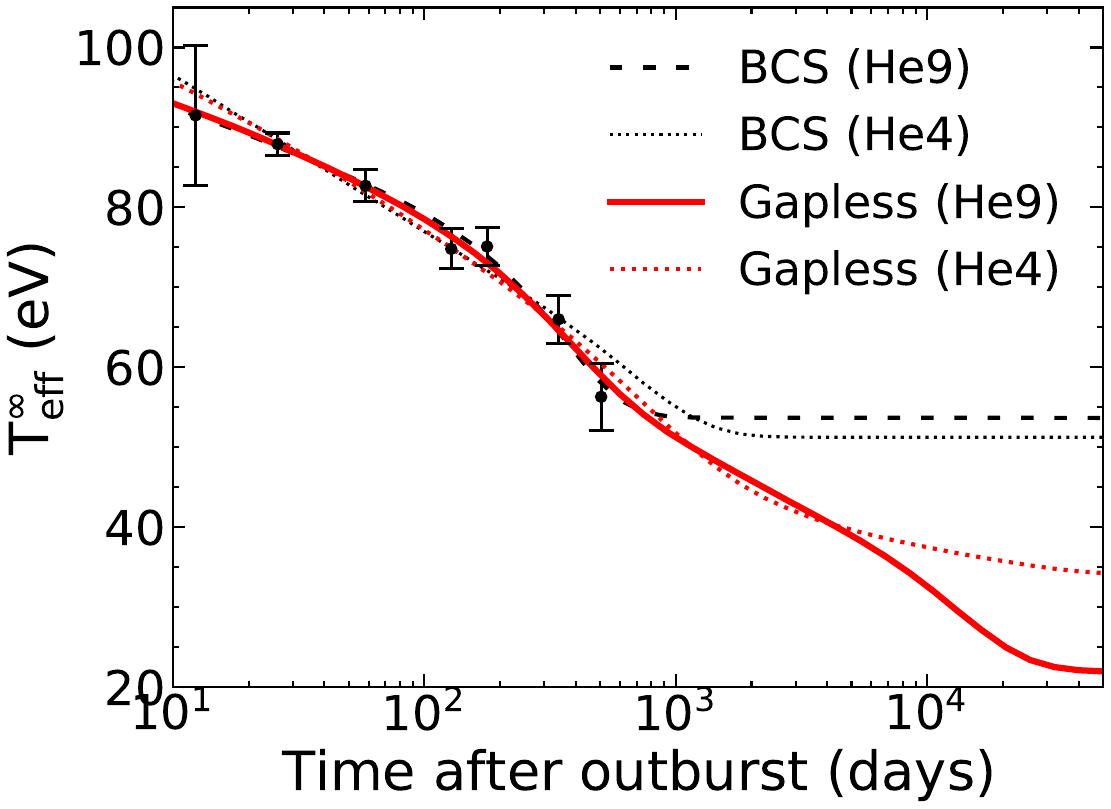}
        \caption{$\MNS=1.4\MSol$ and $\RNS=10$~km.}
    \end{subfigure}
\caption{Evolution of the effective surface temperature of MXB~1659$-$29 in electronvolts (as seen by an observer at infinity) as a function of the time in days after outburst II within the model of Gusakov\&Chugunov~\cite{GusakovChugunov2020,GusakovChugunov2021} and assuming a neutron star with $\MNS=1.62\MSol$ and $\RNS=11.2$~km (left panel) or the canonical values $\MNS=1.4\MSol$ and $\RNS=10$~km (right panel). Symbols represent observational data with error bars. The black and red lines are models considering BCS and gapless superfluidity, respectively. Dotted curves represent results obtained assuming the He9 envelope model while the black-dashed and red-solid curves assumed the He4 envelope model. All the calculations were made using the realistic neutron pairing calculations of Ref.~\cite{Gandolfi2022}.}
\label{fig:MXBOutburstIInHD_Cooling}
\end{figure*}

In 2015, MXB~1659$-$29 reentered into a second accretion phase lasting 1.7~years (referred to as outburst II). The marginalized posterior probability distributions of the model parameters for the subsequent cooling phase are displayed in Figs.~\ref{fig:Corner_MXBOutburstII-BCS} and \ref{fig:Corner_MXBOutburstII-BCS-canonical}. These results have been obtained with the core temperature $\Tcore$ fixed to the value inferred from outburst I since variations of $\Tcore$ between the two outbursts and the subsequent quiescent periods are
expected to lie within the observational uncertainties~\cite{Parikh2019}. The associated cooling curves are displayed in Fig.~\ref{fig:MXBOutburstIInHD_Cooling} and the parameters are given in Table~\ref{tab:MXBII}.

Again, the cooling models with different $\MNS$ and $\RNS$ yield similar results. 
Consistently with the study of Ref.~\cite{Parikh2019} (in which neutron diffusion was not taken into account), no significant change of $\Qimp$ is observed 
between outbursts I and II at variance with $\Tbase$; this parameter is
related to shallow heating and does not need to remain the same. Looking at Fig.~\ref{fig:MXBOutburstIInHD_Cooling}, all the observations are well
reproduced by the models with the He9 envelope, thus corroborating the results obtained for outburst I. 

We find that MXB~1659$-$29 reached thermal equilibrium $\sim$1000~days ($\sim$2000~days) after the end of outburst II for the He9 (He4) envelope model.

\begin{table*}[h]
\caption{Parameters of our best cooling models for outburst  II of  MXB~1659$-$29 within the model of Gusakov\&Chugunov~\cite{GusakovChugunov2020,GusakovChugunov2021} of accreted neutron stars considering BCS (i.e. $\Vn/\VLn=0$) or gapless (i.e. $\Vn/\VLn>1$) superfluidity. Results are displayed with $68\%$ uncertainty level. $\Tcore$ was fixed to the associated value from outburst I. The labels ``He9'' and ``He4'' refer to the envelope composition. See text for details}
\label{tab:MXBII}
\begin{tabular}{@{}ccccccccccccccccc@{}}
\toprule
Envelope && ($\MNS$, $\RNS$) && $\Tcore$ (10$^7$K) && $\Qimp$ && $\Tbase$ (10$^8$K) && $\Vn/\VLn$ \\ \midrule
   He9  && (1.62$\MSol$,~11.2~km)  && \textbf{2.86} && 7.60$^{+1.47}_{-1.26}$ && 1.41$\pm$0.10 && 0  \\
   He4  && (1.62$\MSol$,~11.2~km)  &&  \textbf{5.97} && 11.59$^{+9.90}_{-4.53}$ && 2.83$^{+0.18}_{-0.20}$ && 0  \\
   \midrule
   He9  && (1.4$\MSol$,~10~km)  && \textbf{2.71} && 9.11$^{+1.61}_{-1.35}$ && 1.27$\pm$0.09 && 0  \\
   He4  && (1.4$\MSol$,~10~km)  && \textbf{5.75} && 11.66$^{+7.18}_{-3.79}$ && 2.71$^{+0.16}_{-0.19}$ && 0  \\
   \midrule
   \midrule
   He9  && (1.62$\MSol$,~11.2~km)  && \textbf{0.65} && 26.68$^{+10.84}_{-7.81}$ && 1.38$^{+0.11}_{-0.12}$ && 1.18$\pm$0.12 \\
   He4  && (1.62$\MSol$,~11.2~km)  && \textbf{2.90} && 65.29$^{+21.75}_{-20.36}$ && 2.56$^{+0.16}_{-0.12}$ && 1.18$\pm$0.12 \\
   \midrule
   He9  && (1.4$\MSol$,~10~km)  && \textbf{0.55}&& 30.51$^{+12.43}_{-9.40}$ && 1.26$^{+0.12}_{-0.10}$ && 1.19$\pm$0.12  \\
   He4  && (1.4$\MSol$,~10~km)  && \textbf{2.59} && 64.77$^{+21.20}_{-19.14}$ && 2.45$^{+0.16}_{-0.12}$ && 1.18$\pm$ 0.12  \\
\bottomrule
\end{tabular}
\end{table*}

\subsection{MXB~1659$-$29 with gapless superfluidity}
\subsubsection{Outburst I}

\begin{figure*}
    \centering
    \begin{subfigure}[b]{0.45\textwidth}
        \centering\includegraphics[width=\textwidth]{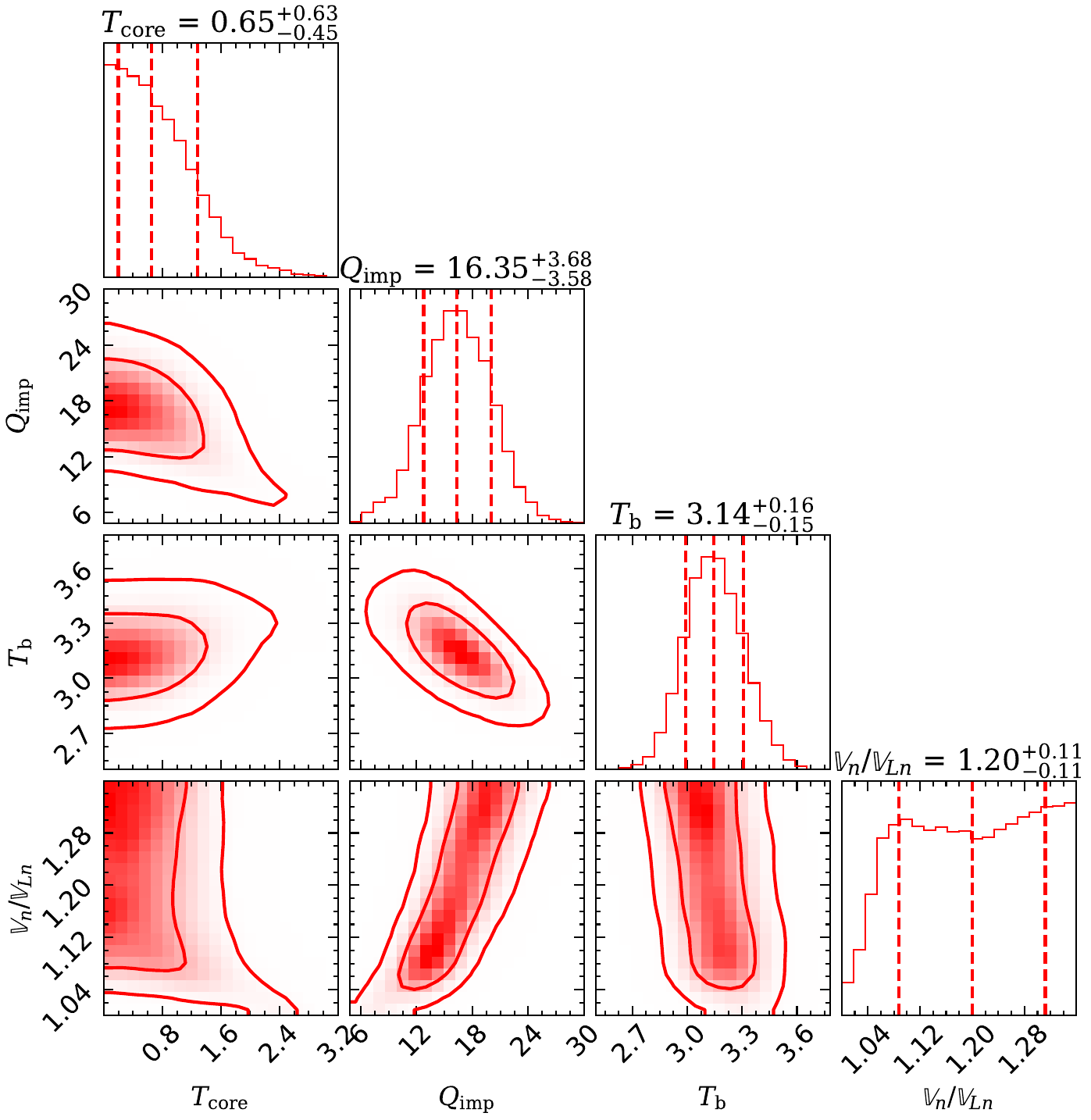}
        \caption{With He9 envelope.}
    \end{subfigure}
    \begin{subfigure}[b]{0.45\textwidth}
        \centering\includegraphics[width=\textwidth]{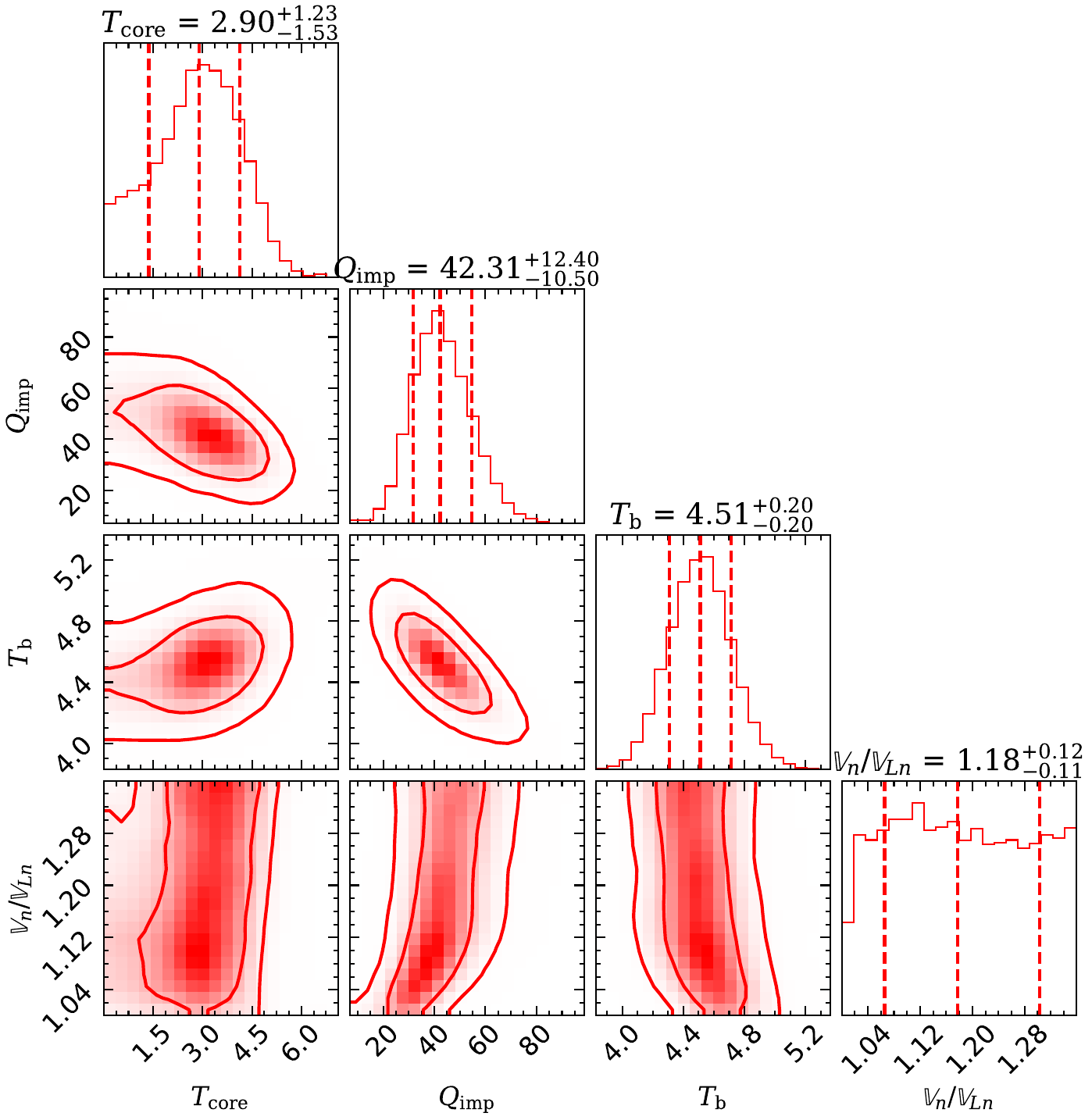}
        \caption{With He4 envelope.}
    \end{subfigure}
    \caption{Marginalized 1-D and 2-D probability distributions for the parameters of our cooling model of MXB~1659$-$29 (outburst I) within the model of Gusakov\&Chugunov~\cite{GusakovChugunov2020,GusakovChugunov2021} of accreted neutron stars but considering gapless superfluidity, using the realistic neutron pairing calculations of Ref.~\cite{Gandolfi2022}. Results were obtained setting $\MNS=1.62\MSol$ and $\RNS=11.2$~km with the envelope model He9 (left panel) and He4 (right panel). $\Tcore$ and $\Tbase$ are expressed in units of $10^7$~K and $10^8$~K respectively. The dotted lines in the histograms mark the median value and the 68\% uncertainty level while the contours in the 2-D probability distributions correspond to 68\% and 95\% confidence ranges. The associated cooling curves, using the median values, are displayed in Fig~\ref{fig:MXBOutburstInHD_Cooling}.}
    \label{fig:Corner_MXBOutburstI-Gapless}
\end{figure*}

\begin{figure*}
    \centering
    \begin{subfigure}[b]{0.45\textwidth}
        \centering\includegraphics[width=\textwidth]{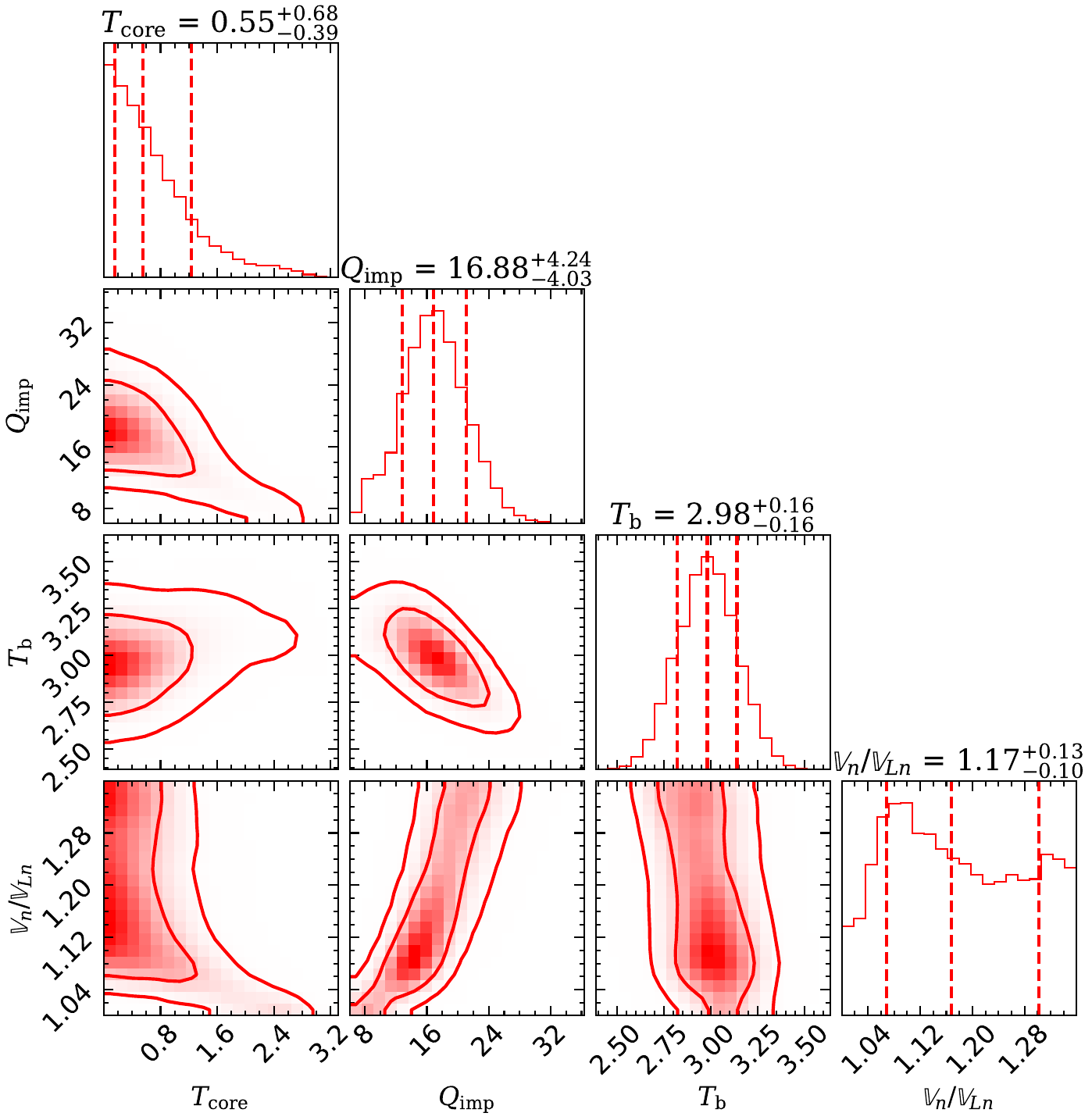}
        \caption{With He9 envelope.}
    \end{subfigure}
    \begin{subfigure}[b]{0.45\textwidth}
        \centering\includegraphics[width=\textwidth]{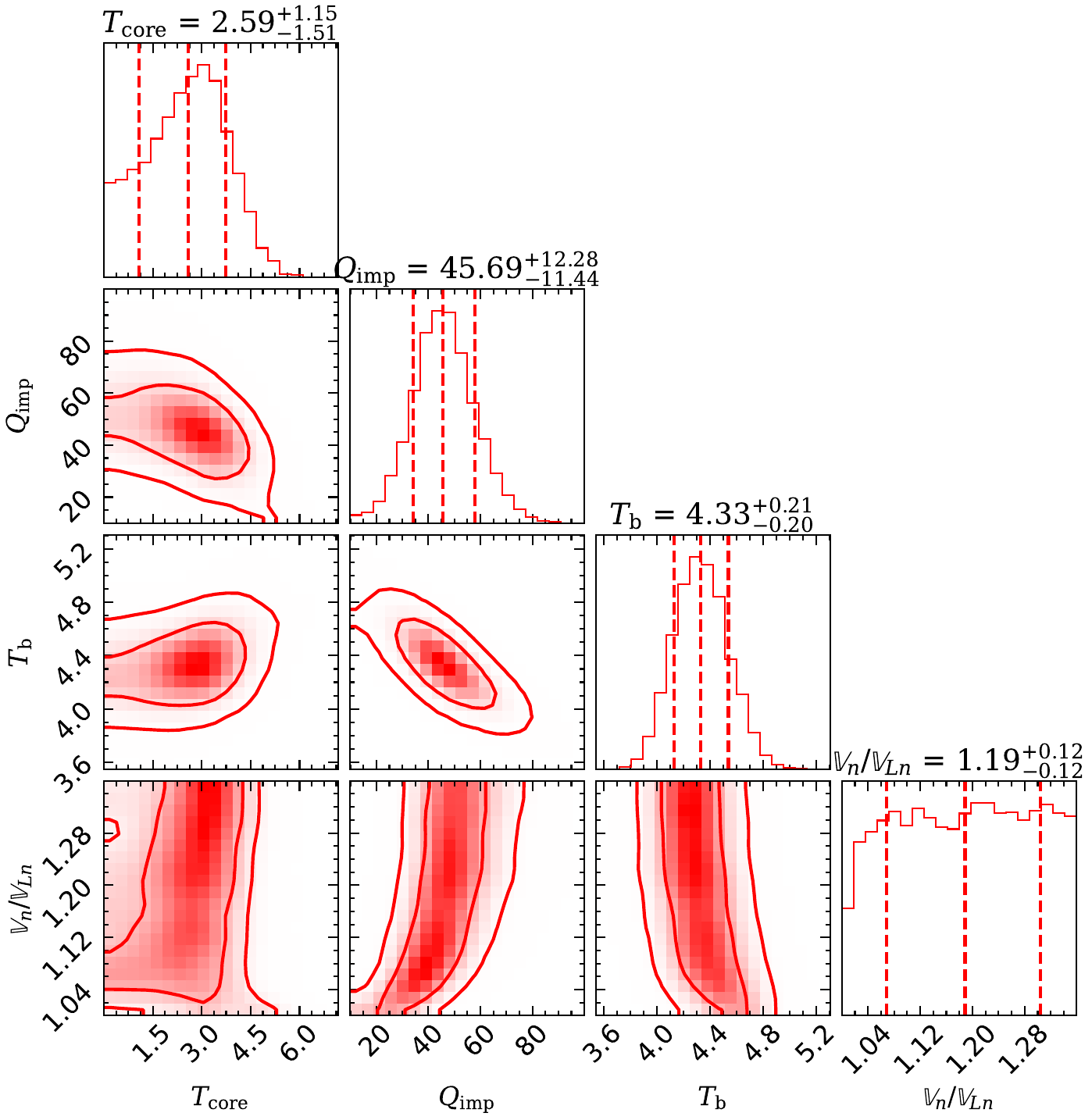}
        \caption{With He4 envelope.}
    \end{subfigure}
    \caption{Same as Fig.~\ref{fig:Corner_MXBOutburstI-Gapless} for $\MNS=1.4\MSol$ and $\RNS=10$~km.}
    \label{fig:Corner_MXBOutburstI-Gapless-canonical}
\end{figure*}

Figures~\ref{fig:Corner_MXBOutburstI-Gapless} and \ref{fig:Corner_MXBOutburstI-Gapless-canonical} show the marginalized posterior probability distributions of the model parameters for the cooling of MXB~1659$-$29 (outburst I) within the accreted crust model of Gusakov and Chugunov~\cite{GusakovChugunov2020,GusakovChugunov2021} allowing for gapless superfluidity. 
The former figure corresponds to models with $\MNS=1.62\MSol$ and $\RNS=11.2$~km as in Ref.~\cite{BrownCumming2009} while 
the latter corresponds to models with the canonical values $\MNS=1.4\MSol$ and $\RNS=10$~km.
The left (right) panels correspond to the He9 (He4) envelope models. 
Associated cooling curves are displayed in Fig.~\ref{fig:MXBOutburstInHD_Cooling} and the parameters are given in Table~\ref{tab:MXBI}. 

As in the models with BCS superfluidity, we do not find any noticeable difference in the cooling curves when changing the global 
structure of the neutron star. Because the temperature $\Tbase$ is related to shallow heating in the outer crust, it is essentially
unaffected by the presence of superflow. However, the increase of $\Qimp$ and the decrease of $\Tcore$ are more significant than in
the case of KS~1731$-$260. This mainly stems from the drop of temperature inferred from the last observation, which can now be 
explained. 

As found for KS~1731$-$260, the cooling data are best reproduced by models with gapless superfluidity although the actual neutron 
effective superfluidity velocity remains uncertain. 

Unlike models with BCS superfluidity, our models with gapless superfluidity indicate that MXB~1659$-$29 was still cooling
before outburst II in 2015 (about 5650~days after the end of outburst I): if this source had remained quiescent, the thermal 
equilibrium would have been reached about 20000 and 40000~days after the end of outburst I for He9 and He4 envelope models
respectively.

\subsubsection{Outburst II}

\begin{figure*}
    \centering
    \begin{subfigure}[b]{0.45\textwidth}
        \centering\includegraphics[width=\textwidth]{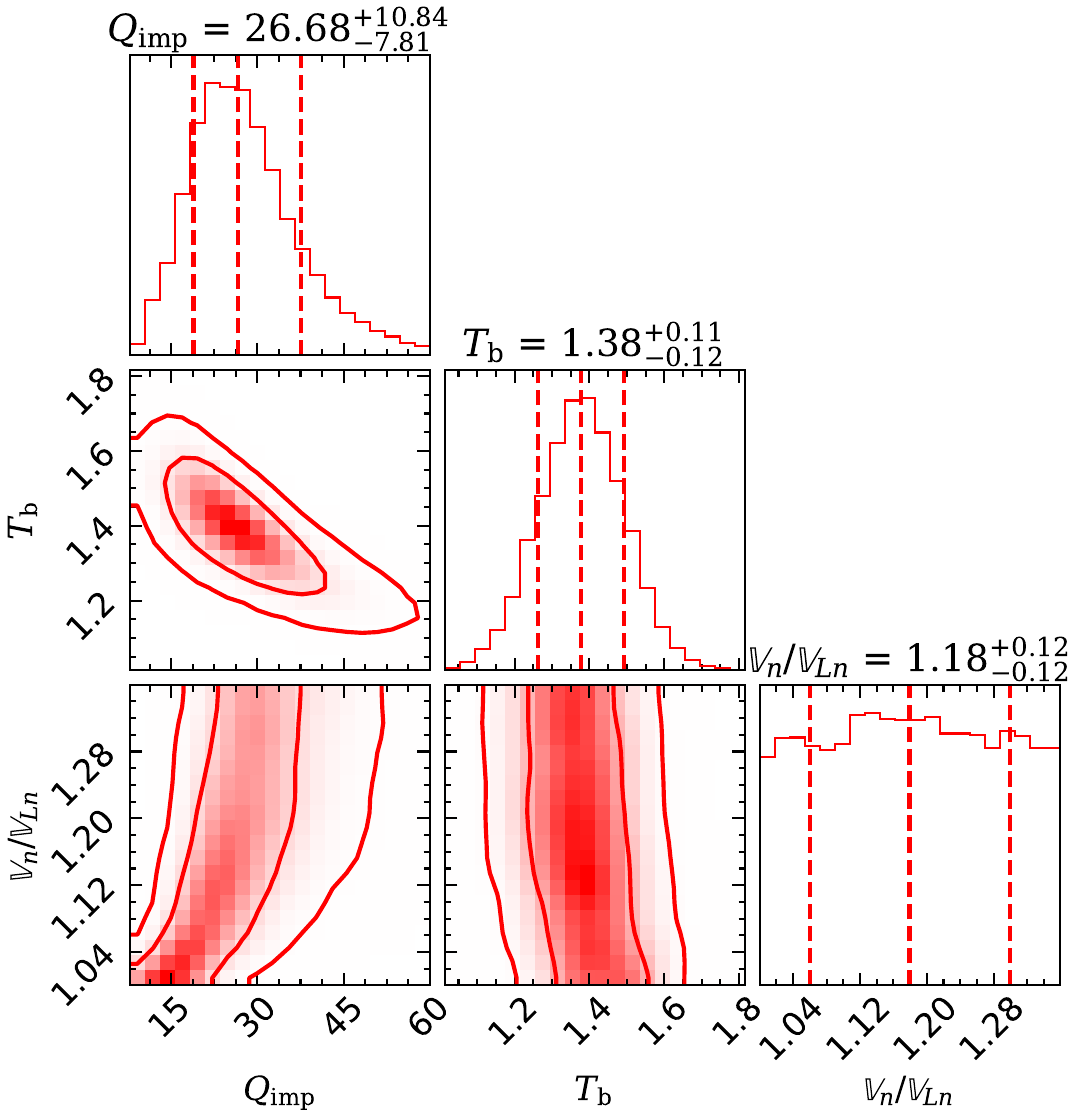}
        \caption{With He9 envelope}
    \end{subfigure}
    \begin{subfigure}[b]{0.45\textwidth}
        \centering\includegraphics[width=\textwidth]{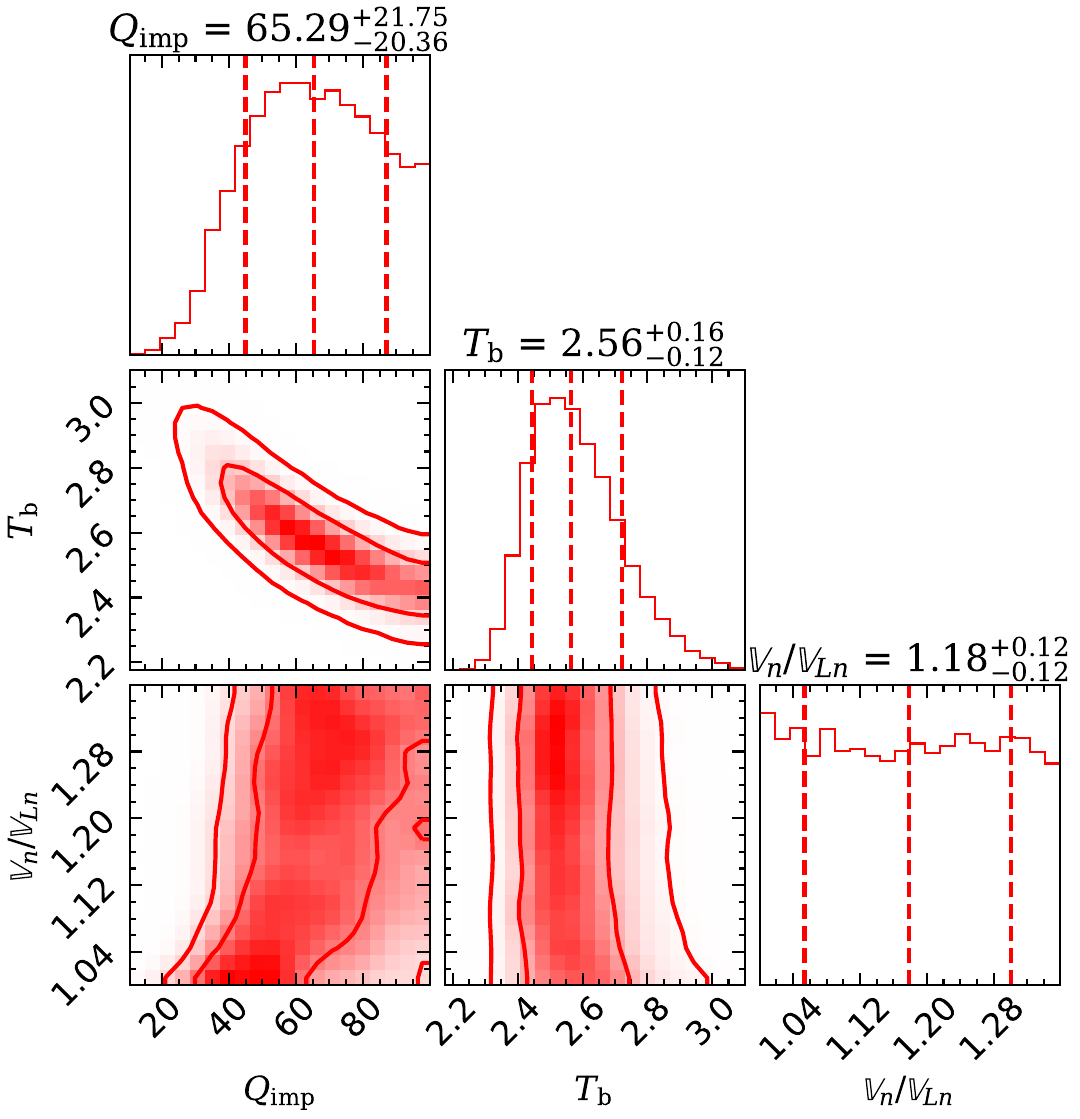}
        \caption{With He4 envelope.}
    \end{subfigure}
    \caption{Marginalized 1-D and 2-D probability distributions for the parameters of our cooling model of MXB~1659$-$29 (outburst II) within the model of Gusakov\&Chugunov~\cite{GusakovChugunov2020,GusakovChugunov2021} of accreted neutron stars but considering gapless superfluidity, using the realistic neutron pairing calculations of Ref.~\cite{Gandolfi2022}. Results were obtained setting $\MNS=1.62\MSol$ and $\RNS=11.2$~km with the envelope model  He9 (left panel) or He4 (right panel). $\Tbase$ are expressed in units of $10^8$~K while $\Tcore$ is fixed to the best-fit value obtained for the associated outburst I. The dotted lines in the histograms mark the median value and the 68\% uncertainty level while the contours in the 2-D probability distributions correspond to 68\% and 95\% confidence ranges. The associated cooling curves, using the median values, are displayed in Fig~\ref{fig:MXBOutburstIInHD_Cooling}.}
    \label{fig:Corner_MXBOutburstII-Gapless}
\end{figure*}

\begin{figure*}
    \centering
    \begin{subfigure}[b]{0.45\textwidth}
        \centering\includegraphics[width=\textwidth]{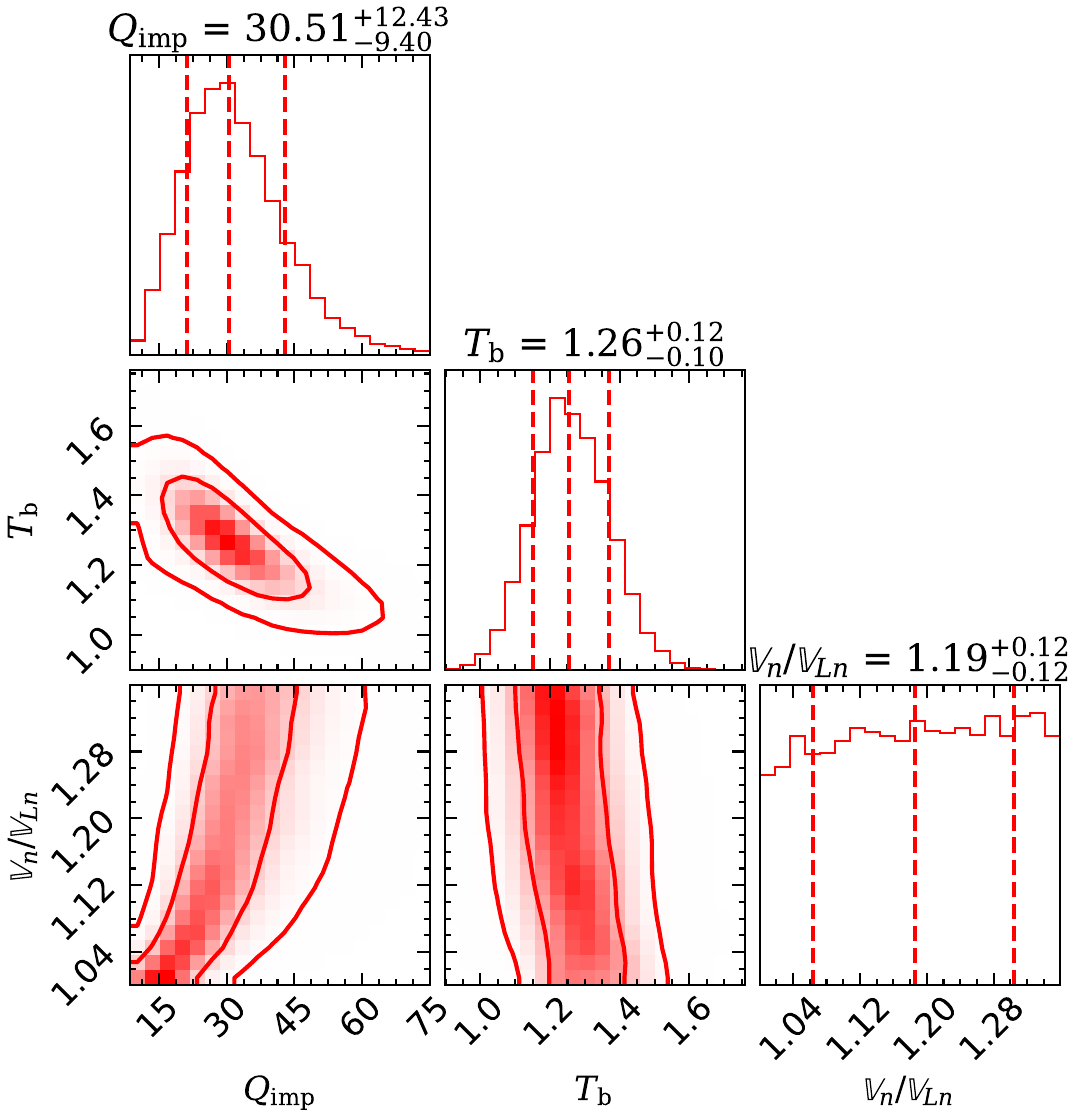}
        \caption{With He9 envelope.}
    \end{subfigure}
    \begin{subfigure}[b]{0.45\textwidth}
        \centering\includegraphics[width=\textwidth]{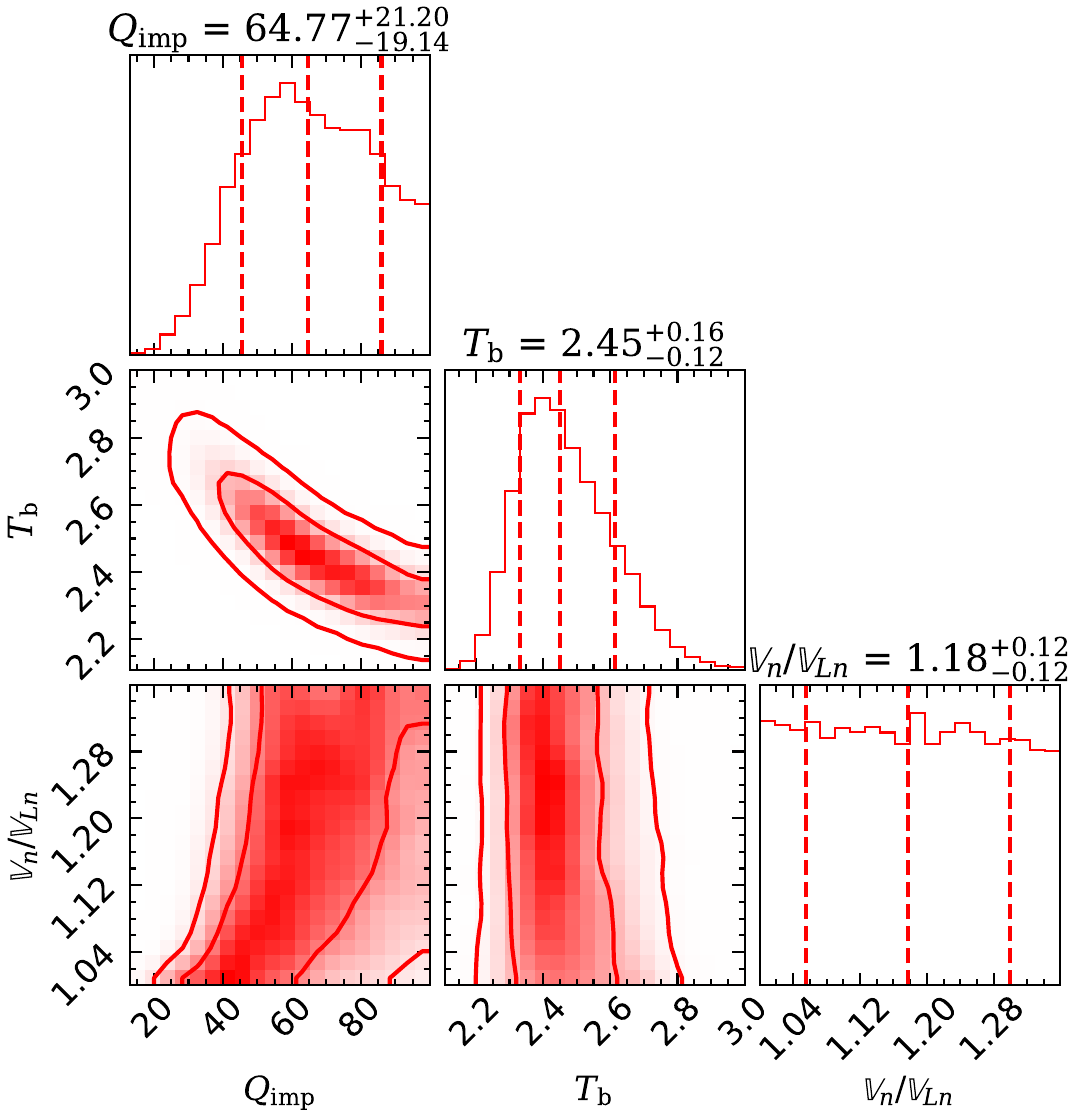}
        \caption{With He4 envelope.}
    \end{subfigure}
    \caption{Same as Fig.~\ref{fig:Corner_MXBOutburstII-Gapless} for $\MNS=1.4\MSol$ and $\RNS=10$~km.}
    \label{fig:Corner_MXBOutburstII-Gapless-canonical}
\end{figure*}

As in the models with BCS superfluidity, we have run cooling simulations for outburst II fixing $\Tcore$ to the values obtained
from the best cooling model of outburst I. The marginalized posterior probability distributions are shown in Figs.~\ref{fig:Corner_MXBOutburstII-Gapless} and \ref{fig:Corner_MXBOutburstII-Gapless-canonical}. The associated cooling curves are plotted in Fig.~\ref{fig:MXBOutburstIInHD_Cooling}
and the parameters are given in Table~\ref{tab:MXBII}. 

The cooling data are all very well reproduced independently of the adopted values for $\MNS$ and $\RNS$ or of the envelope
composition. As found for the models with BCS superfluidity, $\Qimp$ does not vary substantially between the two outbursts
unlike $\Tbase$. The neutron effective superfluid velocity $\Vn/\VLn$ remains essentially unchanged suggesting that no catastrophic unpinning of neutron vortices occurred.

At variance with the models with BCS superfluidity, our models with gapless superfluidity predict that MXB~1659$-$29
is still cooling. These predictions could be tested by future observations. According to our models, thermal 
equilibrium will not be reached before about 30000~days (40000~days) for the He9 (He4) envelope model assuming 
no accretion occurs and vortices remain pinned. 

\subsection{Sensitivity to the mass accretion rate \texorpdfstring{$\dot{m}$}{\textbackslash dot\{m\}}}

In the previous calculations, we adopted the same value $\dot{m}=0.1\mEdd\simeq 10^{17}$~g~s$^{-1}$ for the time-averaged accretion rate for both KS~1731$-$260 and MXB~1659$-$29 (consistent with the value given in Ref.~\cite{Galloway2008}). However, different values have been considered in recent studies. For the first outburst of MXB~1659$-$29, the authors of Ref.~\cite{Iaria2018} estimated the accretion rate as $\dot{m}\approx (0.003-0.02)\mEdd$ whereas, for the second outburst of the same source, the authors of Ref.~\cite{Lu2022} performed cooling simulations with $\dot{m}\approx 0.03\mEdd$. Moreover, the accretion rate may change over time and this may impact the cooling (see Ref.~\cite{ootes2016}). However, the variability in $\dot{m}$ was found to mainly affect the shallowest layers of the inner crust and can be compensated by adjusting the cooling parameters. In this subsection, we study the sensitivity of our results to a change of the accretion rate.

\subsubsection{Results obtained for fixed cooling parameters}

To better assess the influence of the time-averaged accretion rate $\dot{m}$, we have calculated   different cooling curves (for both BCS and gapless regimes) varying  $\dot{m}$ from 0.01$\mEdd$ to 0.2$\mEdd$ while 
keeping the other parameters fixed to their optimal values obtained previously from cooling simulations with $\dot{m}=0.1\mEdd$. 

\begin{figure*}
\centering
\begin{subfigure}[b]{0.45\textwidth}
        \centering
        \includegraphics[width=\textwidth]{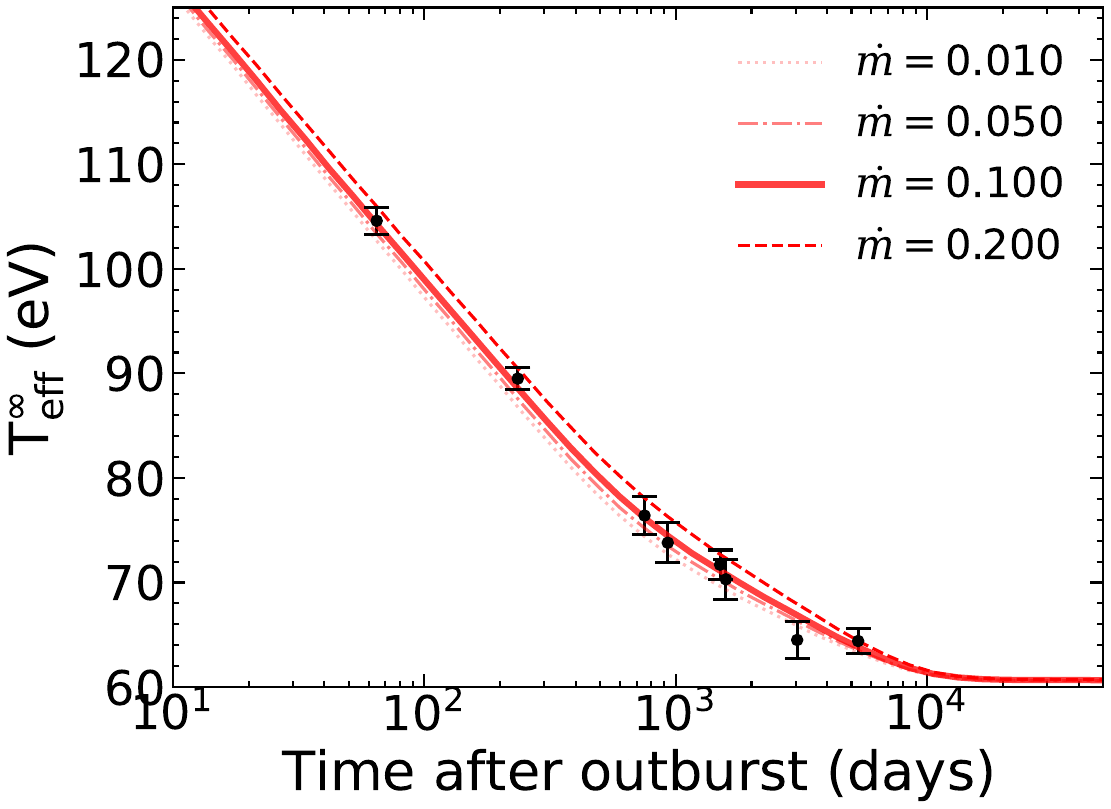}
        \caption{Gapless}
\end{subfigure}
\begin{subfigure}[b]{0.45\textwidth}
        \centering
        \includegraphics[width=\textwidth]{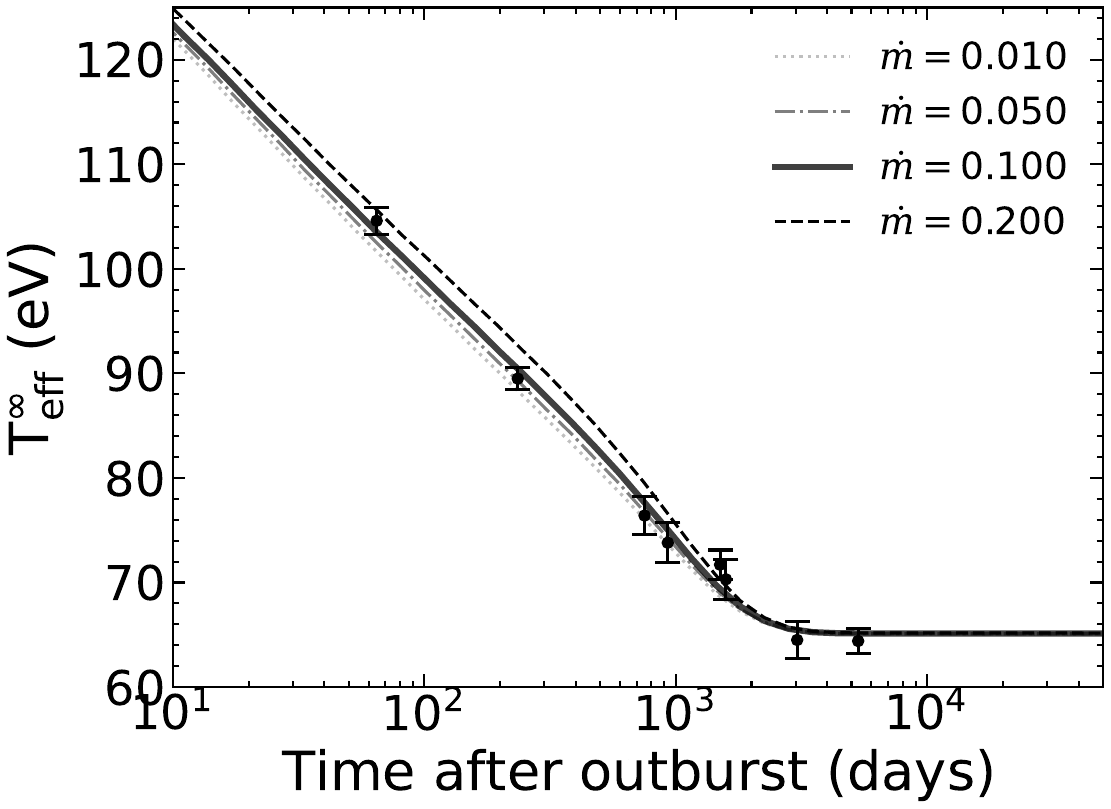}
        \caption{BCS}
\end{subfigure}
\caption{Evolution of the effective surface temperature of KS~1731$-$260 in electronvolts (as seen by an observer at infinity) as a function of the time in days after the end of the outburst within the model of Gusakov\& Chugunov~\cite{GusakovChugunov2020,GusakovChugunov2021} of accreted neutron star with $\MNS=1.4\MSol$ and $\RNS=10$~km, the He4 envelope model and the realistic neutron pairing calculations of~\cite{Gandolfi2022}. Symbols represent observational data with error bars. The red (resp. black) curves correspond to models considering gapless (resp. BCS) superfluidity. All these cooling curves have been obtained by fixing $\Tcore$, $\Qimp$, $\Tbase$ and $\Vn/\VLn$ to their optimal values (given by the median values of the marginalized posterior distributions, see text for details) while varying $\dot{m}$ (in units of $\dot{m}_{\rm Edd}\simeq 1.1\times 10^{18}$~g~s$^{-1}$) between 0.01 and 0.2. Results are shown for neutrons in the gapless (a) or BCS (b) regimes.}
\label{fig:KSmdot}
\end{figure*}

\begin{figure*}
\centering
\begin{subfigure}[b]{0.45\textwidth}
        \centering
        \includegraphics[width=\textwidth]{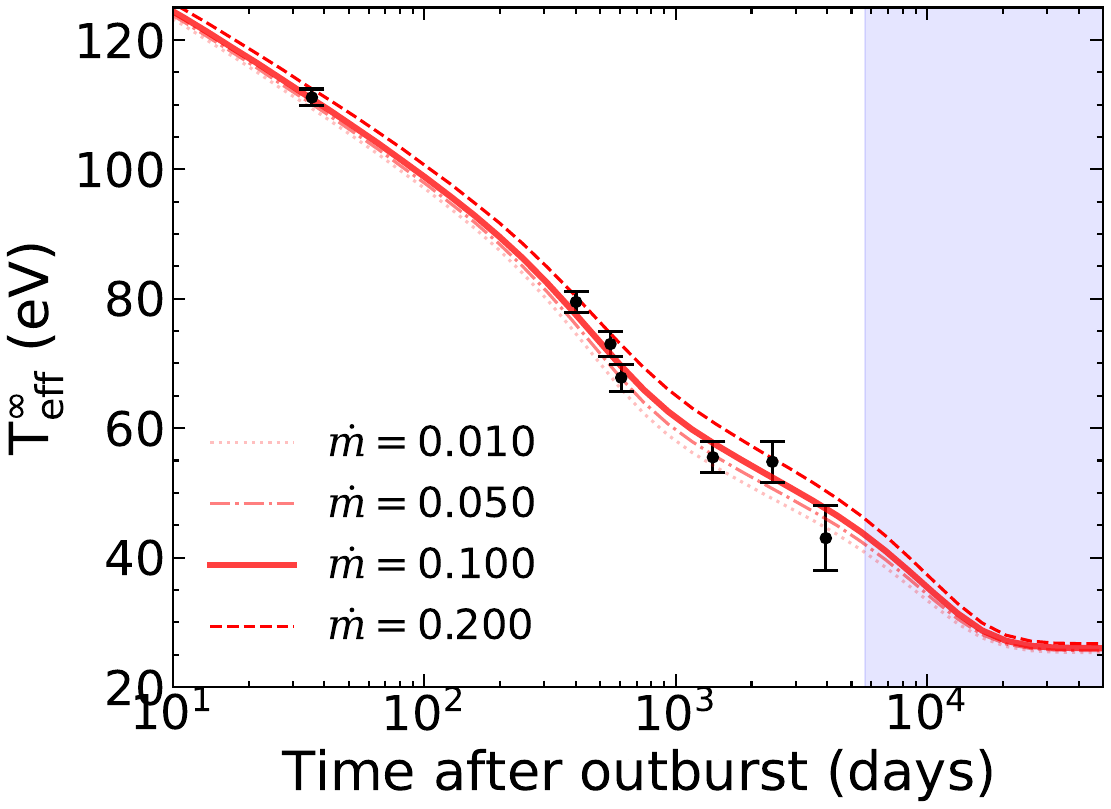}
        \caption{Gapless}
\end{subfigure}
\begin{subfigure}[b]{0.45\textwidth}
        \centering
        \includegraphics[width=\textwidth]{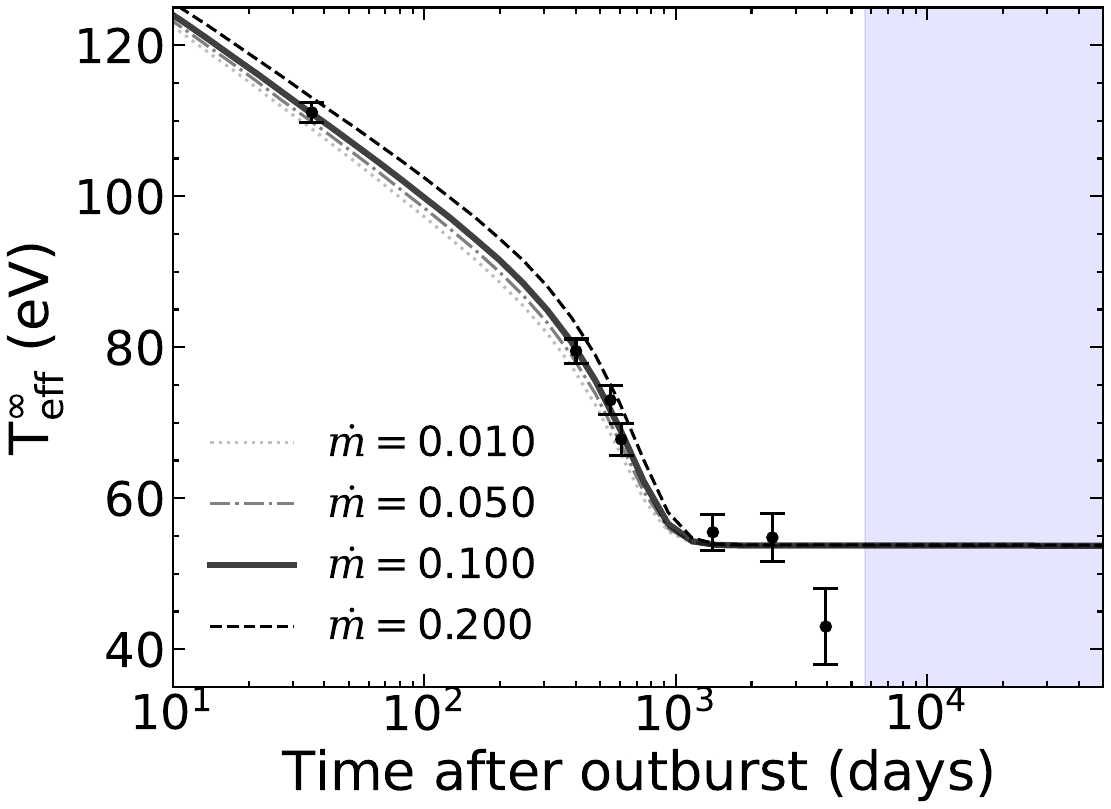}
        \caption{BCS}
\end{subfigure}
\caption{Evolution of the effective surface temperature of MXB~1659$-$29 (outburst I) in electronvolts (as seen by an observer at infinity) as a function of the time in days after the end of the outburst within the model of Gusakov 
 \& Chugunov~\cite{GusakovChugunov2020,GusakovChugunov2021} of accreted neutron star with $\MNS=1.62\MSol$ and $\RNS=11.2$~km, the He9 envelope model and the realistic neutron pairing calculations of~\cite{Gandolfi2022}. Symbols represent observational data with error bars. The red (resp. black) curves correspond to models considering gapless (resp. BCS) superfluidity. All these cooling curves have been obtained by fixing $\Tcore$, $\Qimp$, $\Tbase$ and $\Vn/\VLn$ to their optimal values (given by the median values of the marginalized posterior distributions, see text for details) between 0.01 and 0.2. The shaded area
corresponds to the second outburst (which occurred in 2015) and its subsequent cooling phase: the cooling
curves within this region depict the expected behavior had outburst II not occurred. Results are shown for neutrons in the gapless (a) or BCS (b) regimes.}
\label{fig:MXBmdot1}
\end{figure*}

\begin{figure*}
\centering
\begin{subfigure}[b]{0.45\textwidth}
        \centering
        \includegraphics[width=\textwidth]{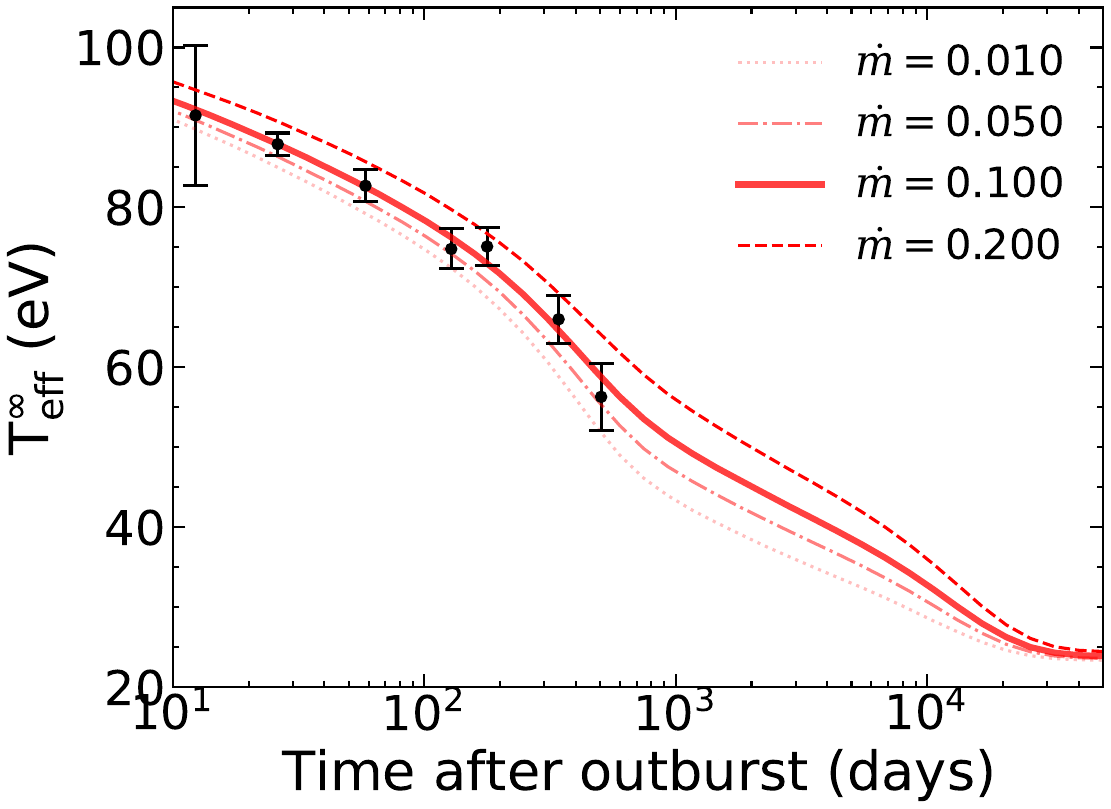}
        \caption{Gapless}
\end{subfigure}
\begin{subfigure}[b]{0.45\textwidth}
        \centering
        \includegraphics[width=\textwidth]{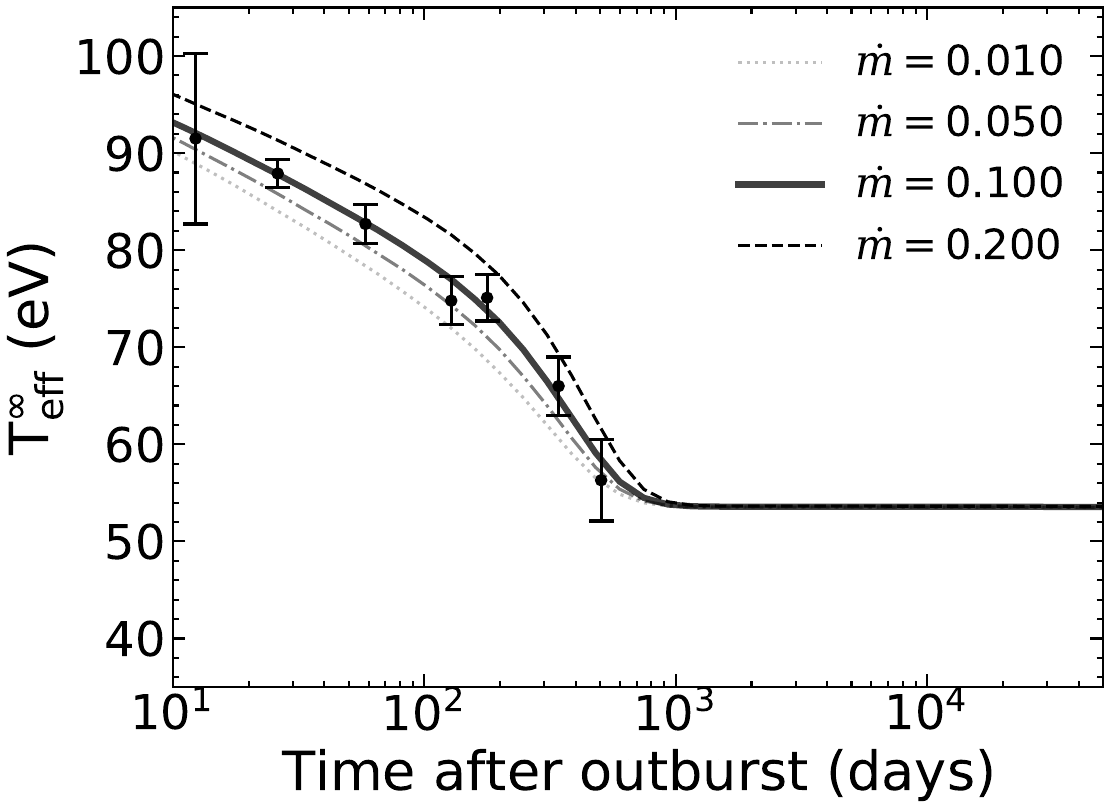}
        \caption{BCS}
\end{subfigure}
\caption{Same as Figure~\ref{fig:MXBmdot1} but for outburst II.}
\label{fig:MXBmdot2}
\end{figure*}

For KS~1731$-$260, the results are displayed in Fig.~\ref{fig:KSmdot}. These cooling curves have been obtained for a neutron star with $\MNS=1.4\MSol$ and $\RNS=$10~km, using the He4 envelope model. The cooling parameters are: $\Tcore=8.96\times 10^7$~K, $\Qimp=10.67$, $\Tbase=4.10\times 10^8$~K for neutrons in BCS regime, and $\Tcore=7.84\times 10^7$~K, $\Qimp=10.68$, $\Tbase=4.50\times 10^8$~K and $\Vn/\VLn=1.21$ for neutron in gapless regime. The cooling curves for MXB~1659$-$29 with $\MNS=1.62\MSol$, $\RNS=$11.2~km and the He9 envelope model are shown in Figs.~\ref{fig:MXBmdot1} and~\ref{fig:MXBmdot2}, following the first and the second outburst, respectively. For the subsequent cooling phase following outburst I, the cooling parameters are: $\Tcore=2.86\times 10^7$~K, $\Qimp=7.77$, $\Tbase=3.12\times 10^8$~K for BCS superfluidity, and $\Tcore=0.65\times 10^7$~K, $\Qimp=16.35$, $\Tbase=3.14\times 10^8$~K and $\Vn/\VLn=1.20$ for gapless superfluidity. The cooling parameters associated with the cooling phase following the second outburst are: $\Tcore=2.86\times 10^7$~K, $\Qimp=7.60$, $\Tbase=1.41\times 10^8$~K for BCS superfluidity, and $\Tcore=0.65\times 10^7$~K, $\Qimp=26.68$, $\Tbase=1.38\times 10^8$~K and $\Vn/\VLn=1.18$ for gapless superfluidity.  

For the same outburst, an increase of $\dot{m}$ corresponds to more accreted material and a more efficient crust heating, resulting in a higher $T_{\rm eff}^{\infty}$, as shown in Figs~\ref{fig:KSmdot},~\ref{fig:MXBmdot1} and~\ref{fig:MXBmdot2}. Consistent with Eq.~\eqref{eq:ThermalTimescaleGapless}, all cooling curves relax towards equilibrium over the same  timescale $\tau$, regardless of the value of the accretion rate $\dot{m}$. For both KS~1731$-$260 and MXB~1659$-$29, the impact of the uncertainties in the time-averaged accretion rate on the cooling curves is of the same order as the errors associated with the observational data, the outburst II of MXB~1659$-$29 being more sensitive to those uncertainties (see Fig.~\ref{fig:MXBmdot2}). Note that, for the case of MXB~1659$-$29 (outburst I) assuming neutrons in BCS regime, changing the value of $\dot{m}$ does not explain the observation reported by Ref.~\cite{Cackett2013} (see left panel of Fig.~\ref{fig:MXBmdot1}).

\subsubsection{Results obtained varying the cooling parameters}

\begin{figure*}
    \centering
    \begin{subfigure}[b]{0.45\textwidth}
        \centering\includegraphics[width=\textwidth]{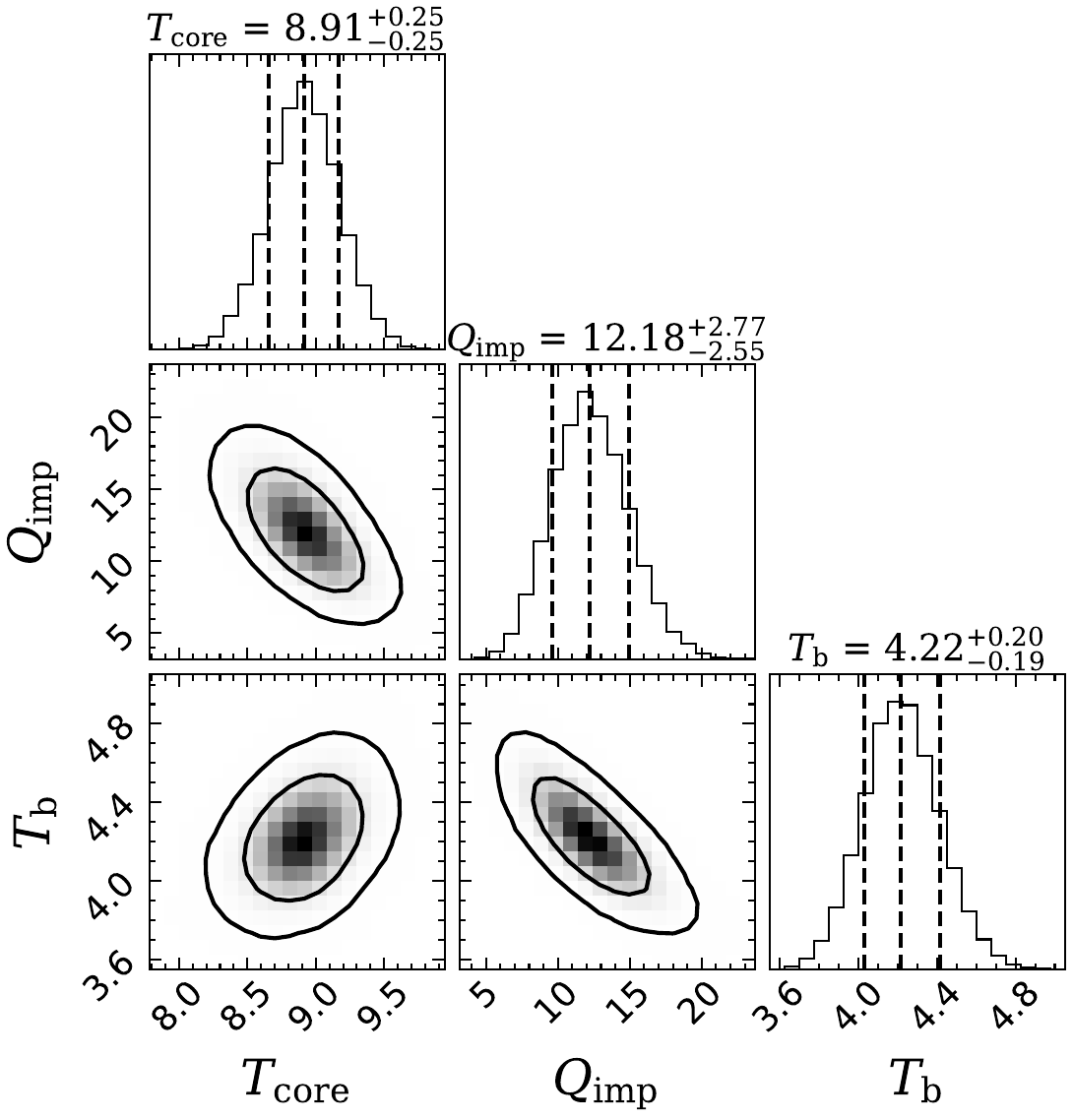}
        \caption{$\dot{m}=0.01\dot{m}_{\rm Edd}$ (BCS).}
    \end{subfigure}
        \begin{subfigure}[b]{0.45\textwidth}
        \centering\includegraphics[width=\textwidth]{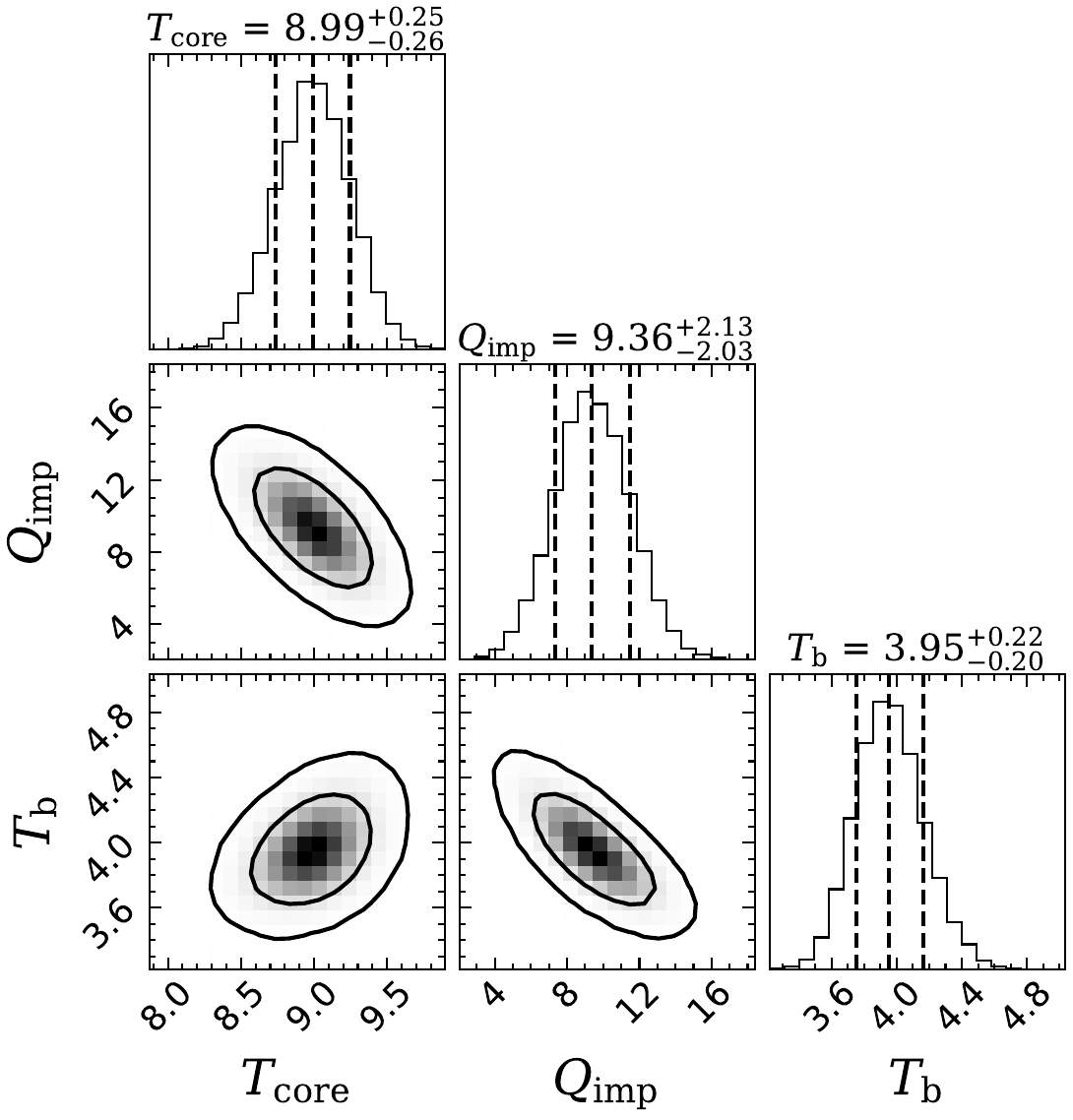}
        \caption{$\dot{m}=0.2\dot{m}_{\rm Edd}$ (BCS).}
    \end{subfigure}
    \vspace{10pt}
    \begin{subfigure}[b]{0.45\textwidth}
        \centering\includegraphics[width=\textwidth]{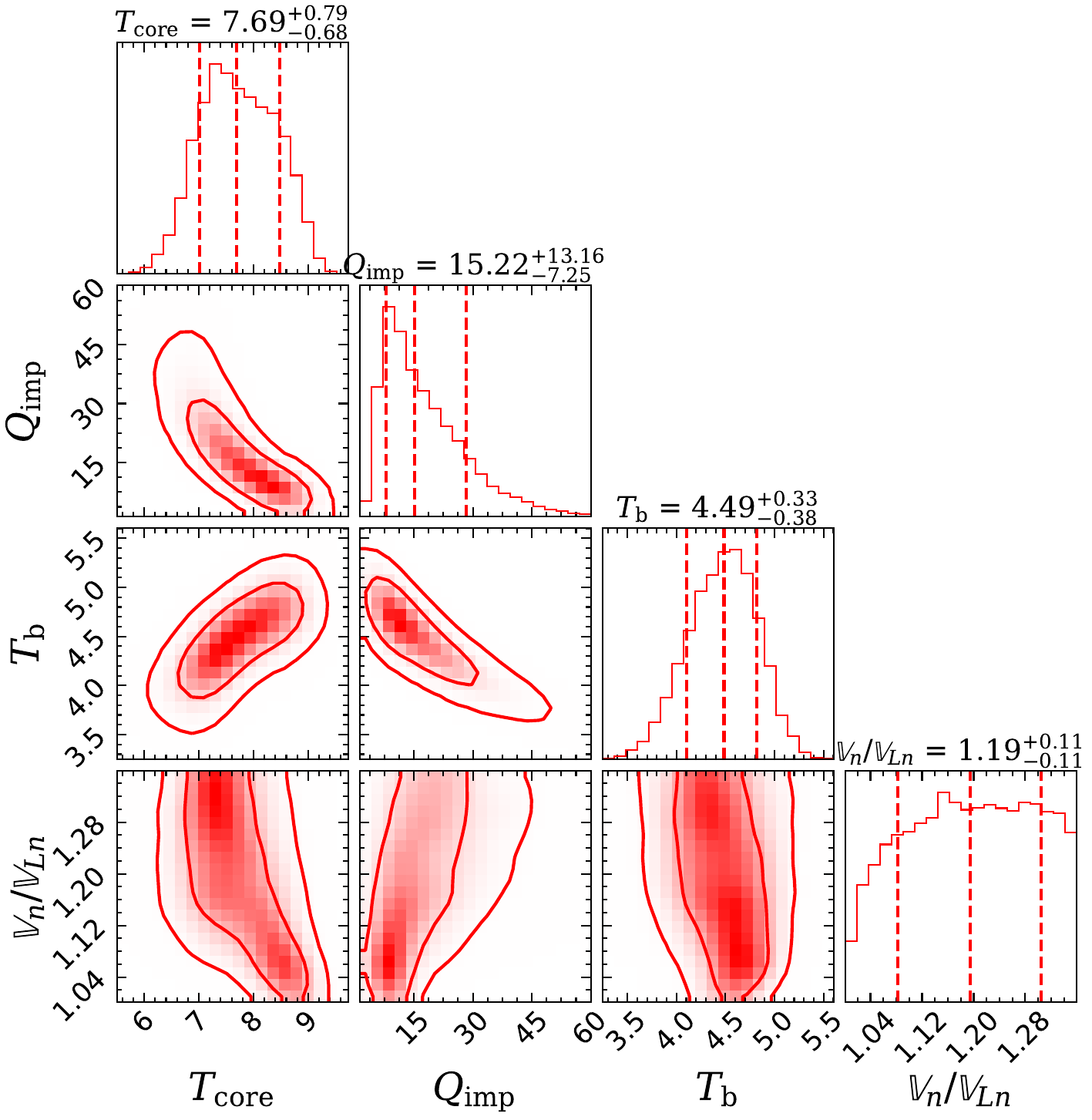}
        \caption{$\dot{m}=0.01\dot{m}_{\rm Edd}$ (Gapless).}
    \end{subfigure}
        \begin{subfigure}[b]{0.45\textwidth}
        \centering\includegraphics[width=\textwidth]{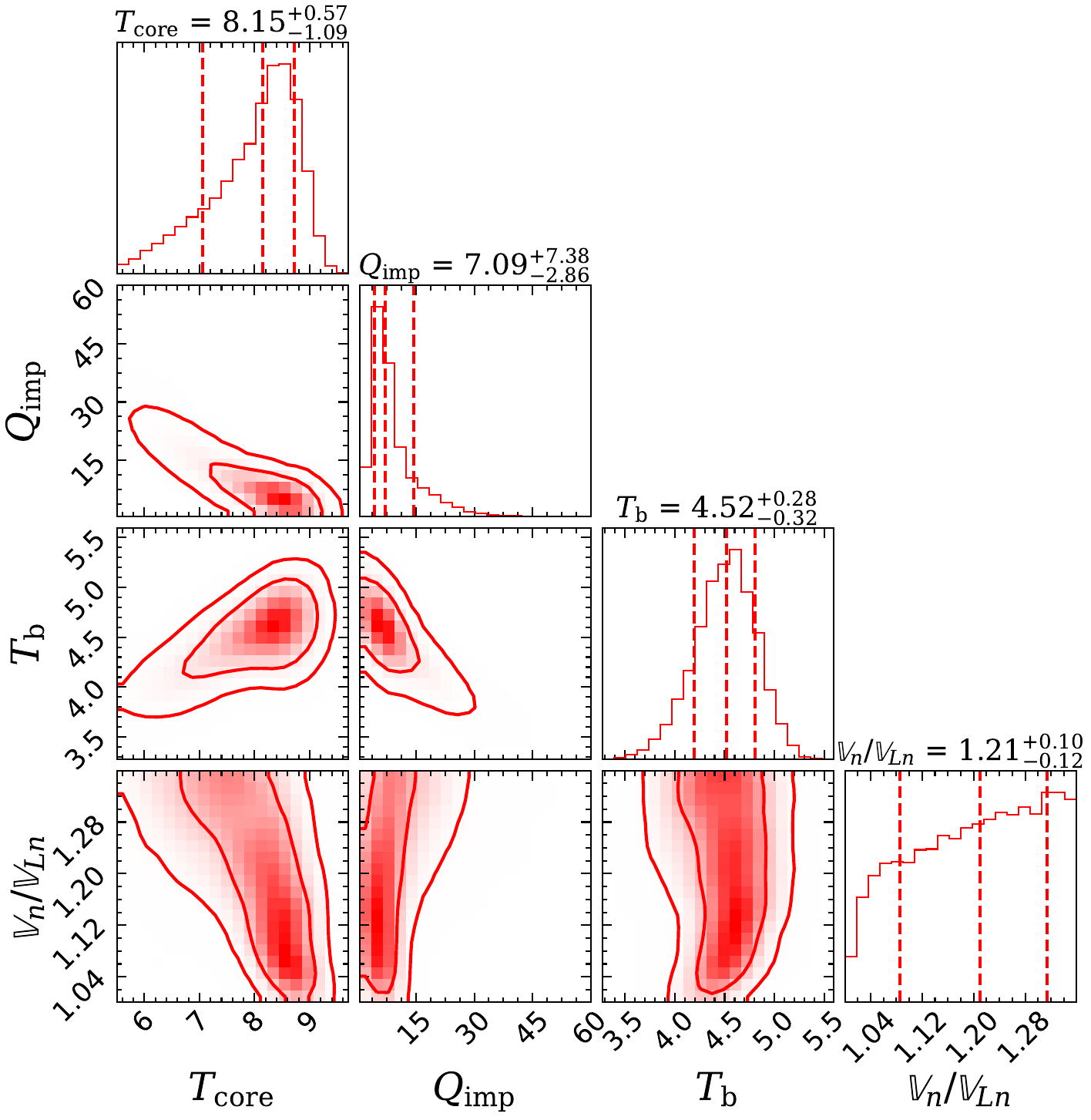}
        \caption{$\dot{m}=0.2\dot{m}_{\rm Edd}$ (Gapless).}
    \end{subfigure}
    \caption{Marginalized 1-D and 2-D probability distributions for the parameters of the cooling model of KS~1731$-$260 within the model of accreted neutron star crusts of Gusakov\& Chugunov~\cite{GusakovChugunov2020,GusakovChugunov2021} for neutrons in BCS (upper panels) or gapless (lower panels) regimes, using the realistic neutron pairing gap of~\cite{Gandolfi2022}. Results were obtained for a neutron star with $\MNS=1.4\MSol$ and $\RNS=10$~km, with the envelope model He4, assuming time-averaged mass-accretion rates of $\dot{m}=0.01\dot{m}_{\rm Edd}$ (left panels) and $\dot{m}=0.2\dot{m}_{\rm Edd}$ (right panels). $\Tcore$ and $\Tbase$ are expressed in units of $10^7$~K and $10^8$~K, respectively. The dotted lines in the histograms correspond to the median value and the 68\% uncertainty level. Contours in the marginalized 2-D probability distributions mark the 68\% and 95\% confidence ranges.}
    \label{fig:Corner_KSmdot}
\end{figure*}

Cooling simulations fixing the accretion rate at its extreme values, i.e. $\dot{m}=0.01\dot{m}_{\rm Edd}$ and $\dot{m}=0.20\dot{m}_{\rm Edd}$, have also been performed. The marginalized posterior distributions of the cooling parameters for KS~1731$-$260 (assuming a neutron star with $\MNS=1.4\MSol$, $\RNS=10$~km and using the He4 envelope) are shown in Fig.~\ref{fig:Corner_KSmdot} for neutrons in BCS (upper panels) and gapless (lower panels) regimes. The resulting cooling parameters obtained from the change of mass-accretion rate are all comparable to those obtained with $\dot{m}=0.1\dot{m}_{\rm Edd}$. The associated cooling curves are indistinguishable from the ones displayed in Fig.~\ref{fig:KSnHD_Cooling} (right panel).

\begin{figure*}
    \centering
    \begin{subfigure}[b]{0.45\textwidth}
        \centering\includegraphics[width=\textwidth]{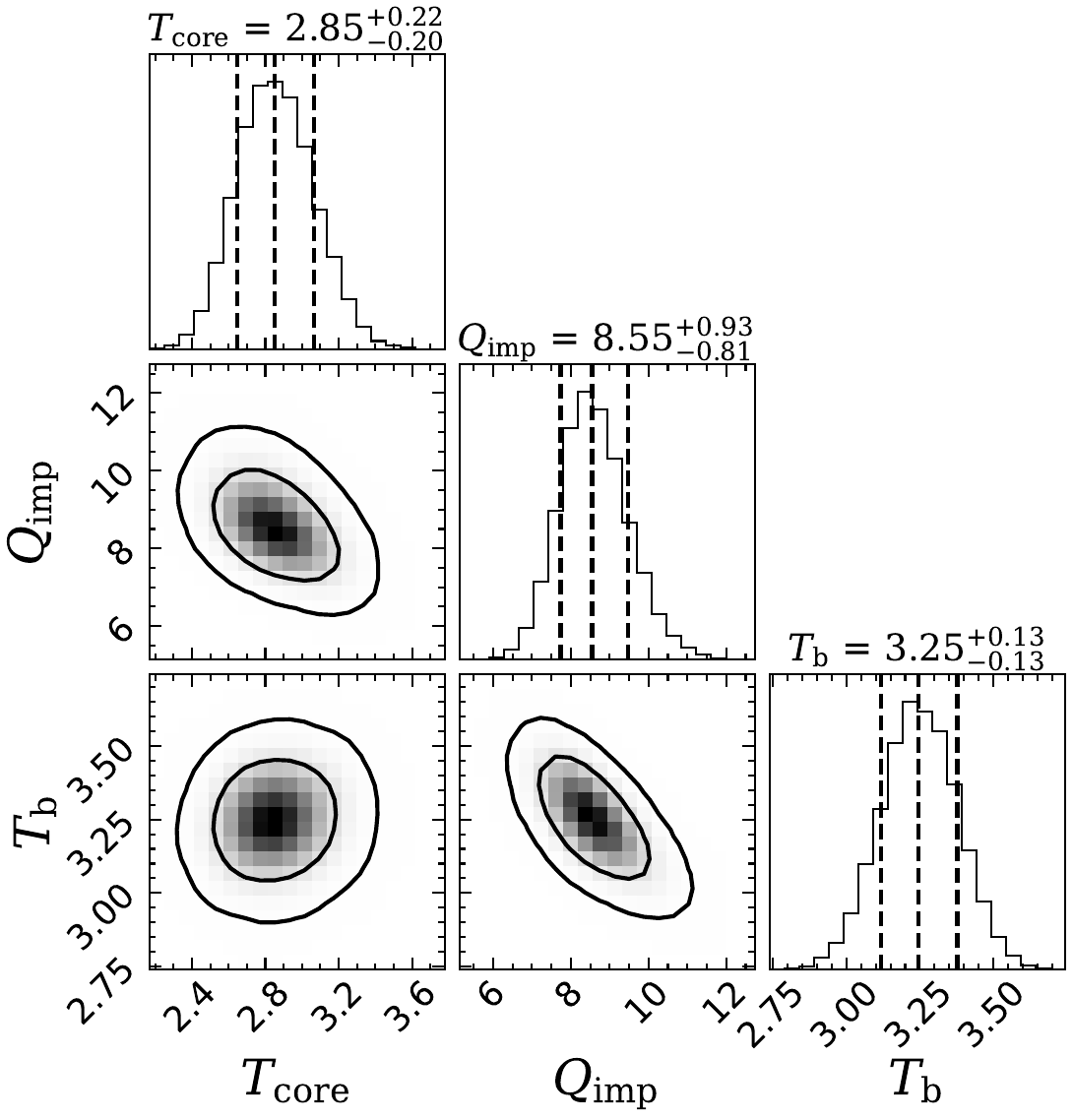}
        \caption{$\dot{m}=0.01\dot{m}_{\rm Edd}$ (BCS).}
    \end{subfigure}
        \begin{subfigure}[b]{0.45\textwidth}
        \centering\includegraphics[width=\textwidth]{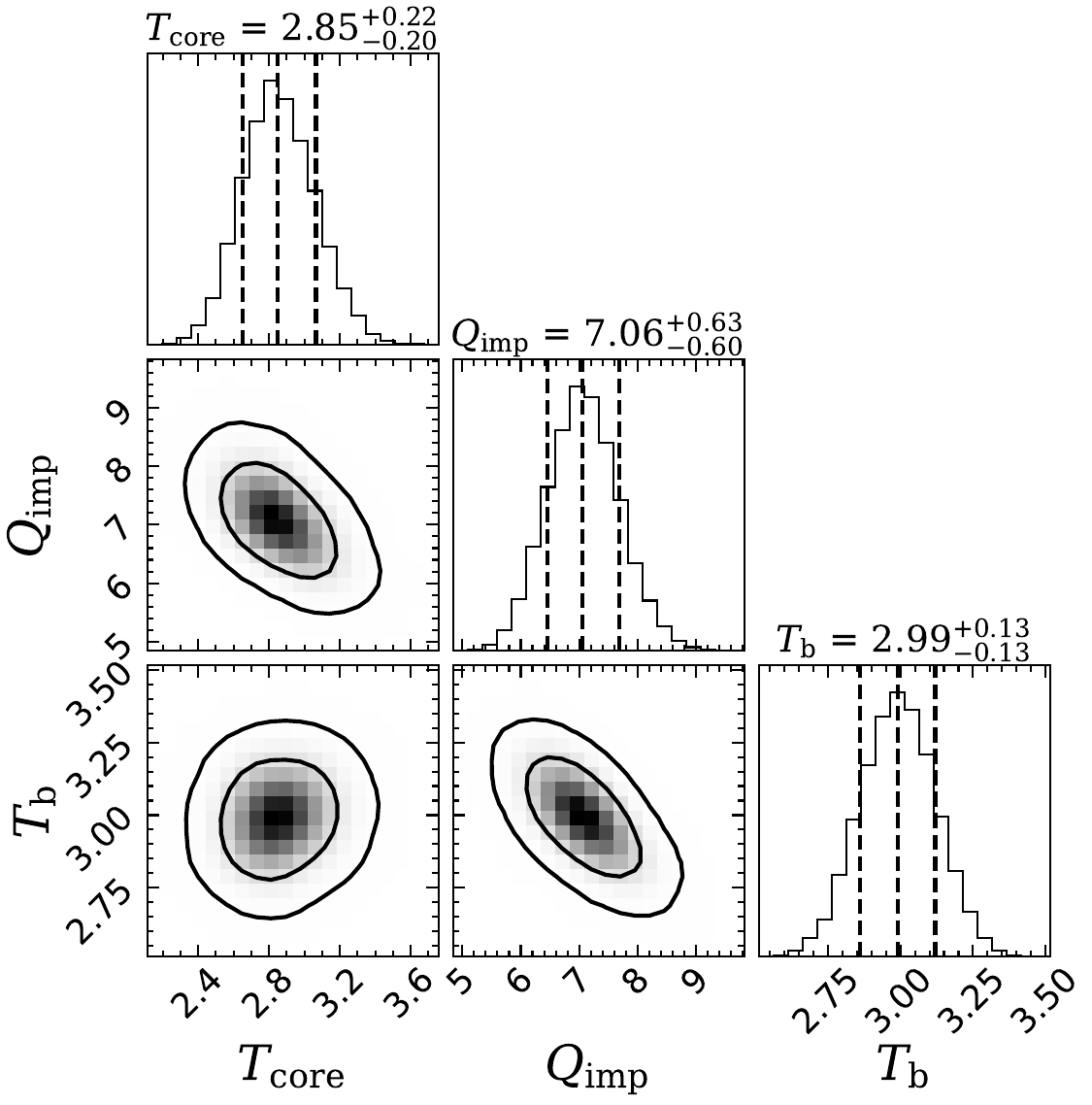}
        \caption{$\dot{m}=0.2\dot{m}_{\rm Edd}$ (BCS).}
    \end{subfigure}
    \vspace{10pt}
    \begin{subfigure}[b]{0.45\textwidth}
        \centering\includegraphics[width=\textwidth]{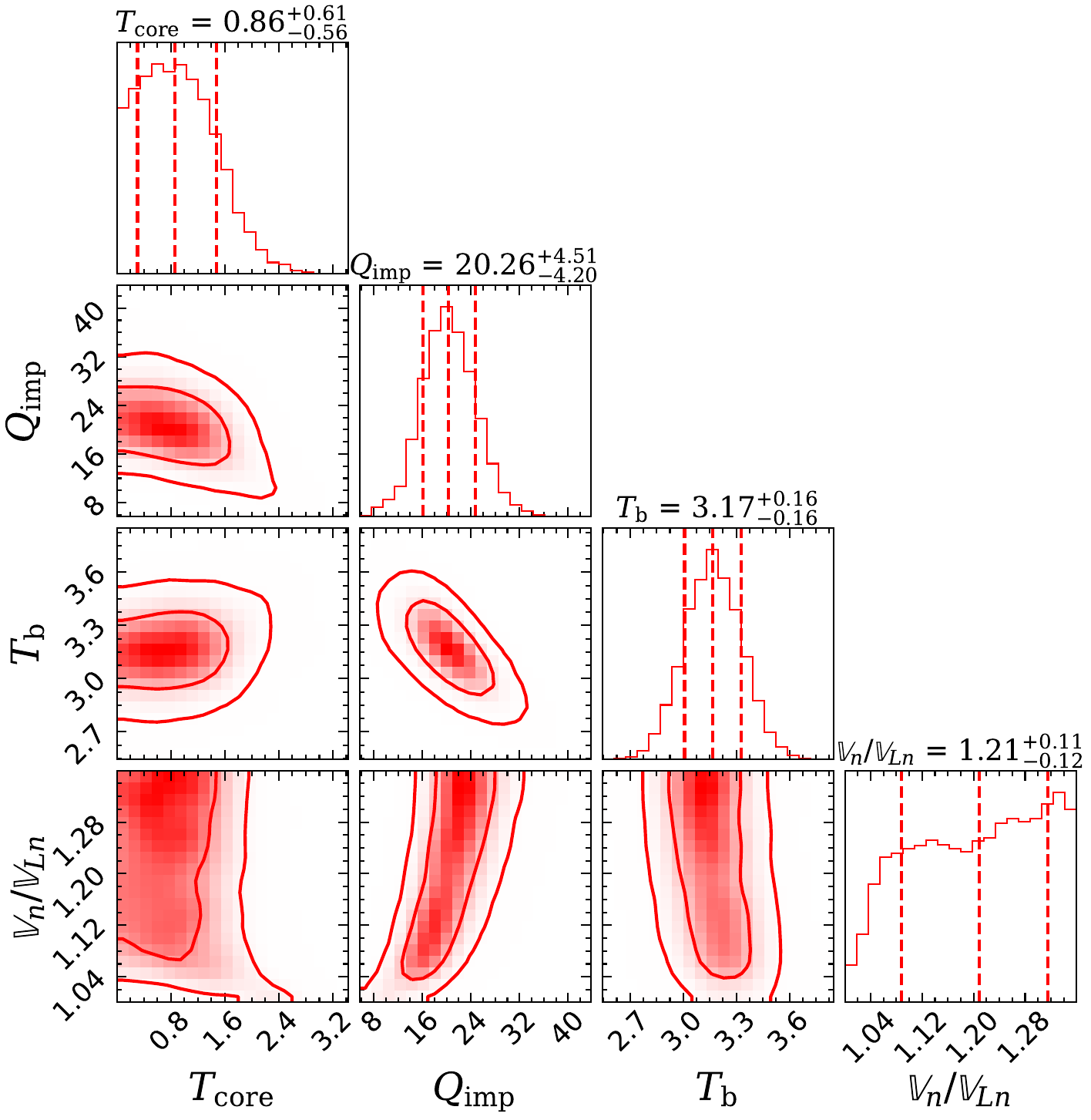}
        \caption{$\dot{m}=0.01\dot{m}_{\rm Edd}$ (Gapless).}
    \end{subfigure}
        \begin{subfigure}[b]{0.45\textwidth}
        \centering\includegraphics[width=\textwidth]{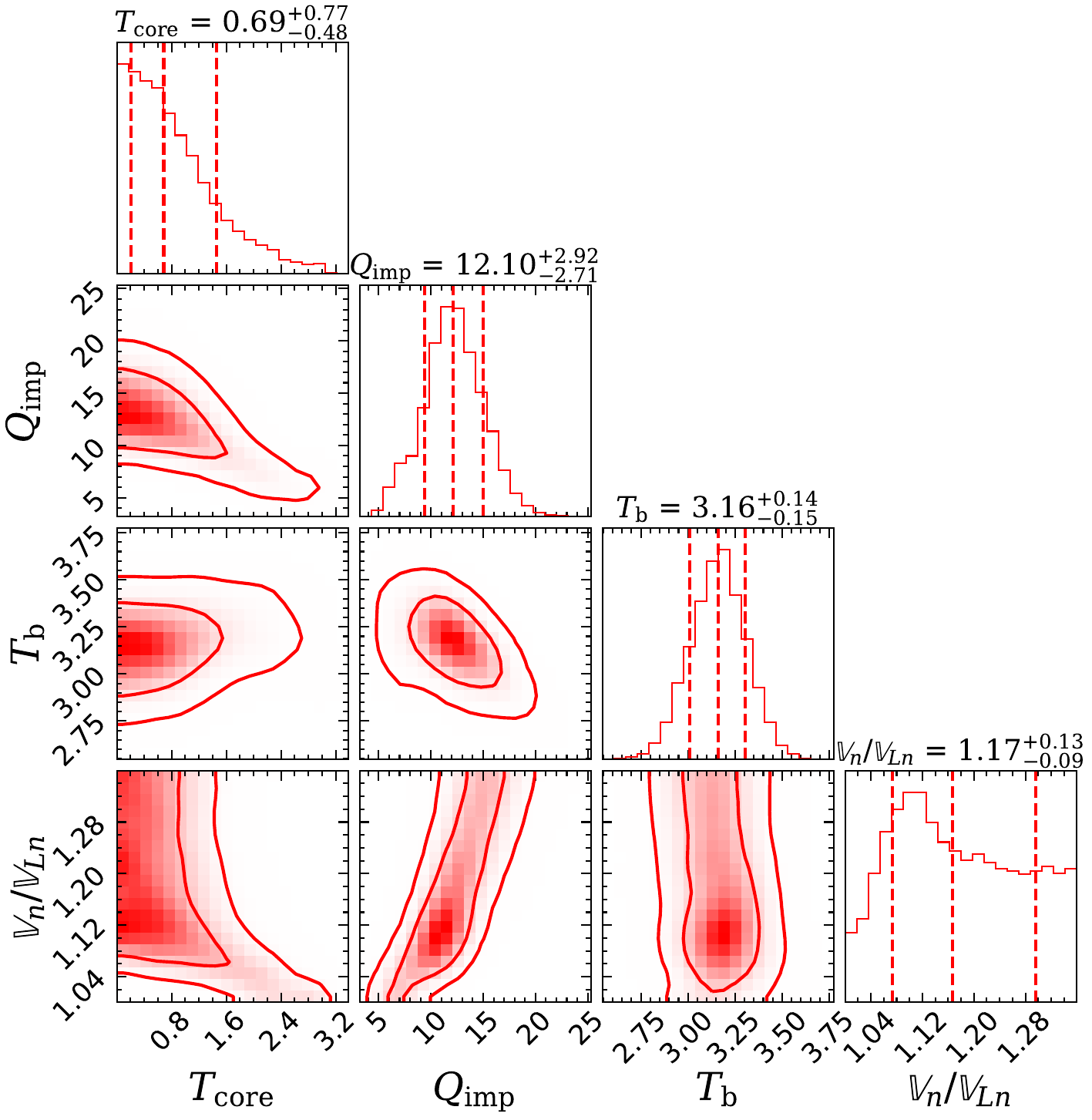}
        \caption{$\dot{m}=0.2\dot{m}_{\rm Edd}$ (Gapless).}
    \end{subfigure}
    \caption{Marginalized 1-D and 2-D probability distributions for the parameters of the cooling model of MXB~1659$-$29 (outburst I) within the model of accreted neutron star crusts of Gusakov\&Chugunov~\cite{GusakovChugunov2020,GusakovChugunov2021} in BCS (upper panels) or gapless regime (lower panels) using the realistic neutron pairing gap of~\cite{Gandolfi2022}. Results were obtained for a neutron star with $\MNS=1.62\MSol$ and $\RNS=11.2$~km, with the envelope model He9, assuming time-averaged mass-accretion rates of $\dot{m}=0.01\dot{m}_{\rm Edd}$ (left panels) and $\dot{m}=0.2\dot{m}_{\rm Edd}$ (right panels). $\Tcore$ and $\Tbase$ are expressed in units of $10^7$~K and $10^8$~K, respectively. The dotted lines in the histograms correspond to the median value and the 68\% uncertainty level. Contours in the marginalized 2-D probability distributions mark the 68\% and 95\% confidence ranges.}
    \label{fig:Corner_MXBmdot}
\end{figure*}

\begin{figure*}
    \centering
    \begin{subfigure}[b]{0.45\textwidth}
        \centering\includegraphics[width=\textwidth]{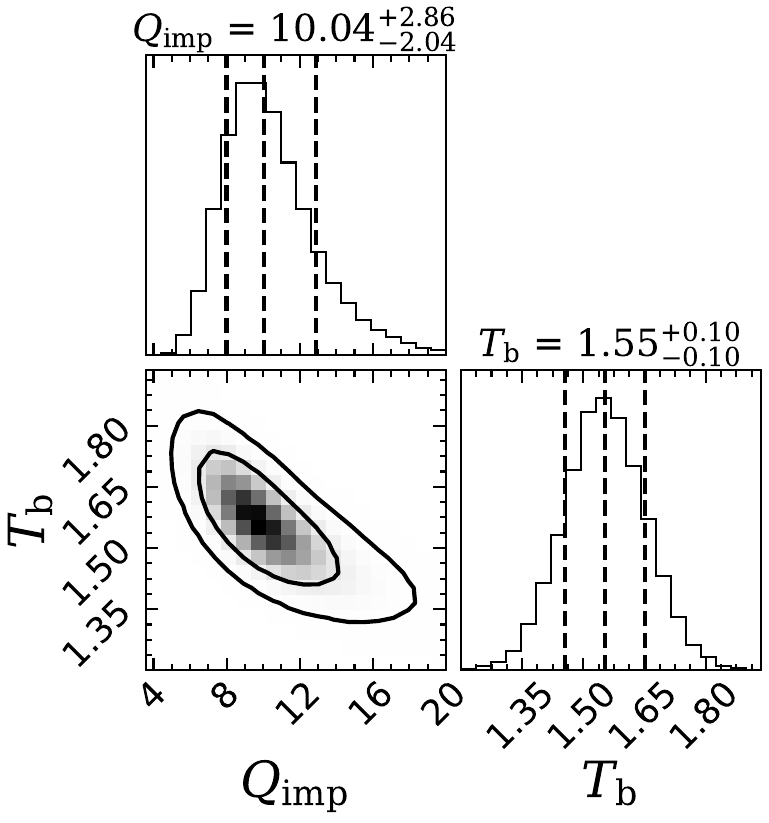}
        \caption{$\dot{m}=0.01\dot{m}_{\rm Edd}$ (BCS).}
    \end{subfigure}
        \begin{subfigure}[b]{0.45\textwidth}
        \centering\includegraphics[width=\textwidth]{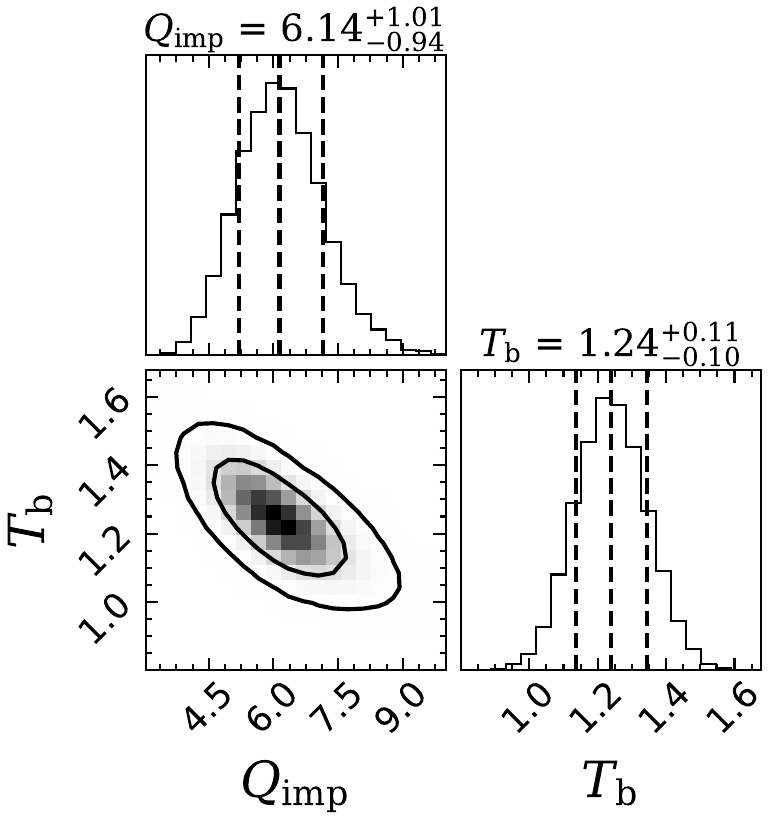}
        \caption{$\dot{m}=0.2\dot{m}_{\rm Edd}$ (BCS).}
    \end{subfigure}
    \vspace{10pt}
    \begin{subfigure}[b]{0.45\textwidth}
        \centering\includegraphics[width=\textwidth]{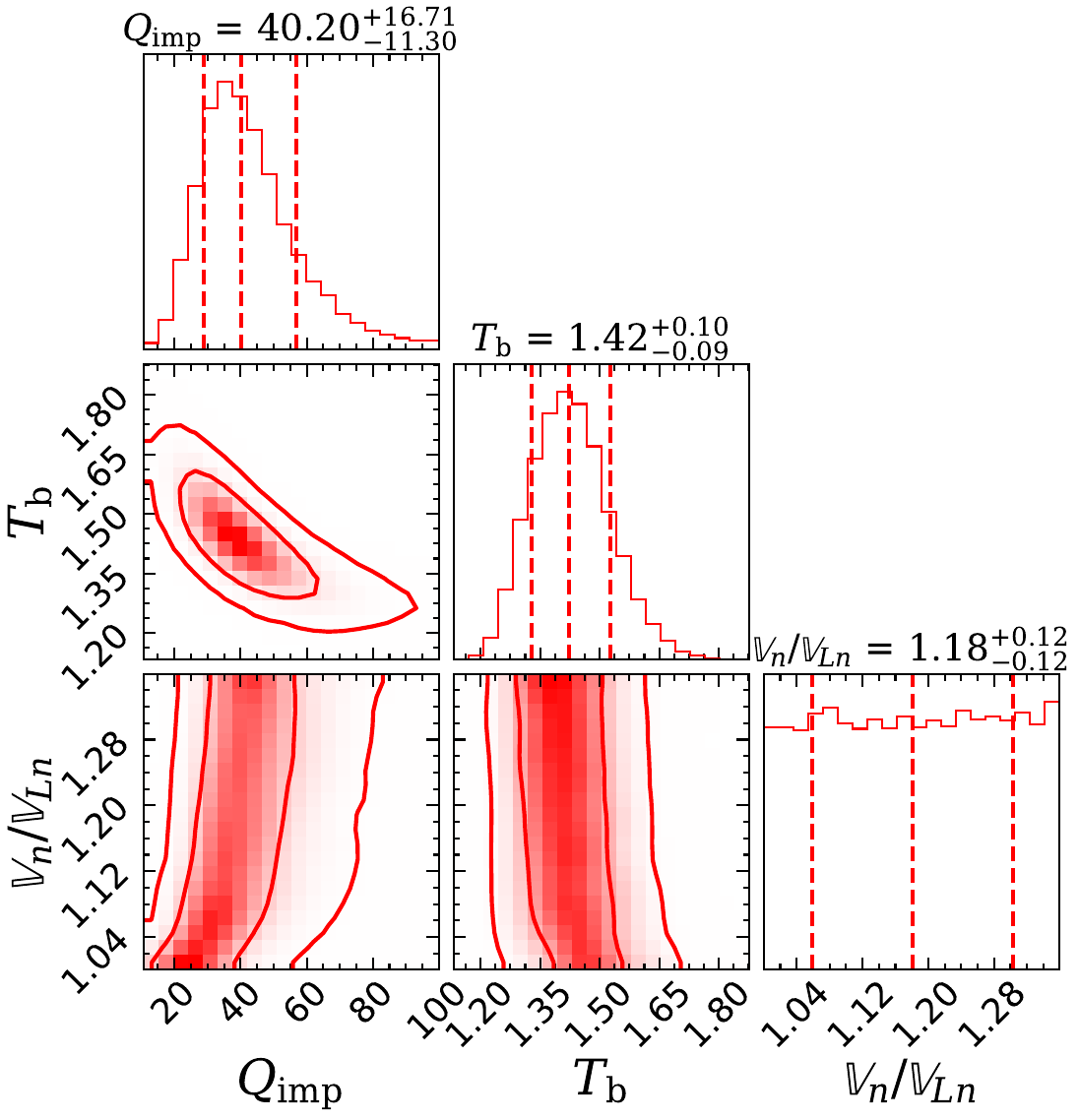}
        \caption{$\dot{m}=0.01\dot{m}_{\rm Edd}$ (Gapless).}
    \end{subfigure}
        \begin{subfigure}[b]{0.45\textwidth}
        \centering\includegraphics[width=\textwidth]{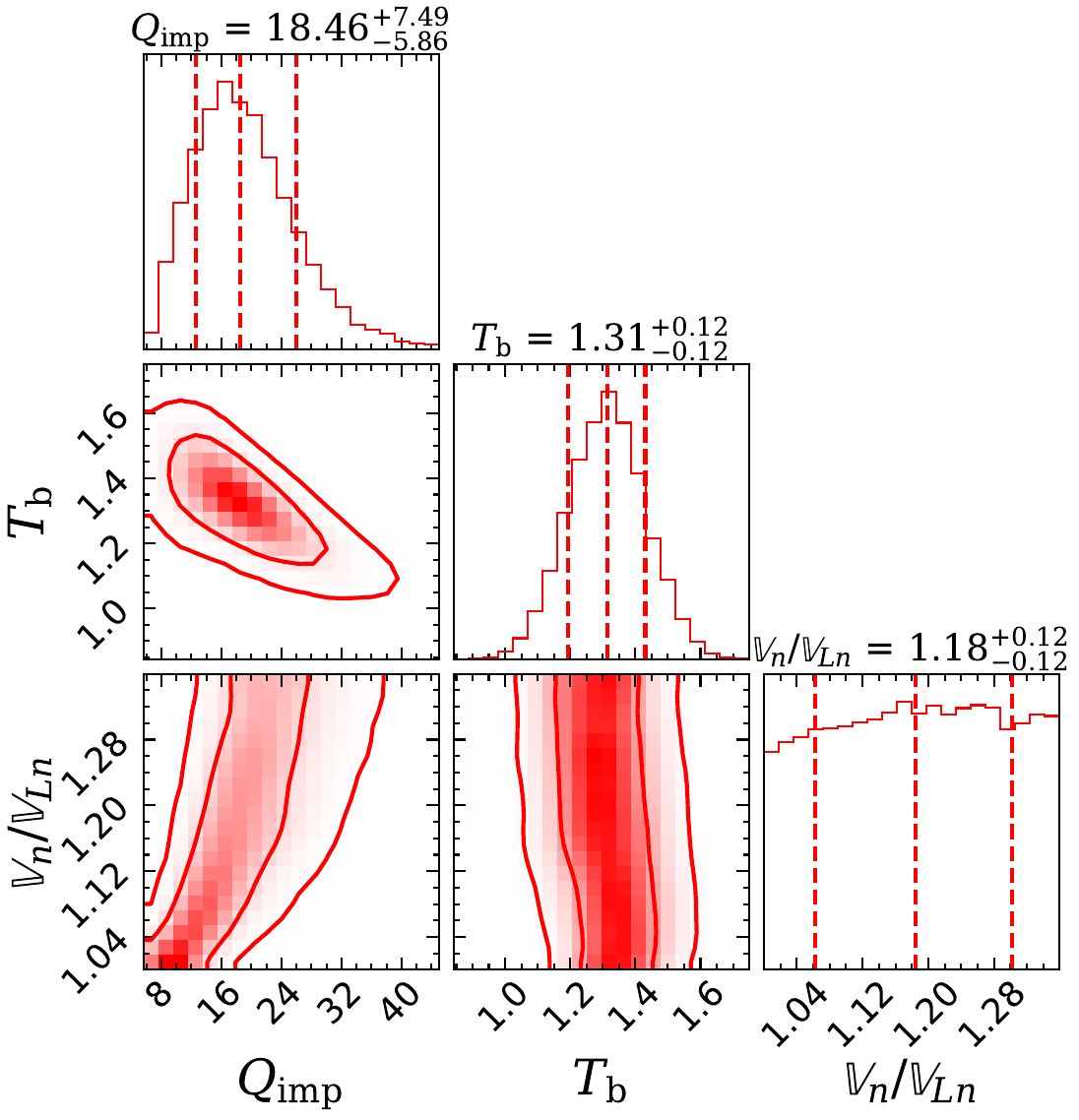}
        \caption{$\dot{m}=0.2\dot{m}_{\rm Edd}$ (Gapless).}
    \end{subfigure}
    \caption{Same as Figure~\ref{fig:Corner_MXBmdot} but for MXB~1659$-$29 (outburst II).}
    \label{fig:Corner_MXBIImdot}
\end{figure*}

Figure~\ref{fig:Corner_MXBmdot} gives the marginalized posterior distributions of the model parameters for the cooling phase following the first outburst of MXB~1659$-$29 (for a neutron star with $\MNS=1.62\MSol$ and $\RNS=11.2$~km and using the He9 envelope), assuming neutron superfluidity in BCS (upper panels) and gapless (lower panels) regimes. The resulting cooling parameters are all  comparable to those obtained with $\dot{m}=0.1\dot{m}_{\rm Edd}$ and the associated cooling curves are indistinguishable from those already shown in Fig.~\ref{fig:MXBOutburstInHD_Cooling}. From these results, we have also run cooling simulations for the subsequent cooling phase following the second outburst of MXB~1659$-$29. Their marginalized posterior distributions are displayed in Fig.~\ref{fig:Corner_MXBIImdot}. Although the cooling curves for outburst II were previously found to be sensitive to the mass accretion rate (see Fig.~\ref{fig:MXBmdot2}), the resulting parameters (after recalculations) do not deviate significantly from those obtained for $\dot{m}=0.1\dot{m}_{\rm Edd}$.

All in all, the uncertainties in the accretion rate (for $\dot{m}$ values between $0.01\dot{m}_{\rm Edd}$ and $0.2\dot{m}_{\rm Edd}$) are contained in the uncertainties of the cooling parameters.

\section{Conclusion}\label{sec:conclusion}

We have further investigated the possible existence of neutron gapless superfluidity in the crust of the SXTs 
MXB~1659$-$29 and KS~1731$-$260 by performing new simulations of their thermal relaxation after the end of their 
outbursts within the thermodynamically consistent approach of Refs.~\cite{GusakovChugunov2020,GusakovChugunov2021} 
but allowing for the presence of neutron superflow in the crust~\cite{AllardChamel2023Letter}. We have made use of the realistic neutron pairing 
calculations of Ref.~\cite{Gandolfi2022} based on quantum Monte Carlo method. We have considered two sets of neutron-star mass 
and radius: $\MNS=1.62\MSol$ and $\RNS=11.2$~km, as in the original study of Ref.~\cite{BrownCumming2009}, and 
the canonical values $\MNS=1.4\MSol$ and $\RNS=10$~km. We have also varied the composition of the heat-blanketing 
envelope by adopting two different models: a light and a heavy envelopes made of He down to the column depth 
$y_{\rm He}=10^9$~g~cm$^{-2}$ (as in Ref.~\cite{BrownCumming2009}) and  $y_{\rm He}=10^4$~g~cm$^{-2}$ respectively, 
the deeper region containing Fe in both cases. We have run MCMC simulations to assess the uncertainties of the 
fitting parameters of our neutron-star crust cooling model, namely the temperature  $\Tcore$ of the  neutron-star core, 
the temperature $\Tbase$ at the bottom of the envelope during outburst, the impurity parameter $\Qimp$, and the neutron 
effective superfluid velocity $\Vn$ expressed in terms of Landau's velocity $\VLn$. We have also investigating the impact of the uncertainties in the mass accretion rate.

The resulting parameters and the associated cooling curves are found to be insensitive to the global structure 
of the neutron star. In contrast, the composition of the envelope has a non-negligible impact on the inferred  
values for $\Tcore$, $\Qimp$ and $\Tbase$. Assuming the neutron superfluid is at rest in the crust as in previous 
studies, the fit to the cooling data favors a light envelope for KS~1731$-$260 and a heavy envelope for MXB~1659$-$29. 
Nonetheless, the late-time cooling of these SXTs can hardly be reproduced by these models. In particular, the drop of 
temperature of MXB~1659$-$29 observed about 4000 days after the outburst I still remains unexplained. Varying the accretion rate does not change the situation. 

This observational puzzle is naturally solved by considering gapless neutron superfluidity in neutron-star crust. This superfluid phase exists whenever
the effective superfluid velocity $\Vn$ exceeds Landau's velocity $\VLn$. The 
presence of such superflow is generally expected from the pinning of quantized 
vortices. The absence of a gap in the spectrum of quasiparticle excitations  leads 
to the huge enhancement of the neutron specific heat by orders of magnitude compared to 
that in the classical BCS phase assumed so far. This translates into a delayed thermal 
relaxation of the crust. We have thus obtained 
excellent fits to all the cooling data for both KS~1731$-$260 and MXB~1659$-$29, independently of the envelope composition and the accretion rate. 
However, the value of $\Vn/\VLn$ is not well constrained by current observations. Here, we have treated this ratio as a constant throughout the crust. Introducing some variations with density $\rho$ would have lead to even better fits to the data and therefore strengthened evidence for gapless superfluidity.

Our models lead to predictions that are 
very different from those obtained within the standard cooling models ignoring the dynamics of the neutron superfluid. 
According to our simulations with gapless superfluidity, the crust of these SXTs has not thermally relaxed yet and their 
thermal evolution will continue during the next decade for KS~1731$-$260 and the next century for MXB~1659$-$29 (if no 
accretion occurs during this interval of time and if neutron vortices remain pinned). The equilibrium temperature will 
depend on the composition of the envelope. These evolutions could be tested with additional observations of these sources.

\backmatter

\bmhead{Acknowledgments}

This work was financially  supported by the Fonds de la Recherche Scientifique (Belgium) under Grants 
No. PDR T.004320 and IISN 4.4502.19.

\section*{Declarations}

\begin{itemize}

\item Availability of data and materials

This manuscript has no associated data
or the data will not be deposited. [Authors’ comment: Our work involves
only theoretical arguments and numerical calculations. All the data and
results are presented as figures and tables in the manuscript.]

\end{itemize}

\begin{appendices}

\section{Thermal relaxation time}
\label{appendix:ThermalRelaxationTime}
 Following the analysis of Deibel et al.~\cite{Deibel2017}, we derive here an analytic estimate for the thermal relaxation time scale of the crust of accreted neutron stars~\eqref{eq:ThermalTime}. 

In the gapless regime, the neutron contribution to the crustal specific heat dominates by one or two orders of magnitude the electronic and ionic contributions in all crustal layers~\cite{AllardChamel2023Letter}. Therefore, we can approximate the crustal specific heat as 
\begin{align}\label{eq:CVnEstimate}
 c_V^{\rm crust}&\simeq c_N^{(n)}(T) R_{00}^{\rm Gapless}(T, \Vn) \notag \\
 & \approx 1\times 10^{18}~{\rm ergs}~{\rm cm}^{-3}~{\rm K}^{-1} \notag \\
 &\qquad\quad \times\left[\left(\rho_{14}Y_{\rm nf}\right)^{1/3} T_7 R_{00}^{\rm Gapless}(T, \Vn)\right]\;,   
\end{align}
with $c_N^{(n)}$ the neutron specific heat in the normal phase~\eqref{eq:CV-normal} and $R_{00}^{\rm Gapless}$ is the reduction factor given by Eq.~\eqref{eq:CV-Landau} or Eq.~\eqref{eq:SpecificHeat-LowT-HighV} (depending on the ratio $\Vn/\VLn$). Here $\rho_{14}$ is the average mass density in units of $10^{14}$~g~cm$^{-3}$, $Y_{\rm nf}$ is the fraction of free neutrons in the inner crust and $T_7$ is the temperature in units of $10^7$~K. 

As shown in Ref.~\cite{Deibel2017}, the thermal conductivity of the crust $\kappa_{\rm crust}$ is mainly determined by electrons even if neutrons are not superfluid (except 
near the crust-core transition where the neutron contribution can become comparable). 
Therefore, we make the following approximation 
\begin{align}
    \kappa_{\rm crust}&\approx \frac{\pi^2}{12}\frac{\varepsilon_{Fe}c}{e^4}\left(\frac{\langle Z\rangle}{\Qimp \Lambda_{eQ}}\right)k_{\rm B}^2 T \notag \\ 
    & \approx 4\times 10^{18}~{\rm ergs}~{\rm s}^{-1}~{\rm cm}^{-1}~{\rm K}^{-1}\notag \\
    & \qquad \quad \times \left[\left(\frac{\rho_{14}Y_e}{0.05}\right)^{1/3}\left( \frac{\langle Z\rangle}{\Qimp \Lambda_{eQ}}\right) T_7 \right]\; , 
\end{align}
where $c$ and $e$ correspond to the speed of light and the elementary charge, respectively, $\varepsilon_{Fe}$ is the electron Fermi energy, $Y_e$ is the 
electron fraction, $\langle Z\rangle$ denotes the average proton number, 
and $\Lambda_{eQ}$ corresponds to the Coulomb logarithm. The ratio 
$\Qimp\Lambda_{eQ}/\langle Z\rangle$ is typically of the order of unity~\cite{Deibel2017}. 

The thermal relaxation time scale~\eqref{eq:ThermalTime} is approximately given 
by
\begin{align}
    \tau 
    & \approx 3\times 10^4~{\rm days}\left(\frac{Y_e}{0.05 Y_{\rm nf}}\right)^{-1/3} \left(\frac{\Qimp \Lambda_{eQ}}{\langle Z\rangle}\right)\left(\frac{\Delta R}{1~{\rm km}}\right)^2 \notag\\
    &\qquad\qquad \times  R_{00}^{\rm Gapless}(T,\Vn/\VLn)\; . 
\end{align}

In the BCS regime, the neutron specific heat is suppressed. The specific heat of the crust is thus given only 
by that of electrons and ions and the thermal relaxation time is therefore shorter.

\end{appendices}



\begin{thebibliography}{109}
\ifx \bisbn   \undefined \def \bisbn  #1{ISBN #1}\fi
\ifx \binits  \undefined \def \binits#1{#1}\fi
\ifx \bauthor  \undefined \def \bauthor#1{#1}\fi
\ifx \batitle  \undefined \def \batitle#1{#1}\fi
\ifx \bjtitle  \undefined \def \bjtitle#1{#1}\fi
\ifx \bvolume  \undefined \def \bvolume#1{\textbf{#1}}\fi
\ifx \byear  \undefined \def \byear#1{#1}\fi
\ifx \bissue  \undefined \def \bissue#1{#1}\fi
\ifx \bfpage  \undefined \def \bfpage#1{#1}\fi
\ifx \blpage  \undefined \def \blpage #1{#1}\fi
\ifx \burl  \undefined \def \burl#1{\textsf{#1}}\fi
\ifx \doiurl  \undefined \def \doiurl#1{\url{https://doi.org/#1}}\fi
\ifx \betal  \undefined \def \betal{\textit{et al.}}\fi
\ifx \binstitute  \undefined \def \binstitute#1{#1}\fi
\ifx \binstitutionaled  \undefined \def \binstitutionaled#1{#1}\fi
\ifx \bctitle  \undefined \def \bctitle#1{#1}\fi
\ifx \beditor  \undefined \def \beditor#1{#1}\fi
\ifx \bpublisher  \undefined \def \bpublisher#1{#1}\fi
\ifx \bbtitle  \undefined \def \bbtitle#1{#1}\fi
\ifx \bedition  \undefined \def \bedition#1{#1}\fi
\ifx \bseriesno  \undefined \def \bseriesno#1{#1}\fi
\ifx \blocation  \undefined \def \blocation#1{#1}\fi
\ifx \bsertitle  \undefined \def \bsertitle#1{#1}\fi
\ifx \bsnm \undefined \def \bsnm#1{#1}\fi
\ifx \bsuffix \undefined \def \bsuffix#1{#1}\fi
\ifx \bparticle \undefined \def \bparticle#1{#1}\fi
\ifx \barticle \undefined \def \barticle#1{#1}\fi
\bibcommenthead
\ifx \bconfdate \undefined \def \bconfdate #1{#1}\fi
\ifx \botherref \undefined \def \botherref #1{#1}\fi
\ifx \url \undefined \def \url#1{\textsf{#1}}\fi
\ifx \bchapter \undefined \def \bchapter#1{#1}\fi
\ifx \bbook \undefined \def \bbook#1{#1}\fi
\ifx \bcomment \undefined \def \bcomment#1{#1}\fi
\ifx \oauthor \undefined \def \oauthor#1{#1}\fi
\ifx \citeauthoryear \undefined \def \citeauthoryear#1{#1}\fi
\ifx \endbibitem  \undefined \def \endbibitem {}\fi
\ifx \bconflocation  \undefined \def \bconflocation#1{#1}\fi
\ifx \arxivurl  \undefined \def \arxivurl#1{\textsf{#1}}\fi
\csname PreBibitemsHook\endcsname

\bibitem[\protect\citeauthoryear{{Bahramian} and {Degenaar}}{2023}]{Bahramian2023}
\begin{bchapter}
\bauthor{\bsnm{{Bahramian}}, \binits{A.}},
\bauthor{\bsnm{{Degenaar}}, \binits{N.}}:
\bctitle{{Low-Mass X-ray Binaries}}.
In: \bbtitle{Handbook of X-ray and Gamma-ray Astrophysics. Edited by Cosimo Bambi and Andrea Santangelo},
p. \bfpage{120}
(\byear{2023}).
\doiurl{10.1007/978-981-16-4544-0_94-1}
\end{bchapter}
\endbibitem

\bibitem[\protect\citeauthoryear{{Haensel} and {Zdunik}}{1990}]{HaenselZdunik1990}
\begin{barticle}
\bauthor{\bsnm{{Haensel}}, \binits{P.}},
\bauthor{\bsnm{{Zdunik}}, \binits{J.L.}}:
\batitle{{Non-equilibrium processes in the crust of an accreting neutron star}}.
\bjtitle{\aap}
\bvolume{227}(\bissue{2}),
\bfpage{431}--\blpage{436}
(\byear{1990})
\end{barticle}
\endbibitem

\bibitem[\protect\citeauthoryear{{Brown} et~al.}{1998}]{Brown1998}
\begin{barticle}
\bauthor{\bsnm{{Brown}}, \binits{E.F.}},
\bauthor{\bsnm{{Bildsten}}, \binits{L.}},
\bauthor{\bsnm{{Rutledge}}, \binits{R.E.}}:
\batitle{{Crustal Heating and Quiescent Emission from Transiently Accreting Neutron Stars}}.
\bjtitle{\apjl}
\bvolume{504}(\bissue{2}),
\bfpage{95}--\blpage{98}
(\byear{1998})
\doiurl{10.1086/311578}
\end{barticle}
\endbibitem

\bibitem[\protect\citeauthoryear{Wijnands et~al.}{2001}]{Wijnands2001}
\begin{barticle}
\bauthor{\bsnm{Wijnands}, \binits{R.}},
\bauthor{\bsnm{Strohmayer}, \binits{T.}},
\bauthor{\bsnm{Franco}, \binits{L.M.}}:
\batitle{Discovery of nearly coherent oscillations with a frequency of $\sim$567 {H}z during type {I} {X}-ray bursts of the {X}-ray transient and eclipsing binary {X}1658-298}.
\bjtitle{\apj}
\bvolume{549}(\bissue{1}),
\bfpage{71}--\blpage{75}
(\byear{2001})
\doiurl{10.1086/319128}
\end{barticle}
\endbibitem

\bibitem[\protect\citeauthoryear{Wijnands et~al.}{2002}]{wijnands2002}
\begin{barticle}
\bauthor{\bsnm{Wijnands}, \binits{R.}},
\bauthor{\bsnm{Guainazzi}, \binits{M.}},
\bauthor{\bsnm{Klis}, \binits{M.}},
\bauthor{\bsnm{Méndez}, \binits{M.}}:
\batitle{{XMM}-{N}ewton observations of the neutron star {X}-ray transient {KS}~1731–260 in quiescence}.
\bjtitle{\apj}
\bvolume{573}(\bissue{1}),
\bfpage{45}
(\byear{2002})
\doiurl{10.1086/341960}
\end{barticle}
\endbibitem

\bibitem[\protect\citeauthoryear{{Cackett} et~al.}{2006}]{cackett2006}
\begin{barticle}
\bauthor{\bsnm{{Cackett}}, \binits{E.M.}},
\bauthor{\bsnm{{Wijnands}}, \binits{R.}},
\bauthor{\bsnm{{Linares}}, \binits{M.}},
\bauthor{\bsnm{{Miller}}, \binits{J.M.}},
\bauthor{\bsnm{{Homan}}, \binits{J.}},
\bauthor{\bsnm{{Lewin}}, \binits{W.H.G.}}:
\batitle{{Cooling of the quasi-persistent neutron star X-ray transients KS 1731-260 and MXB 1659-29}}.
\bjtitle{\mnras}
\bvolume{372}(\bissue{1}),
\bfpage{479}--\blpage{488}
(\byear{2006})
\doiurl{10.1111/j.1365-2966.2006.10895.x}
\end{barticle}
\endbibitem

\bibitem[\protect\citeauthoryear{{Rutledge} et~al.}{2002}]{rutledge2002}
\begin{barticle}
\bauthor{\bsnm{{Rutledge}}, \binits{R.E.}},
\bauthor{\bsnm{{Bildsten}}, \binits{L.}},
\bauthor{\bsnm{{Brown}}, \binits{E.F.}},
\bauthor{\bsnm{{Pavlov}}, \binits{G.G.}},
\bauthor{\bsnm{{Zavlin}}, \binits{V.E.}},
\bauthor{\bsnm{{Ushomirsky}}, \binits{G.}}:
\batitle{{Crustal Emission and the Quiescent Spectrum of the Neutron Star in KS 1731-260}}.
\bjtitle{\apj}
\bvolume{580}(\bissue{1}),
\bfpage{413}--\blpage{422}
(\byear{2002})
\doiurl{10.1086/342745}
\end{barticle}
\endbibitem

\bibitem[\protect\citeauthoryear{{Shternin} et~al.}{2007}]{shternin2007}
\begin{barticle}
\bauthor{\bsnm{{Shternin}}, \binits{P.S.}},
\bauthor{\bsnm{{Yakovlev}}, \binits{D.G.}},
\bauthor{\bsnm{{Haensel}}, \binits{P.}},
\bauthor{\bsnm{{Potekhin}}, \binits{A.Y.}}:
\batitle{{Neutron star cooling after deep crustal heating in the X-ray transient KS 1731-260}}.
\bjtitle{\mnras}
\bvolume{382}(\bissue{1}),
\bfpage{43}--\blpage{47}
(\byear{2007})
\doiurl{10.1111/j.1745-3933.2007.00386.x}
\end{barticle}
\endbibitem

\bibitem[\protect\citeauthoryear{Brown and Cumming}{2009}]{BrownCumming2009}
\begin{barticle}
\bauthor{\bsnm{Brown}, \binits{E.F.}},
\bauthor{\bsnm{Cumming}, \binits{A.}}:
\batitle{Mapping crustal heating with the cooling light curves of quasi-persistent transients}.
\bjtitle{\apj}
\bvolume{698}(\bissue{2}),
\bfpage{1020}--\blpage{1032}
(\byear{2009})
\doiurl{10.1088/0004-637x/698/2/1020}
\end{barticle}
\endbibitem

\bibitem[\protect\citeauthoryear{Cackett et~al.}{2010}]{Cackett2010}
\begin{barticle}
\bauthor{\bsnm{Cackett}, \binits{E.M.}},
\bauthor{\bsnm{Brown}, \binits{E.F.}},
\bauthor{\bsnm{Cumming}, \binits{A.}},
\bauthor{\bsnm{Degenaar}, \binits{N.}},
\bauthor{\bsnm{Miller}, \binits{J.M.}},
\bauthor{\bsnm{Wijnands}, \binits{R.}}:
\batitle{Continued cooling of the crust in the neutron star low-mass {X}-ray binary {KS} 1731-260}.
\bjtitle{\apjl}
\bvolume{722}(\bissue{2}),
\bfpage{137}
(\byear{2010})
\doiurl{10.1088/2041-8205/722/2/l137}
\end{barticle}
\endbibitem

\bibitem[\protect\citeauthoryear{Merritt et~al.}{2016}]{Merritt2016}
\begin{barticle}
\bauthor{\bsnm{Merritt}, \binits{R.L.}},
\bauthor{\bsnm{Cackett}, \binits{E.M.}},
\bauthor{\bsnm{Brown}, \binits{E.F.}},
\bauthor{\bsnm{Page}, \binits{D.}},
\bauthor{\bsnm{Cumming}, \binits{A.}},
\bauthor{\bsnm{Degenaar}, \binits{N.}},
\bauthor{\bsnm{Deibel}, \binits{A.}},
\bauthor{\bsnm{Homan}, \binits{J.}},
\bauthor{\bsnm{Miller}, \binits{J.M.}},
\bauthor{\bsnm{Wijnands}, \binits{R.}}:
\batitle{The thermal state of {KS} 1731-260 after 14.5 years in quiescence}.
\bjtitle{\apj}
\bvolume{833}(\bissue{2}),
\bfpage{186}
(\byear{2016})
\doiurl{10.3847/1538-4357/833/2/186}
\end{barticle}
\endbibitem

\bibitem[\protect\citeauthoryear{Wijnands et~al.}{2003}]{Wijnands2003}
\begin{barticle}
\bauthor{\bsnm{Wijnands}, \binits{R.}},
\bauthor{\bsnm{Nowak}, \binits{M.}},
\bauthor{\bsnm{Miller}, \binits{J.M.}},
\bauthor{\bsnm{Homan}, \binits{J.}},
\bauthor{\bsnm{Wachter}, \binits{S.}},
\bauthor{\bsnm{Lewin}, \binits{W.H.G.}}:
\batitle{A \textit{Chandra} observation of the neutron star {X}-ray transient and eclipsing binary {MXB} 1659–29 in quiescence}.
\bjtitle{\apj}
\bvolume{594}(\bissue{2}),
\bfpage{952}--\blpage{960}
(\byear{2003})
\doiurl{10.1086/377122}
\end{barticle}
\endbibitem

\bibitem[\protect\citeauthoryear{{Wijnands} et~al.}{2004}]{wijnands2004}
\begin{barticle}
\bauthor{\bsnm{{Wijnands}}, \binits{R.}},
\bauthor{\bsnm{{Homan}}, \binits{J.}},
\bauthor{\bsnm{{Miller}}, \binits{J.M.}},
\bauthor{\bsnm{{Lewin}}, \binits{W.H.G.}}:
\batitle{{Monitoring Chandra Observations of the Quasi-persistent Neutron Star X-Ray Transient MXB 1659-29 in Quiescence: The Cooling Curve of the Heated Neutron Star Crust}}.
\bjtitle{\apjl}
\bvolume{606}(\bissue{1}),
\bfpage{61}--\blpage{64}
(\byear{2004})
\doiurl{10.1086/421081}
\end{barticle}
\endbibitem

\bibitem[\protect\citeauthoryear{{Cackett} et~al.}{2008}]{cackett2008}
\begin{barticle}
\bauthor{\bsnm{{Cackett}}, \binits{E.M.}},
\bauthor{\bsnm{{Wijnands}}, \binits{R.}},
\bauthor{\bsnm{{Miller}}, \binits{J.M.}},
\bauthor{\bsnm{{Brown}}, \binits{E.F.}},
\bauthor{\bsnm{{Degenaar}}, \binits{N.}}:
\batitle{{Cooling of the Crust in the Neutron Star Low-Mass X-Ray Binary MXB 1659-29}}.
\bjtitle{\apjl}
\bvolume{687}(\bissue{2}),
\bfpage{87}
(\byear{2008})
\doiurl{10.1086/593703}
\end{barticle}
\endbibitem

\bibitem[\protect\citeauthoryear{Cackett et~al.}{2013}]{Cackett2013}
\begin{barticle}
\bauthor{\bsnm{Cackett}, \binits{E.}},
\bauthor{\bsnm{Brown}, \binits{E.}},
\bauthor{\bsnm{Cumming}, \binits{A.}},
\bauthor{\bsnm{Degenaar}, \binits{N.}},
\bauthor{\bsnm{Fridriksson}, \binits{J.K.}},
\bauthor{\bsnm{Homan}, \binits{J.}},
\bauthor{\bsnm{Miller}, \binits{J.}},
\bauthor{\bsnm{Wijnands}, \binits{R.}}:
\batitle{A change in the quiescent {X}-ray spectrum of the neutron star low-mass {X}-ray binary {MXB} 1659-29}.
\bjtitle{\apj}
\bvolume{774}(\bissue{2}),
\bfpage{131}
(\byear{2013})
\doiurl{10.1088/0004-637X/774/2/131}
\end{barticle}
\endbibitem

\bibitem[\protect\citeauthoryear{{Negoro} et~al.}{2015}]{negoro2015}
\begin{barticle}
\bauthor{\bsnm{{Negoro}}, \binits{H.}},
\bauthor{\bsnm{{Furuya}}, \binits{K.}},
\bauthor{\bsnm{{Ueno}}, \binits{S.}},
\bauthor{\bsnm{{Tomida}}, \binits{H.}},
\bauthor{\bsnm{{Nakahira}}, \binits{S.}},
\bauthor{\bsnm{{Kimura}}, \binits{M.}},
\bauthor{\bsnm{{Ishikawa}}, \binits{M.}},
\bauthor{\bsnm{{Nakagawa}}, \binits{Y.E.}},
\bauthor{\bsnm{{Mihara}}, \binits{T.}},
\bauthor{\bsnm{{Sugizaki}}, \binits{M.}},
\bauthor{\bsnm{{Serino}}, \binits{M.}},
\bauthor{\bsnm{{Shidatsu}}, \binits{M.}},
\bauthor{\bsnm{{Sugimoto}}, \binits{J.}},
\bauthor{\bsnm{{Takagi}}, \binits{T.}},
\bauthor{\bsnm{{Matsuoka}}, \binits{M.}},
\bauthor{\bsnm{{Kawai}}, \binits{N.}},
\bauthor{\bsnm{{Tachibana}}, \binits{Y.}},
\bauthor{\bsnm{{Yoshii}}, \binits{T.}},
\bauthor{\bsnm{{Yoshida}}, \binits{A.}},
\bauthor{\bsnm{{Sakamoto}}, \binits{T.}},
\bauthor{\bsnm{{Kawakubo}}, \binits{Y.}},
\bauthor{\bsnm{{Ohtsuki}}, \binits{H.}},
\bauthor{\bsnm{{Tsunemi}}, \binits{H.}},
\bauthor{\bsnm{{Imatani}}, \binits{R.}},
\bauthor{\bsnm{{Nakajima}}, \binits{M.}},
\bauthor{\bsnm{{Masumitsu}}, \binits{T.}},
\bauthor{\bsnm{{Tanaka}}, \binits{K.}},
\bauthor{\bsnm{{Ueda}}, \binits{Y.}},
\bauthor{\bsnm{{Kawamuro}}, \binits{T.}},
\bauthor{\bsnm{{Hori}}, \binits{T.}},
\bauthor{\bsnm{{Tsuboi}}, \binits{Y.}},
\bauthor{\bsnm{{Kanetou}}, \binits{S.}},
\bauthor{\bsnm{{Yamauchi}}, \binits{M.}},
\bauthor{\bsnm{{Itoh}}, \binits{D.}},
\bauthor{\bsnm{{Yamaoka}}, \binits{K.}},
\bauthor{\bsnm{{Morii}}, \binits{M.}}:
\batitle{{MAXI/GSC detection of renewed X-ray activities of SAX J1324.5-6313/MAXI J1327-627 and H 1658-298/MAXI J1702-301}}.
\bjtitle{The Astronomer's Telegram}
\bvolume{7943},
\bfpage{1}
(\byear{2015})
\end{barticle}
\endbibitem

\bibitem[\protect\citeauthoryear{{Sanchez-Fernandez} et~al.}{2015}]{sanchez2015}
\begin{barticle}
\bauthor{\bsnm{{Sanchez-Fernandez}}, \binits{C.}},
\bauthor{\bsnm{{Eckert}}, \binits{D.}},
\bauthor{\bsnm{{Bozzo}}, \binits{E.}},
\bauthor{\bsnm{{Kajava}}, \binits{J.}},
\bauthor{\bsnm{{Kuulkers}}, \binits{E.}},
\bauthor{\bsnm{{Chenevez}}, \binits{J.}}:
\batitle{{INTEGRAL confirms the detection of renewed activity from the NS transient H 1658-298}}.
\bjtitle{The Astronomer's Telegram}
\bvolume{7946},
\bfpage{1}
(\byear{2015})
\end{barticle}
\endbibitem

\bibitem[\protect\citeauthoryear{{Parikh, A. S.} et~al.}{2019}]{Parikh2019}
\begin{barticle}
\bauthor{\bsnm{{Parikh, A. S.}}},
\bauthor{\bsnm{{Wijnands, R.}}},
\bauthor{\bsnm{{Ootes, L. S.}}},
\bauthor{\bsnm{{Page, D.}}},
\bauthor{\bsnm{{Degenaar, N.}}},
\bauthor{\bsnm{{Bahramian, A.}}},
\bauthor{\bsnm{{Brown, E. F.}}},
\bauthor{\bsnm{{Cackett, E. M.}}},
\bauthor{\bsnm{{Cumming, A.}}},
\bauthor{\bsnm{{Heinke, C.}}},
\bauthor{\bsnm{{Homan, J.}}},
\bauthor{\bsnm{{Rouco Escorial, A.}}},
\bauthor{\bsnm{{Wijngaarden, M. J. P.}}}:
\batitle{Consistent accretion-induced heating of the neutron-star crust in {MXB} 1659-29 during two different outbursts}.
\bjtitle{\aap}
\bvolume{624},
\bfpage{84}
(\byear{2019})
\doiurl{10.1051/0004-6361/201834412}
\end{barticle}
\endbibitem

\bibitem[\protect\citeauthoryear{Wijnands et~al.}{2017}]{Wijnands2017}
\begin{barticle}
\bauthor{\bsnm{Wijnands}, \binits{R.}},
\bauthor{\bsnm{Degenaar}, \binits{N.}},
\bauthor{\bsnm{Page}, \binits{D.}}:
\batitle{Cooling of accretion-heated neutron stars}.
\bjtitle{\jaa}
\bvolume{38}(\bissue{3}),
\bfpage{1}--\blpage{16}
(\byear{2017})
\doiurl{10.1007/s12036-017-9466-5}
\end{barticle}
\endbibitem

\bibitem[\protect\citeauthoryear{{Page} and {Reddy}}{2012}]{page2012}
\begin{bchapter}
\bauthor{\bsnm{{Page}}, \binits{D.}},
\bauthor{\bsnm{{Reddy}}, \binits{S.}}:
\bctitle{{Thermal and transport properties of the neutron star inner crust}}.
In: \beditor{\bsnm{{Bertulani}}, \binits{C.A.}},
\beditor{\bsnm{{Piekarewicz}}, \binits{J.}} (eds.)
\bbtitle{Neutron Star Crust},
pp. \bfpage{281}--\blpage{308}.
\bpublisher{Nova Science Publishers},
\blocation{New York}
(\byear{2012})
\end{bchapter}
\endbibitem

\bibitem[\protect\citeauthoryear{{Chaikin} et~al.}{2018}]{Chaikin2018}
\begin{barticle}
\bauthor{\bsnm{{Chaikin}}, \binits{E.A.}},
\bauthor{\bsnm{{Kaminker}}, \binits{A.D.}},
\bauthor{\bsnm{{Yakovlev}}, \binits{D.G.}}:
\batitle{{Afterburst thermal relaxation in neutron star crusts}}.
\bjtitle{\apss}
\bvolume{363}(\bissue{10}),
\bfpage{209}
(\byear{2018})
\doiurl{10.1007/s10509-018-3393-z}
\end{barticle}
\endbibitem

\bibitem[\protect\citeauthoryear{Singh et~al.}{2020}]{Singh2020}
\begin{barticle}
\bauthor{\bsnm{Singh}, \binits{N.}},
\bauthor{\bsnm{Haskell}, \binits{B.}},
\bauthor{\bsnm{Mukherjee}, \binits{D.}},
\bauthor{\bsnm{Bulik}, \binits{T.}}:
\batitle{{Asymmetric accretion and thermal ‘mountains’ in magnetized neutron star crusts}}.
\bjtitle{\mnras}
\bvolume{493}(\bissue{3}),
\bfpage{3866}--\blpage{3878}
(\byear{2020})
\doiurl{10.1093/mnras/staa442}
\end{barticle}
\endbibitem

\bibitem[\protect\citeauthoryear{{Chamel} et~al.}{2020}]{ChamelFantina2020}
\begin{barticle}
\bauthor{\bsnm{{Chamel}}, \binits{N.}},
\bauthor{\bsnm{{Fantina}}, \binits{A.F.}},
\bauthor{\bsnm{{Zdunik}}, \binits{J.L.}},
\bauthor{\bsnm{{Haensel}}, \binits{P.}}:
\batitle{{Experimental constraints on shallow heating in accreting neutron-star crusts}}.
\bjtitle{\prc}
\bvolume{102}(\bissue{1}),
\bfpage{015804}
(\byear{2020})
\doiurl{10.1103/PhysRevC.102.015804}
\end{barticle}
\endbibitem

\bibitem[\protect\citeauthoryear{{Horowitz} et~al.}{2015}]{horowitz2015}
\begin{barticle}
\bauthor{\bsnm{{Horowitz}}, \binits{C.J.}},
\bauthor{\bsnm{{Berry}}, \binits{D.K.}},
\bauthor{\bsnm{{Briggs}}, \binits{C.M.}},
\bauthor{\bsnm{{Caplan}}, \binits{M.E.}},
\bauthor{\bsnm{{Cumming}}, \binits{A.}},
\bauthor{\bsnm{{Schneider}}, \binits{A.S.}}:
\batitle{{Disordered Nuclear Pasta, Magnetic Field Decay, and Crust Cooling in Neutron Stars}}.
\bjtitle{\prl}
\bvolume{114}(\bissue{3}),
\bfpage{031102}
(\byear{2015})
\doiurl{10.1103/PhysRevLett.114.031102}
\end{barticle}
\endbibitem

\bibitem[\protect\citeauthoryear{{Nandi} and {Schramm}}{2018}]{nandi2018}
\begin{barticle}
\bauthor{\bsnm{{Nandi}}, \binits{R.}},
\bauthor{\bsnm{{Schramm}}, \binits{S.}}:
\batitle{{Transport Properties of the Nuclear Pasta Phase with Quantum Molecular Dynamics}}.
\bjtitle{\apj}
\bvolume{852}(\bissue{2}),
\bfpage{135}
(\byear{2018})
\doiurl{10.3847/1538-4357/aa9f12}
\end{barticle}
\endbibitem

\bibitem[\protect\citeauthoryear{{Turlione, A.} et~al.}{2015}]{Turlione2015}
\begin{barticle}
\bauthor{\bsnm{{Turlione, A.}}},
\bauthor{\bsnm{{Aguilera, D. N.}}},
\bauthor{\bsnm{{Pons, J. A.}}}:
\batitle{Quiescent thermal emission from neutron stars in low-mass {X}-ray binaries}.
\bjtitle{\aap}
\bvolume{577},
\bfpage{5}
(\byear{2015})
\doiurl{10.1051/0004-6361/201322690}
\end{barticle}
\endbibitem

\bibitem[\protect\citeauthoryear{Deibel et~al.}{2017}]{Deibel2017}
\begin{barticle}
\bauthor{\bsnm{Deibel}, \binits{A.}},
\bauthor{\bsnm{Cumming}, \binits{A.}},
\bauthor{\bsnm{Brown}, \binits{E.F.}},
\bauthor{\bsnm{Reddy}, \binits{S.}}:
\batitle{Late-time cooling of neutron star transients and the physics of the inner crust}.
\bjtitle{\apj}
\bvolume{839}(\bissue{2}),
\bfpage{95}
(\byear{2017})
\doiurl{10.3847/1538-4357/aa6a19}
\end{barticle}
\endbibitem

\bibitem[\protect\citeauthoryear{Gandolfi et~al.}{2008}]{Gandolfi2008}
\begin{barticle}
\bauthor{\bsnm{Gandolfi}, \binits{S.}},
\bauthor{\bsnm{Illarionov}, \binits{A.Y.}},
\bauthor{\bsnm{Fantoni}, \binits{S.}},
\bauthor{\bsnm{Pederiva}, \binits{F.}},
\bauthor{\bsnm{Schmidt}, \binits{K.E.}}:
\batitle{Equation of state of superfluid neutron matter and the calculation of the $^{1}{S}_{0}$ pairing gap}.
\bjtitle{\prl}
\bvolume{101},
\bfpage{132501}
(\byear{2008})
\doiurl{10.1103/PhysRevLett.101.132501}
\end{barticle}
\endbibitem

\bibitem[\protect\citeauthoryear{Gandolfi et~al.}{2022}]{Gandolfi2022}
\begin{botherref}
\oauthor{\bsnm{Gandolfi}, \binits{S.}},
\oauthor{\bsnm{Palkanoglou}, \binits{G.}},
\oauthor{\bsnm{Carlson}, \binits{J.}},
\oauthor{\bsnm{Gezerlis}, \binits{A.}},
\oauthor{\bsnm{Schmidt}, \binits{K.E.}}:
The $^{1}{S}_{0}$ pairing gap in neutron matter.
Condensed Matter
\textbf{7}(1)
(2022)
\doiurl{10.3390/condmat7010019}
\end{botherref}
\endbibitem

\bibitem[\protect\citeauthoryear{{Drissi} and {Rios}}{2022}]{drissi2022}
\begin{barticle}
\bauthor{\bsnm{{Drissi}}, \binits{M.}},
\bauthor{\bsnm{{Rios}}, \binits{A.}}:
\batitle{{Many-body approximations to the superfluid gap and critical temperature in pure neutron matter}}.
\bjtitle{\epja}
\bvolume{58}(\bissue{5}),
\bfpage{90}
(\byear{2022})
\doiurl{10.1140/epja/s10050-022-00738-2}
\end{barticle}
\endbibitem

\bibitem[\protect\citeauthoryear{Krotscheck et~al.}{2023}]{Krotscheck2023}
\begin{barticle}
\bauthor{\bsnm{Krotscheck}, \binits{E.}},
\bauthor{\bsnm{Papakonstantinou}, \binits{P.}},
\bauthor{\bsnm{Wang}, \binits{J.}}:
\batitle{Triplet pairing in neutron matter}.
\bjtitle{\apj}
\bvolume{955}(\bissue{1}),
\bfpage{76}
(\byear{2023})
\doiurl{10.3847/1538-4357/acee7c}
\end{barticle}
\endbibitem

\bibitem[\protect\citeauthoryear{{Potekhin} and {Chabrier}}{2021}]{potekhin2021}
\begin{barticle}
\bauthor{\bsnm{{Potekhin}}, \binits{A.Y.}},
\bauthor{\bsnm{{Chabrier}}, \binits{G.}}:
\batitle{{Crust structure and thermal evolution of neutron stars in soft X-ray transients}}.
\bjtitle{\aap}
\bvolume{645},
\bfpage{102}
(\byear{2021})
\doiurl{10.1051/0004-6361/202039006}
\end{barticle}
\endbibitem

\bibitem[\protect\citeauthoryear{{Lu} et~al.}{2022}]{Lu2022}
\begin{barticle}
\bauthor{\bsnm{{Lu}}, \binits{X.Y.}},
\bauthor{\bsnm{{L\"u}}, \binits{G.L.}},
\bauthor{\bsnm{{Liu}}, \binits{H.L.}},
\bauthor{\bsnm{{Zhu}}, \binits{C.H.}},
\bauthor{\bsnm{{Wang}}, \binits{Z.J.}}:
\batitle{{Crust Cooling of Soft X-Ray Transients{\textemdash}the Uncertainties of Shallow Heating}}.
\bjtitle{Res. Astron. Astrophys.}
\bvolume{22}(\bissue{5}),
\bfpage{055018}
(\byear{2022})
\doiurl{10.1088/1674-4527/ac630f}
\end{barticle}
\endbibitem

\bibitem[\protect\citeauthoryear{{Haensel} and {Zdunik}}{2003}]{HaenselZdunik2003}
\begin{barticle}
\bauthor{\bsnm{{Haensel}}, \binits{P.}},
\bauthor{\bsnm{{Zdunik}}, \binits{J.L.}}:
\batitle{{Nuclear composition and heating in accreting neutron-star crusts}}.
\bjtitle{\aap}
\bvolume{404},
\bfpage{33}--\blpage{36}
(\byear{2003})
\doiurl{10.1051/0004-6361:20030708}
\end{barticle}
\endbibitem

\bibitem[\protect\citeauthoryear{{Haensel, P.} and {Zdunik, J. L.}}{2008}]{HaenselZdunik2008}
\begin{barticle}
\bauthor{\bsnm{{Haensel, P.}}},
\bauthor{\bsnm{{Zdunik, J. L.}}}:
\batitle{Models of crustal heating in accreting neutron stars}.
\bjtitle{\aap}
\bvolume{480}(\bissue{2}),
\bfpage{459}--\blpage{464}
(\byear{2008})
\doiurl{10.1051/0004-6361:20078578}
\end{barticle}
\endbibitem

\bibitem[\protect\citeauthoryear{{Fantina} et~al.}{2018}]{fantina2018}
\begin{barticle}
\bauthor{\bsnm{{Fantina}}, \binits{A.F.}},
\bauthor{\bsnm{{Zdunik}}, \binits{J.L.}},
\bauthor{\bsnm{{Chamel}}, \binits{N.}},
\bauthor{\bsnm{{Pearson}}, \binits{J.M.}},
\bauthor{\bsnm{{Haensel}}, \binits{P.}},
\bauthor{\bsnm{{Goriely}}, \binits{S.}}:
\batitle{{Crustal heating in accreting neutron stars from the nuclear energy-density functional theory. I. Proton shell effects and neutron-matter constraint}}.
\bjtitle{\aap}
\bvolume{620},
\bfpage{105}
(\byear{2018})
\doiurl{10.1051/0004-6361/201833605}
\end{barticle}
\endbibitem

\bibitem[\protect\citeauthoryear{Steiner}{2012}]{Steiner2012}
\begin{barticle}
\bauthor{\bsnm{Steiner}, \binits{A.W.}}:
\batitle{Deep crustal heating in a multicomponent accreted neutron star crust}.
\bjtitle{\prc}
\bvolume{85},
\bfpage{055804}
(\byear{2012})
\doiurl{10.1103/PhysRevC.85.055804}
\end{barticle}
\endbibitem

\bibitem[\protect\citeauthoryear{{Chamel} et~al.}{2015}]{ChamelFantina2015}
\begin{barticle}
\bauthor{\bsnm{{Chamel}}, \binits{N.}},
\bauthor{\bsnm{{Fantina}}, \binits{A.F.}},
\bauthor{\bsnm{{Zdunik}}, \binits{J.L.}},
\bauthor{\bsnm{{Haensel}}, \binits{P.}}:
\batitle{{Neutron drip transition in accreting and nonaccreting neutron star crusts}}.
\bjtitle{\prc}
\bvolume{91}(\bissue{5}),
\bfpage{055803}
(\byear{2015})
\doiurl{10.1103/PhysRevC.91.055803}
\end{barticle}
\endbibitem

\bibitem[\protect\citeauthoryear{{Chugunov} and {Shchechilin}}{2020}]{ChugunovShchechilin2020}
\begin{barticle}
\bauthor{\bsnm{{Chugunov}}, \binits{A.I.}},
\bauthor{\bsnm{{Shchechilin}}, \binits{N.N.}}:
\batitle{{Crucial role of neutron diffusion in the crust of accreting neutron stars}}.
\bjtitle{\mnras}
\bvolume{495}(\bissue{1}),
\bfpage{32}--\blpage{36}
(\byear{2020})
\doiurl{10.1093/mnrasl/slaa055}
\end{barticle}
\endbibitem

\bibitem[\protect\citeauthoryear{Gusakov and Chugunov}{2020}]{GusakovChugunov2020}
\begin{barticle}
\bauthor{\bsnm{Gusakov}, \binits{M.E.}},
\bauthor{\bsnm{Chugunov}, \binits{A.I.}}:
\batitle{Thermodynamically consistent equation of state for an accreted neutron star crust}.
\bjtitle{\prl}
\bvolume{124},
\bfpage{191101}
(\byear{2020})
\doiurl{10.1103/PhysRevLett.124.191101}
\end{barticle}
\endbibitem

\bibitem[\protect\citeauthoryear{Gusakov and Chugunov}{2021}]{GusakovChugunov2021}
\begin{barticle}
\bauthor{\bsnm{Gusakov}, \binits{M.E.}},
\bauthor{\bsnm{Chugunov}, \binits{A.I.}}:
\batitle{Heat release in accreting neutron stars}.
\bjtitle{Phys. Rev. D}
\bvolume{103},
\bfpage{101301}
(\byear{2021})
\doiurl{10.1103/PhysRevD.103.L101301}
\end{barticle}
\endbibitem

\bibitem[\protect\citeauthoryear{{Gusakov} et~al.}{2021}]{GusakovKantor2021}
\begin{barticle}
\bauthor{\bsnm{{Gusakov}}, \binits{M.E.}},
\bauthor{\bsnm{{Kantor}}, \binits{E.M.}},
\bauthor{\bsnm{{Chugunov}}, \binits{A.I.}}:
\batitle{{Nonequilibrium thermodynamics of accreted neutron-star crust}}.
\bjtitle{\prd}
\bvolume{104}(\bissue{8}),
\bfpage{081301}
(\byear{2021})
\doiurl{10.1103/PhysRevD.104.L081301}
\end{barticle}
\endbibitem

\bibitem[\protect\citeauthoryear{{Shchechilin} et~al.}{2021}]{Shchechilin2021}
\begin{barticle}
\bauthor{\bsnm{{Shchechilin}}, \binits{N.N.}},
\bauthor{\bsnm{{Gusakov}}, \binits{M.E.}},
\bauthor{\bsnm{{Chugunov}}, \binits{A.I.}}:
\batitle{{Deep crustal heating for realistic compositions of thermonuclear ashes}}.
\bjtitle{\mnras}
\bvolume{507}(\bissue{3}),
\bfpage{3860}--\blpage{3870}
(\byear{2021})
\doiurl{10.1093/mnras/stab2415}
\end{barticle}
\endbibitem

\bibitem[\protect\citeauthoryear{{Shchechilin} et~al.}{2022}]{Shchechilin2022}
\begin{barticle}
\bauthor{\bsnm{{Shchechilin}}, \binits{N.N.}},
\bauthor{\bsnm{{Gusakov}}, \binits{M.E.}},
\bauthor{\bsnm{{Chugunov}}, \binits{A.I.}}:
\batitle{{Accreting neutron stars: heating of the upper layers of the inner crust}}.
\bjtitle{\mnras}
\bvolume{515}(\bissue{1}),
\bfpage{6}--\blpage{10}
(\byear{2022})
\doiurl{10.1093/mnrasl/slac059}
\end{barticle}
\endbibitem

\bibitem[\protect\citeauthoryear{{Shchechilin} et~al.}{2023}]{Shchechilin2023}
\begin{barticle}
\bauthor{\bsnm{{Shchechilin}}, \binits{N.N.}},
\bauthor{\bsnm{{Gusakov}}, \binits{M.E.}},
\bauthor{\bsnm{{Chugunov}}, \binits{A.I.}}:
\batitle{{Accreting neutron stars: composition of the upper layers of the inner crust}}.
\bjtitle{\mnras}
\bvolume{523}(\bissue{3}),
\bfpage{4643}--\blpage{4653}
(\byear{2023})
\doiurl{10.1093/mnras/stad1731}
\end{barticle}
\endbibitem

\bibitem[\protect\citeauthoryear{Potekhin et~al.}{2023}]{potekhin2023}
\begin{barticle}
\bauthor{\bsnm{Potekhin}, \binits{A.Y.}},
\bauthor{\bsnm{Gusakov}, \binits{M.E.}},
\bauthor{\bsnm{Chugunov}, \binits{A.I.}}:
\batitle{{Thermal evolution of neutron stars in soft X-ray transients with thermodynamically consistent models of the accreted crust}}.
\bjtitle{\mnras}
\bvolume{522}(\bissue{4}),
\bfpage{4830}--\blpage{4840}
(\byear{2023})
\doiurl{10.1093/mnras/stad1309}
\end{barticle}
\endbibitem

\bibitem[\protect\citeauthoryear{{Alpar} et~al.}{1984}]{alpar1984}
\begin{barticle}
\bauthor{\bsnm{{Alpar}}, \binits{M.A.}},
\bauthor{\bsnm{{Langer}}, \binits{S.A.}},
\bauthor{\bsnm{{Sauls}}, \binits{J.A.}}:
\batitle{{Rapid postglitch spin-up of the superfluid core in pulsars}}.
\bjtitle{\apj}
\bvolume{282},
\bfpage{533}--\blpage{541}
(\byear{1984})
\doiurl{10.1086/162232}
\end{barticle}
\endbibitem

\bibitem[\protect\citeauthoryear{Smith et~al.}{1997}]{Smith1997}
\begin{barticle}
\bauthor{\bsnm{Smith}, \binits{D.A.}},
\bauthor{\bsnm{Morgan}, \binits{E.H.}},
\bauthor{\bsnm{Bradt}, \binits{H.}}:
\batitle{\textit{Rossi {X}-Ray Timing Explorer} discovery of coherent millisecond pulsations during an {X}-ray burst from {KS} 1731-260}.
\bjtitle{\apj}
\bvolume{479}(\bissue{2}),
\bfpage{137}--\blpage{140}
(\byear{1997})
\doiurl{10.1086/310604}
\end{barticle}
\endbibitem

\bibitem[\protect\citeauthoryear{Galloway et~al.}{2008}]{Galloway2008}
\begin{barticle}
\bauthor{\bsnm{Galloway}, \binits{D.K.}},
\bauthor{\bsnm{Muno}, \binits{M.P.}},
\bauthor{\bsnm{Hartman}, \binits{J.M.}},
\bauthor{\bsnm{Psaltis}, \binits{D.}},
\bauthor{\bsnm{Chakrabarty}, \binits{D.}}:
\batitle{Thermonuclear (type {I}) {X}-ray bursts observed by the \textit{{R}ossi {X}-{R}ay {T}iming {E}xplorer}}.
\bjtitle{\apjs}
\bvolume{179}(\bissue{2}),
\bfpage{360}
(\byear{2008})
\doiurl{10.1086/592044}
\end{barticle}
\endbibitem

\bibitem[\protect\citeauthoryear{{Alpar} et~al.}{1982}]{alpar1982}
\begin{barticle}
\bauthor{\bsnm{{Alpar}}, \binits{M.A.}},
\bauthor{\bsnm{{Cheng}}, \binits{A.F.}},
\bauthor{\bsnm{{Ruderman}}, \binits{M.A.}},
\bauthor{\bsnm{{Shaham}}, \binits{J.}}:
\batitle{{A new class of radio pulsars}}.
\bjtitle{\nat}
\bvolume{300}(\bissue{5894}),
\bfpage{728}--\blpage{730}
(\byear{1982})
\doiurl{10.1038/300728a0}
\end{barticle}
\endbibitem

\bibitem[\protect\citeauthoryear{{Radhakrishnan} and {Srinivasan}}{1982}]{radhakrishnan1982}
\begin{barticle}
\bauthor{\bsnm{{Radhakrishnan}}, \binits{V.}},
\bauthor{\bsnm{{Srinivasan}}, \binits{G.}}:
\batitle{{On the origin of the recently discovered ultra-rapid pulsar}}.
\bjtitle{\currsc}
\bvolume{51},
\bfpage{1096}--\blpage{1099}
(\byear{1982})
\end{barticle}
\endbibitem

\bibitem[\protect\citeauthoryear{{Patruno} and {Watts}}{2021}]{patruno2021}
\begin{bchapter}
\bauthor{\bsnm{{Patruno}}, \binits{A.}},
\bauthor{\bsnm{{Watts}}, \binits{A.L.}}:
\bctitle{{Accreting Millisecond X-ray Pulsars}}.
In: \beditor{\bsnm{{Belloni}}, \binits{T.M.}},
\beditor{\bsnm{{M{\'e}ndez}}, \binits{M.}},
\beditor{\bsnm{{Zhang}}, \binits{C.}} (eds.)
\bbtitle{Astrophysics and Space Science Library}.
\bsertitle{Astrophysics and Space Science Library},
vol. \bseriesno{461},
pp. \bfpage{143}--\blpage{208}
(\byear{2021}).
\doiurl{10.1007/978-3-662-62110-3_4}
\end{bchapter}
\endbibitem

\bibitem[\protect\citeauthoryear{{Salvo} and {Sanna}}{2022}]{Salvo2022}
\begin{bbook}
\bauthor{\bsnm{{Salvo}}, \binits{T.D.}},
\bauthor{\bsnm{{Sanna}}, \binits{A.}}:
In: \beditor{\bsnm{{Bhattacharyya}}, \binits{S.}},
\beditor{\bsnm{{Papitto}}, \binits{A.}},
\beditor{\bsnm{{Bhattacharya}}, \binits{D.}} (eds.)
\bbtitle{{Accretion Powered X-ray Millisecond Pulsars}},
pp. \bfpage{87}--\blpage{124}.
\bpublisher{Springer},
\blocation{Cham}
(\byear{2022}).
\doiurl{10.1007/978-3-030-85198-9_4}
\end{bbook}
\endbibitem

\bibitem[\protect\citeauthoryear{{Papitto} and {de Martino}}{2022}]{papitto2022}
\begin{bchapter}
\bauthor{\bsnm{{Papitto}}, \binits{A.}},
\bauthor{\bsnm{{de Martino}}, \binits{D.}}:
\bctitle{{Transitional Millisecond Pulsars}}.
In: \beditor{\bsnm{{Bhattacharyya}}, \binits{S.}},
\beditor{\bsnm{{Papitto}}, \binits{A.}},
\beditor{\bsnm{{Bhattacharya}}, \binits{D.}} (eds.)
\bbtitle{Astrophysics and Space Science Library}.
\bsertitle{Astrophysics and Space Science Library},
vol. \bseriesno{465},
pp. \bfpage{157}--\blpage{200}
(\byear{2022}).
\doiurl{10.1007/978-3-030-85198-9_6}
\end{bchapter}
\endbibitem

\bibitem[\protect\citeauthoryear{{Sourie} and {Chamel}}{2020}]{SourieChamel2020}
\begin{barticle}
\bauthor{\bsnm{{Sourie}}, \binits{A.}},
\bauthor{\bsnm{{Chamel}}, \binits{N.}}:
\batitle{{Generalization of the Kutta-Joukowski theorem for the hydrodynamic forces acting on a quantized vortex}}.
\bjtitle{International Journal of Modern Physics B}
\bvolume{34}(\bissue{10}),
\bfpage{2050099}--\blpage{137}
(\byear{2020})
\doiurl{10.1142/S021797922050099X}
\end{barticle}
\endbibitem

\bibitem[\protect\citeauthoryear{{Antonopoulou} et~al.}{2022}]{antonopoulou2022}
\begin{barticle}
\bauthor{\bsnm{{Antonopoulou}}, \binits{D.}},
\bauthor{\bsnm{{Haskell}}, \binits{B.}},
\bauthor{\bsnm{{Espinoza}}, \binits{C.M.}}:
\batitle{{Pulsar glitches: observations and physical interpretation}}.
\bjtitle{\RepProgPhys}
\bvolume{85}(\bissue{12}),
\bfpage{126901}
(\byear{2022})
\doiurl{10.1088/1361-6633/ac9ced}
\end{barticle}
\endbibitem

\bibitem[\protect\citeauthoryear{{Anderson} and {Itoh}}{1975}]{anderson1975}
\begin{barticle}
\bauthor{\bsnm{{Anderson}}, \binits{P.W.}},
\bauthor{\bsnm{{Itoh}}, \binits{N.}}:
\batitle{{Pulsar glitches and restlessness as a hard superfluidity phenomenon}}.
\bjtitle{\nat}
\bvolume{256}(\bissue{5512}),
\bfpage{25}--\blpage{27}
(\byear{1975})
\doiurl{10.1038/256025a0}
\end{barticle}
\endbibitem

\bibitem[\protect\citeauthoryear{{Pines} and {Alpar}}{1985}]{alpar1985}
\begin{barticle}
\bauthor{\bsnm{{Pines}}, \binits{D.}},
\bauthor{\bsnm{{Alpar}}, \binits{M.A.}}:
\batitle{{Superfluidity in neutron stars}}.
\bjtitle{\nat}
\bvolume{316}(\bissue{6023}),
\bfpage{27}--\blpage{32}
(\byear{1985})
\doiurl{10.1038/316027a0}
\end{barticle}
\endbibitem

\bibitem[\protect\citeauthoryear{Galloway et~al.}{2004}]{Galloway2004}
\begin{barticle}
\bauthor{\bsnm{Galloway}, \binits{D.K.}},
\bauthor{\bsnm{Morgan}, \binits{E.H.}},
\bauthor{\bsnm{Levine}, \binits{A.M.}}:
\batitle{A frequency glitch in an accreting pulsar}.
\bjtitle{\apj}
\bvolume{613}(\bissue{2}),
\bfpage{1164}--\blpage{1172}
(\byear{2004})
\doiurl{10.1086/423265}
\end{barticle}
\endbibitem

\bibitem[\protect\citeauthoryear{{Serim} et~al.}{2017}]{serim2017}
\begin{barticle}
\bauthor{\bsnm{{Serim}}, \binits{M.M.}},
\bauthor{\bsnm{{{\c{S}}ahiner}}, \binits{{\c{S}}.}},
\bauthor{\bsnm{{{\c{C}}erri-Serim}}, \binits{D.}},
\bauthor{\bsnm{{Inam}}, \binits{S.{\c{C}}.}},
\bauthor{\bsnm{{Baykal}}, \binits{A.}}:
\batitle{{Discovery of a glitch in the accretion-powered pulsar SXP 1062}}.
\bjtitle{\mnras}
\bvolume{471}(\bissue{4}),
\bfpage{4982}--\blpage{4989}
(\byear{2017})
\doiurl{10.1093/mnras/stx1771}
\end{barticle}
\endbibitem

\bibitem[\protect\citeauthoryear{{Allard} and {Chamel}}{2023}]{AllardChamel2023PartI}
\begin{barticle}
\bauthor{\bsnm{{Allard}}, \binits{V.}},
\bauthor{\bsnm{{Chamel}}, \binits{N.}}:
\batitle{{Gapless superfluidity in neutron stars: Thermal properties}}.
\bjtitle{\prc}
\bvolume{108}(\bissue{1}),
\bfpage{015801}
(\byear{2023})
\doiurl{10.1103/PhysRevC.108.015801}
\end{barticle}
\endbibitem

\bibitem[\protect\citeauthoryear{{Allard} and {Chamel}}{2024}]{AllardChamel2023Letter}
\begin{botherref}
\oauthor{\bsnm{{Allard}}, \binits{V.}},
\oauthor{\bsnm{{Chamel}}, \binits{N.}}:
{Gapless neutron superfluidity can explain the late time cooling of transiently accreting neutron stars}.
arXiv e-prints,
2403--07740
(2024)
\doiurl{10.48550/arXiv.2403.07740} .
in press
\end{botherref}
\endbibitem

\bibitem[\protect\citeauthoryear{Allard and Chamel}{2021}]{ChamelAllard2020}
\begin{barticle}
\bauthor{\bsnm{Allard}, \binits{V.}},
\bauthor{\bsnm{Chamel}, \binits{N.}}:
\batitle{Entrainment effects in neutron-proton mixtures within the nuclear energy-density functional theory. {II}. finite temperatures and arbitrary currents}.
\bjtitle{\prc}
\bvolume{103},
\bfpage{025804}
(\byear{2021})
\doiurl{10.1103/PhysRevC.103.025804}
\end{barticle}
\endbibitem

\bibitem[\protect\citeauthoryear{Bulgac and Yu}{2002}]{bulgac2002}
\begin{barticle}
\bauthor{\bsnm{Bulgac}, \binits{A.}},
\bauthor{\bsnm{Yu}, \binits{Y.}}:
\batitle{Renormalization of the {H}artree-{F}ock-{B}ogoliubov equations in the case of a zero range pairing interaction}.
\bjtitle{\prl}
\bvolume{88},
\bfpage{042504}
(\byear{2002})
\doiurl{10.1103/PhysRevLett.88.042504}
\end{barticle}
\endbibitem

\bibitem[\protect\citeauthoryear{{Schunck}}{2019}]{schunck2019}
\begin{bbook}
\beditor{\bsnm{{Schunck}}, \binits{N.}} (ed.):
\bbtitle{Energy Density Functional Methods for Atomic Nuclei}.
\bsertitle{2053-2563}.
\bpublisher{IOP Publishing},
\blocation{Bristol, UK}
(\byear{2019}).
\doiurl{10.1088/2053-2563/aae0ed}
\end{bbook}
\endbibitem

\bibitem[\protect\citeauthoryear{Landau}{1941}]{Landau1941}
\begin{barticle}
\bauthor{\bsnm{Landau}, \binits{L.}}:
\batitle{Theory of the superfluidity of helium {II}}.
\bjtitle{Physical Review}
\bvolume{60}(\bissue{4}),
\bfpage{356}
(\byear{1941})
\doiurl{10.1103/PhysRev.60.356}
\end{barticle}
\endbibitem

\bibitem[\protect\citeauthoryear{Allard and Chamel}{2021}]{ChamelAllard2021}
\begin{barticle}
\bauthor{\bsnm{Allard}, \binits{V.}},
\bauthor{\bsnm{Chamel}, \binits{N.}}:
\batitle{$^{1}{S}_{0}$ pairing gaps, chemical potentials and entrainment matrix in superfluid neutron-star cores for the {B}russels-{M}ontreal functionals}.
\bjtitle{Universe}
\bvolume{7}(\bissue{12}),
\bfpage{470}
(\byear{2021})
\doiurl{10.3390/universe7120470}
\end{barticle}
\endbibitem

\bibitem[\protect\citeauthoryear{Parmenter}{1962}]{Parmenter1962}
\begin{barticle}
\bauthor{\bsnm{Parmenter}, \binits{R.H.}}:
\batitle{Nonlinear electrodynamics of superconductors with a very small coherence distance}.
\bjtitle{RCA Rev. (United States)}
\bvolume{23},
\bfpage{323}--\blpage{352}
(\byear{1962})
\end{barticle}
\endbibitem

\bibitem[\protect\citeauthoryear{{Bardeen}}{1962}]{bardeen1962}
\begin{barticle}
\bauthor{\bsnm{{Bardeen}}, \binits{J.}}:
\batitle{{Critical Fields and Currents in Superconductors}}.
\bjtitle{\revmodphys}
\bvolume{34}(\bissue{4}),
\bfpage{667}--\blpage{681}
(\byear{1962})
\doiurl{10.1103/RevModPhys.34.667}
\end{barticle}
\endbibitem

\bibitem[\protect\citeauthoryear{Allard and Chamel}{2023}]{AllardChamel2023PartII}
\begin{barticle}
\bauthor{\bsnm{Allard}, \binits{V.}},
\bauthor{\bsnm{Chamel}, \binits{N.}}:
\batitle{Gapless superfluidity in neutron stars: Normal-fluid fraction}.
\bjtitle{\prc}
\bvolume{108},
\bfpage{045801}
(\byear{2023})
\doiurl{10.1103/PhysRevC.108.045801}
\end{barticle}
\endbibitem

\bibitem[\protect\citeauthoryear{{Cao} et~al.}{2006}]{cao2006}
\begin{barticle}
\bauthor{\bsnm{{Cao}}, \binits{L.G.}},
\bauthor{\bsnm{{Lombardo}}, \binits{U.}},
\bauthor{\bsnm{{Shen}}, \binits{C.W.}},
\bauthor{\bsnm{{Giai}}, \binits{N.V.}}:
\batitle{{From Brueckner approach to Skyrme-type energy density functional}}.
\bjtitle{\prc}
\bvolume{73}(\bissue{1}),
\bfpage{014313}
(\byear{2006})
\doiurl{10.1103/PhysRevC.73.014313}
\end{barticle}
\endbibitem

\bibitem[\protect\citeauthoryear{{Abrikosov}}{1988}]{Abrikosov}
\begin{bbook}
\bauthor{\bsnm{{Abrikosov}}, \binits{A.A.}}:
\bbtitle{{Fundamentals of the Theory of Metals}}.
\bpublisher{North Holland},
\blocation{The Netherlands}
(\byear{1988})
\end{bbook}
\endbibitem

\bibitem[\protect\citeauthoryear{{Carter} and {Chamel}}{2005}]{CarterChamel2005}
\begin{barticle}
\bauthor{\bsnm{{Carter}}, \binits{B.}},
\bauthor{\bsnm{{Chamel}}, \binits{N.}}:
\batitle{{Covariant Analysis of Newtonian Multi-Fluid Models for Neutron Stars Iii:}}.
\bjtitle{International Journal of Modern Physics D}
\bvolume{14}(\bissue{5}),
\bfpage{749}--\blpage{774}
(\byear{2005})
\doiurl{10.1142/S0218271805006845}
\end{barticle}
\endbibitem

\bibitem[\protect\citeauthoryear{Melatos and Millhouse}{2023}]{Melatos2023}
\begin{botherref}
\oauthor{\bsnm{Melatos}, \binits{A.}},
\oauthor{\bsnm{Millhouse}, \binits{M.}}:
Measuring the vortex-nucleus pinning force from pulsar glitch rates.
arXiv preprint arXiv:2302.11079
(2023)
\end{botherref}
\endbibitem

\bibitem[\protect\citeauthoryear{Pizzochero}{2011}]{pizzochero2011}
\begin{barticle}
\bauthor{\bsnm{Pizzochero}, \binits{P.M.}}:
\batitle{Angular momentum transfer in vela-like pulsar glitches}.
\bjtitle{\apjl}
\bvolume{743}(\bissue{1}),
\bfpage{20}
(\byear{2011})
\doiurl{10.1088/2041-8205/743/1/L20}
\end{barticle}
\endbibitem

\bibitem[\protect\citeauthoryear{P\c{e}cak et~al.}{2021}]{Pecak2021}
\begin{barticle}
\bauthor{\bsnm{P\c{e}cak}, \binits{D.}},
\bauthor{\bsnm{Chamel}, \binits{N.}},
\bauthor{\bsnm{Magierski}, \binits{P.}},
\bauthor{\bsnm{Wlaz\l{}owski}, \binits{G.}}:
\batitle{Properties of a quantum vortex in neutron matter at finite temperatures}.
\bjtitle{\prc}
\bvolume{104},
\bfpage{055801}
(\byear{2021})
\doiurl{10.1103/PhysRevC.104.055801}
\end{barticle}
\endbibitem

\bibitem[\protect\citeauthoryear{Klausner et~al.}{2023}]{klausner2023}
\begin{barticle}
\bauthor{\bsnm{Klausner}, \binits{P.}},
\bauthor{\bsnm{Barranco}, \binits{F.}},
\bauthor{\bsnm{Pizzochero}, \binits{P.M.}},
\bauthor{\bsnm{Roca-Maza}, \binits{X.}},
\bauthor{\bsnm{Vigezzi}, \binits{E.}}:
\batitle{Microscopic calculation of the pinning energy of a vortex in the inner crust of a neutron star}.
\bjtitle{\prc}
\bvolume{108},
\bfpage{035808}
(\byear{2023})
\doiurl{10.1103/PhysRevC.108.035808}
\end{barticle}
\endbibitem

\bibitem[\protect\citeauthoryear{{Wlaz{\l}owski} et~al.}{2016}]{wlazlowski2016}
\begin{barticle}
\bauthor{\bsnm{{Wlaz{\l}owski}}, \binits{G.}},
\bauthor{\bsnm{{Sekizawa}}, \binits{K.}},
\bauthor{\bsnm{{Magierski}}, \binits{P.}},
\bauthor{\bsnm{{Bulgac}}, \binits{A.}},
\bauthor{\bsnm{{Forbes}}, \binits{M.M.}}:
\batitle{{Vortex Pinning and Dynamics in the Neutron Star Crust}}.
\bjtitle{\prl}
\bvolume{117}(\bissue{23}),
\bfpage{232701}
(\byear{2016})
\doiurl{10.1103/PhysRevLett.117.232701}
\end{barticle}
\endbibitem

\bibitem[\protect\citeauthoryear{{Seveso} et~al.}{2016}]{seveso2016}
\begin{barticle}
\bauthor{\bsnm{{Seveso}}, \binits{S.}},
\bauthor{\bsnm{{Pizzochero}}, \binits{P.M.}},
\bauthor{\bsnm{{Grill}}, \binits{F.}},
\bauthor{\bsnm{{Haskell}}, \binits{B.}}:
\batitle{{Mesoscopic pinning forces in neutron star crusts}}.
\bjtitle{\mnras}
\bvolume{455}(\bissue{4}),
\bfpage{3952}--\blpage{3967}
(\byear{2016})
\doiurl{10.1093/mnras/stv2579}
\end{barticle}
\endbibitem

\bibitem[\protect\citeauthoryear{Link and Levin}{2022}]{link2022}
\begin{barticle}
\bauthor{\bsnm{Link}, \binits{B.}},
\bauthor{\bsnm{Levin}, \binits{Y.}}:
\batitle{Vortex pinning in neutron stars, slipstick dynamics, and the origin of spin glitches}.
\bjtitle{\apj}
\bvolume{941}(\bissue{2}),
\bfpage{148}
(\byear{2022})
\doiurl{10.3847/1538-4357/ac9b29}
\end{barticle}
\endbibitem

\bibitem[\protect\citeauthoryear{Zhang and Pethick}{2021}]{ZhaoWen2021}
\begin{barticle}
\bauthor{\bsnm{Zhang}, \binits{Z.-W.}},
\bauthor{\bsnm{Pethick}, \binits{C.J.}}:
\batitle{Proton superconductivity in pasta phases in neutron star crusts}.
\bjtitle{\prc}
\bvolume{103},
\bfpage{055807}
(\byear{2021})
\doiurl{10.1103/PhysRevC.103.055807}
\end{barticle}
\endbibitem

\bibitem[\protect\citeauthoryear{{Muslimov} and {Tsygan}}{1985}]{muslimov1985vortex}
\begin{barticle}
\bauthor{\bsnm{{Muslimov}}, \binits{A.G.}},
\bauthor{\bsnm{{Tsygan}}, \binits{A.I.}}:
\batitle{{Vortex lines in neutron star superfluids and decay of pulsar magnetic fields}}.
\bjtitle{\apss}
\bvolume{115},
\bfpage{43}--\blpage{49}
(\byear{1985})
\doiurl{10.1007/BF00653825}
\end{barticle}
\endbibitem

\bibitem[\protect\citeauthoryear{{Srinivasan} et~al.}{1990}]{srinivasan1990novel}
\begin{barticle}
\bauthor{\bsnm{{Srinivasan}}, \binits{G.}},
\bauthor{\bsnm{{Bhattacharya}}, \binits{D.}},
\bauthor{\bsnm{{Muslimov}}, \binits{A.G.}},
\bauthor{\bsnm{{Tsygan}}, \binits{A.J.}}:
\batitle{{A novel mechanism for the decay of neutron star magnetic fields}}.
\bjtitle{Curr. Sci.}
\bvolume{59},
\bfpage{31}--\blpage{38}
(\byear{1990})
\end{barticle}
\endbibitem

\bibitem[\protect\citeauthoryear{Sauls}{1989}]{sauls1989superfluidity}
\begin{bbook}
\bauthor{\bsnm{Sauls}, \binits{J.A.}}:
In: \beditor{\bsnm{{\"O}gelman}, \binits{H.}},
\beditor{\bsnm{Heuvel}, \binits{E.P.J.}} (eds.)
\bbtitle{Superfluidity in the Interiors of Neutron Stars}.
\bsertitle{NATO Advanced Science Institutes (ASI) Series C},
pp. \bfpage{457}--\blpage{490}.
\bpublisher{Springer},
\blocation{Dordrecht}
(\byear{1989})
\end{bbook}
\endbibitem

\bibitem[\protect\citeauthoryear{{Ruderman} et~al.}{1998}]{ruderman1998neutron}
\begin{barticle}
\bauthor{\bsnm{{Ruderman}}, \binits{M.}},
\bauthor{\bsnm{{Zhu}}, \binits{T.}},
\bauthor{\bsnm{{Chen}}, \binits{K.}}:
\batitle{{Neutron Star Magnetic Field Evolution, Crust Movement, and Glitches}}.
\bjtitle{\apj}
\bvolume{492},
\bfpage{267}--\blpage{280}
(\byear{1998})
\doiurl{10.1086/305026}
\end{barticle}
\endbibitem

\bibitem[\protect\citeauthoryear{{Alpar}}{2017}]{alpar2017}
\begin{barticle}
\bauthor{\bsnm{{Alpar}}, \binits{M.A.}}:
\batitle{{Flux-Vortex Pinning and Neutron Star Evolution}}.
\bjtitle{\japa}
\bvolume{38}(\bissue{3}),
\bfpage{44}
(\byear{2017})
\doiurl{10.1007/s12036-017-9473-6}
\end{barticle}
\endbibitem

\bibitem[\protect\citeauthoryear{{Baym} et~al.}{1969}]{baym1969}
\begin{barticle}
\bauthor{\bsnm{{Baym}}, \binits{G.}},
\bauthor{\bsnm{{Pethick}}, \binits{C.}},
\bauthor{\bsnm{{Pines}}, \binits{D.}}:
\batitle{{Superfluidity in Neutron Stars}}.
\bjtitle{\nat}
\bvolume{224}(\bissue{5220}),
\bfpage{673}--\blpage{674}
(\byear{1969})
\doiurl{10.1038/224673a0}
\end{barticle}
\endbibitem

\bibitem[\protect\citeauthoryear{{Charbonneau} and {Zhitnitsky}}{2007}]{charbonneau2007}
\begin{barticle}
\bauthor{\bsnm{{Charbonneau}}, \binits{J.}},
\bauthor{\bsnm{{Zhitnitsky}}, \binits{A.}}:
\batitle{{Novel mechanism for type I superconductivity in neutron stars}}.
\bjtitle{\prc}
\bvolume{76}(\bissue{1}),
\bfpage{015801}
(\byear{2007})
\doiurl{10.1103/PhysRevC.76.015801}
\end{barticle}
\endbibitem

\bibitem[\protect\citeauthoryear{{Alford} and {Good}}{2008}]{alford2008}
\begin{barticle}
\bauthor{\bsnm{{Alford}}, \binits{M.G.}},
\bauthor{\bsnm{{Good}}, \binits{G.}}:
\batitle{{Flux tubes and the type-I/type-II transition in a superconductor coupled to a superfluid}}.
\bjtitle{\prb}
\bvolume{78}(\bissue{2}),
\bfpage{024510}
(\byear{2008})
\doiurl{10.1103/PhysRevB.78.024510}
\end{barticle}
\endbibitem

\bibitem[\protect\citeauthoryear{{Wood} and {Graber}}{2022}]{wood2022}
\begin{barticle}
\bauthor{\bsnm{{Wood}}, \binits{T.S.}},
\bauthor{\bsnm{{Graber}}, \binits{V.}}:
\batitle{{Superconducting Phases in Neutron Star Cores}}.
\bjtitle{Universe}
\bvolume{8}(\bissue{4}),
\bfpage{228}
(\byear{2022})
\doiurl{10.3390/universe8040228}
\end{barticle}
\endbibitem

\bibitem[\protect\citeauthoryear{{Sourie} and {Chamel}}{2020a}]{Sourie2020}
\begin{barticle}
\bauthor{\bsnm{{Sourie}}, \binits{A.}},
\bauthor{\bsnm{{Chamel}}, \binits{N.}}:
\batitle{{Vortex pinning in the superfluid core of neutron stars and the rise of pulsar glitches}}.
\bjtitle{\mnras}
\bvolume{493}(\bissue{1}),
\bfpage{98}--\blpage{102}
(\byear{2020})
\doiurl{10.1093/mnrasl/slaa015}
\end{barticle}
\endbibitem

\bibitem[\protect\citeauthoryear{{Sourie} and {Chamel}}{2020b}]{SourieChamel2020KJ}
\begin{barticle}
\bauthor{\bsnm{{Sourie}}, \binits{A.}},
\bauthor{\bsnm{{Chamel}}, \binits{N.}}:
\batitle{{Generalization of the Kutta-Joukowski theorem for the hydrodynamic forces acting on a quantized vortex}}.
\bjtitle{International Journal of Modern Physics B}
\bvolume{34}(\bissue{10}),
\bfpage{2050099}--\blpage{137}
(\byear{2020})
\doiurl{10.1142/S021797922050099X}
{\href{https://arxiv.org/abs/2202.06011}{{arXiv:2202.06011}}}
{[cond-mat.quant-gas]}
\end{barticle}
\endbibitem

\bibitem[\protect\citeauthoryear{{Antonelli} and {Pizzochero}}{2017}]{Antonelli2017}
\begin{barticle}
\bauthor{\bsnm{{Antonelli}}, \binits{M.}},
\bauthor{\bsnm{{Pizzochero}}, \binits{P.M.}}:
\batitle{{Axially symmetric equations for differential pulsar rotation with superfluid entrainment}}.
\bjtitle{\mnras}
\bvolume{464}(\bissue{1}),
\bfpage{721}--\blpage{733}
(\byear{2017})
\doiurl{10.1093/mnras/stw2376}
\end{barticle}
\endbibitem

\bibitem[\protect\citeauthoryear{{Carter} et~al.}{2006}]{carterchamelhaensel2006}
\begin{barticle}
\bauthor{\bsnm{{Carter}}, \binits{B.}},
\bauthor{\bsnm{{Chamel}}, \binits{N.}},
\bauthor{\bsnm{{Haensel}}, \binits{P.}}:
\batitle{{Entrainment Coefficient and Effective Mass for Conduction Neutrons in Neutron Star Crust:. Macroscopic Treatment}}.
\bjtitle{International Journal of Modern Physics D}
\bvolume{15}(\bissue{5}),
\bfpage{777}--\blpage{803}
(\byear{2006})
\doiurl{10.1142/S0218271806008504}
\end{barticle}
\endbibitem

\bibitem[\protect\citeauthoryear{{Chamel}}{2012}]{chamel2012}
\begin{barticle}
\bauthor{\bsnm{{Chamel}}, \binits{N.}}:
\batitle{{Neutron conduction in the inner crust of a neutron star in the framework of the band theory of solids}}.
\bjtitle{\prc}
\bvolume{85}(\bissue{3}),
\bfpage{035801}
(\byear{2012})
\doiurl{10.1103/PhysRevC.85.035801}
\end{barticle}
\endbibitem

\bibitem[\protect\citeauthoryear{{Carter} et~al.}{2005}]{CarterChamelhaensel2005b}
\begin{barticle}
\bauthor{\bsnm{{Carter}}, \binits{B.}},
\bauthor{\bsnm{{Chamel}}, \binits{N.}},
\bauthor{\bsnm{{Haensel}}, \binits{P.}}:
\batitle{{Effect of BCS pairing on entrainment in neutron superfluid current in neutron star crust}}.
\bjtitle{\nphysa}
\bvolume{759},
\bfpage{441}--\blpage{464}
(\byear{2005})
\end{barticle}
\endbibitem

\bibitem[\protect\citeauthoryear{Miller et~al.}{2007}]{miller2007}
\begin{barticle}
\bauthor{\bsnm{Miller}, \binits{D.E.}},
\bauthor{\bsnm{Chin}, \binits{J.K.}},
\bauthor{\bsnm{Stan}, \binits{C.A.}},
\bauthor{\bsnm{Liu}, \binits{Y.}},
\bauthor{\bsnm{Setiawan}, \binits{W.}},
\bauthor{\bsnm{Sanner}, \binits{C.}},
\bauthor{\bsnm{Ketterle}, \binits{W.}}:
\batitle{Critical velocity for superfluid flow across the bec-bcs crossover}.
\bjtitle{\prl}
\bvolume{99},
\bfpage{070402}
(\byear{2007})
\doiurl{10.1103/PhysRevLett.99.070402}
\end{barticle}
\endbibitem

\bibitem[\protect\citeauthoryear{Page and Reddy}{2013}]{Page2013}
\begin{barticle}
\bauthor{\bsnm{Page}, \binits{D.}},
\bauthor{\bsnm{Reddy}, \binits{S.}}:
\batitle{Forecasting neutron star temperatures: Predictability and variability}.
\bjtitle{\prl}
\bvolume{111},
\bfpage{241102}
(\byear{2013})
\doiurl{10.1103/PhysRevLett.111.241102}
\end{barticle}
\endbibitem

\bibitem[\protect\citeauthoryear{{Douchin} and {Haensel}}{2001}]{douchinhaensel01}
\begin{barticle}
\bauthor{\bsnm{{Douchin}}, \binits{F.}},
\bauthor{\bsnm{{Haensel}}, \binits{P.}}:
\batitle{{A unified equation of state of dense matter and neutron star structure}}.
\bjtitle{\aap}
\bvolume{380},
\bfpage{151}--\blpage{167}
(\byear{2001})
\doiurl{10.1051/0004-6361:20011402}
\end{barticle}
\endbibitem

\bibitem[\protect\citeauthoryear{{Levenfish} and {Yakovlev}}{1994}]{YakovlevandLevenfish1994}
\begin{barticle}
\bauthor{\bsnm{{Levenfish}}, \binits{K.P.}},
\bauthor{\bsnm{{Yakovlev}}, \binits{D.G.}}:
\batitle{{Specific heat of neutron star cores with superfluid nucleons}}.
\bjtitle{\arep}
\bvolume{38}(\bissue{2}),
\bfpage{247}--\blpage{251}
(\byear{1994})
\end{barticle}
\endbibitem

\bibitem[\protect\citeauthoryear{Schwenk et~al.}{2003}]{Schwenk2003}
\begin{barticle}
\bauthor{\bsnm{Schwenk}, \binits{A.}},
\bauthor{\bsnm{Friman}, \binits{B.}},
\bauthor{\bsnm{Brown}, \binits{G.E.}}:
\batitle{Renormalization group approach to neutron matter: quasiparticle interactions, superfluid gaps and the equation of state}.
\bjtitle{\nphysa}
\bvolume{713}(\bissue{1}),
\bfpage{191}--\blpage{216}
(\byear{2003})
\doiurl{10.1016/S0375-9474(02)01290-3}
\end{barticle}
\endbibitem

\bibitem[\protect\citeauthoryear{Brown et~al.}{2002}]{brown2002}
\begin{barticle}
\bauthor{\bsnm{Brown}, \binits{E.F.}},
\bauthor{\bsnm{Bildsten}, \binits{L.}},
\bauthor{\bsnm{Chang}, \binits{P.}}:
\batitle{Variability in the thermal emission from accreting neutron star transients}.
\bjtitle{\apj}
\bvolume{574}(\bissue{2}),
\bfpage{920}
(\byear{2002})
\doiurl{10.1086/341066}
\end{barticle}
\endbibitem

\bibitem[\protect\citeauthoryear{Gudmundsson et~al.}{1983}]{gudmundsson1983}
\begin{barticle}
\bauthor{\bsnm{Gudmundsson}, \binits{E.H.}},
\bauthor{\bsnm{Pethick}, \binits{C.}},
\bauthor{\bsnm{Epstein}, \binits{R.I.}}:
\batitle{Structure of neutron star envelopes}.
\bjtitle{Astrophysical Journal, Part 1 (ISSN 0004-637X), vol. 272, Sept. 1, 1983, p. 286-300.}
\bvolume{272},
\bfpage{286}--\blpage{300}
(\byear{1983})
\doiurl{10.1086/161292}
\end{barticle}
\endbibitem

\bibitem[\protect\citeauthoryear{{Potekhin} et~al.}{1997}]{potekhin1997}
\begin{barticle}
\bauthor{\bsnm{{Potekhin}}, \binits{A.Y.}},
\bauthor{\bsnm{{Chabrier}}, \binits{G.}},
\bauthor{\bsnm{{Yakovlev}}, \binits{D.G.}}:
\batitle{{Internal temperatures and cooling of neutron stars with accreted envelopes.}}
\bjtitle{\aap}
\bvolume{323},
\bfpage{415}--\blpage{428}
(\byear{1997})
\doiurl{10.48550/arXiv.astro-ph/9706148}
\end{barticle}
\endbibitem

\bibitem[\protect\citeauthoryear{{Cumming} et~al.}{2017}]{cumming2017}
\begin{barticle}
\bauthor{\bsnm{{Cumming}}, \binits{A.}},
\bauthor{\bsnm{{Brown}}, \binits{E.F.}},
\bauthor{\bsnm{{Fattoyev}}, \binits{F.J.}},
\bauthor{\bsnm{{Horowitz}}, \binits{C.J.}},
\bauthor{\bsnm{{Page}}, \binits{D.}},
\bauthor{\bsnm{{Reddy}}, \binits{S.}}:
\batitle{{Lower limit on the heat capacity of the neutron star core}}.
\bjtitle{\prc}
\bvolume{95}(\bissue{2}),
\bfpage{025806}
(\byear{2017})
\doiurl{10.1103/PhysRevC.95.025806}
\end{barticle}
\endbibitem

\bibitem[\protect\citeauthoryear{{Brown} et~al.}{2018}]{brown2018}
\begin{barticle}
\bauthor{\bsnm{{Brown}}, \binits{E.F.}},
\bauthor{\bsnm{{Cumming}}, \binits{A.}},
\bauthor{\bsnm{{Fattoyev}}, \binits{F.J.}},
\bauthor{\bsnm{{Horowitz}}, \binits{C.J.}},
\bauthor{\bsnm{{Page}}, \binits{D.}},
\bauthor{\bsnm{{Reddy}}, \binits{S.}}:
\batitle{{Rapid Neutrino Cooling in the Neutron Star MXB 1659-29}}.
\bjtitle{\prl}
\bvolume{120}(\bissue{18}),
\bfpage{182701}
(\byear{2018})
\doiurl{10.1103/PhysRevLett.120.182701}
\end{barticle}
\endbibitem

\bibitem[\protect\citeauthoryear{Deibel et~al.}{2015}]{Deibel2015}
\begin{barticle}
\bauthor{\bsnm{Deibel}, \binits{A.}},
\bauthor{\bsnm{Cumming}, \binits{A.}},
\bauthor{\bsnm{Brown}, \binits{E.F.}},
\bauthor{\bsnm{Page}, \binits{D.}}:
\batitle{A strong shallow heat source in the accreting neutron star {MAXI} {J}0556-332}.
\bjtitle{\apj}
\bvolume{809}(\bissue{2}),
\bfpage{31}
(\byear{2015})
\doiurl{10.1088/2041-8205/809/2/l31}
\end{barticle}
\endbibitem

\bibitem[\protect\citeauthoryear{{Iaria} et~al.}{2018}]{Iaria2018}
\begin{barticle}
\bauthor{\bsnm{{Iaria}}, \binits{R.}},
\bauthor{\bsnm{{Gambino}}, \binits{A.F.}},
\bauthor{\bsnm{{Di Salvo}}, \binits{T.}},
\bauthor{\bsnm{{Burderi}}, \binits{L.}},
\bauthor{\bsnm{{Matranga}}, \binits{M.}},
\bauthor{\bsnm{{Riggio}}, \binits{A.}},
\bauthor{\bsnm{{Sanna}}, \binits{A.}},
\bauthor{\bsnm{{Scarano}}, \binits{F.}},
\bauthor{\bsnm{{D'A{\`\i}}}, \binits{A.}}:
\batitle{{A possible solution of the puzzling variation of the orbital period of MXB 1659-298}}.
\bjtitle{\mnras}
\bvolume{473}(\bissue{3}),
\bfpage{3490}--\blpage{3499}
(\byear{2018})
\doiurl{10.1093/mnras/stx2529}
\end{barticle}
\endbibitem

\bibitem[\protect\citeauthoryear{{Ootes} et~al.}{2016}]{ootes2016}
\begin{barticle}
\bauthor{\bsnm{{Ootes}}, \binits{L.S.}},
\bauthor{\bsnm{{Page}}, \binits{D.}},
\bauthor{\bsnm{{Wijnands}}, \binits{R.}},
\bauthor{\bsnm{{Degenaar}}, \binits{N.}}:
\batitle{{Neutron star crust cooling in KS 1731-260: the influence of accretion outburst variability on the crustal temperature evolution}}.
\bjtitle{\mnras}
\bvolume{461}(\bissue{4}),
\bfpage{4400}--\blpage{4405}
(\byear{2016})
\doiurl{10.1093/mnras/stw1799}
\end{barticle}
\endbibitem
\end{thebibliography}
\end{document}